\documentclass[review,11pt,3p]{elsarticle}
\usepackage{amsmath} 
\usepackage{amssymb}
\usepackage{textcomp}
\usepackage{graphicx}		
\usepackage{float}
\usepackage{color}
\usepackage{url}
\usepackage{array}
\usepackage{booktabs,ctable,multirow}
\usepackage{bm}
\usepackage[T1]{fontenc}

\DeclareFontFamily{U}{msb}{}
\DeclareFontShape{U}{msb}{m}{n}{ <5> <6> <7> <8> <9> gen * msbm
  <10> <10.95> <12> <14.4> <17.28> <20.74> <24.88> msbm10}{} 
\DeclareSymbolFont{AMSb}{U}{msb}{m}{n}
\DeclareMathSymbol{\E}{\mathalpha}{AMSb}{"45}
\DeclareMathSymbol{\I}{\mathalpha}{AMSb}{"49}
\DeclareMathSymbol{\N}{\mathalpha}{AMSb}{"4E}
\DeclareMathSymbol{\R}{\mathalpha}{AMSb}{"52}
\DeclareMathSymbol{\Z}{\mathalpha}{AMSb}{"5A}
\DeclareRobustCommand{\ihat}{{\mathpalette\rothat\relax}}
\newcommand{\rothat}[2]{\rotatebox[origin=c]{90}{$#1\!\!\bm{\langle}$}}
\def\widecheck#1{\overset{\ihat}{#1}}
\newcommand{\red}{\color[rgb]{1,0,0}}
\newcommand{\blue}{\color[rgb]{0,0,1}}
\newcommand{\SF}{\bfseries \sffamily}

\newcommand{\bn}{\bm{n}}

\newcommand{\bx}{\bm{x}}

\newcommand{\bq}{\bm{q}}
\newcommand{\bD}{\bm{D}}
\newcommand{\bH}{\bm{H}}
\newcommand{\bI}{\bm{I}}
\newcommand{\bL}{\bm{L}}

\newcommand{\bu}{\bm{u}}

\newcommand{\bW}{\bm{W}}

\newcommand{\bfi}{\bm{\phi}}
\newcommand{\cfi}{\widecheck{\bm{\phi}}}
\newcommand{\hfi}{\widehat{\bm{\phi}}}
\newcommand{\tfi}{\widetilde{\bm{\phi}}}
\newcommand{\ftf}{{\cal F}{\bm{\phi}}}
\newcommand{\ftt}{{\cal F}\widetilde{\bm{\phi}}}

\newcommand{\hu}{{\widehat{\bu}}}
\newcommand{\tu}{{\widetilde{\bu}}}

\newcommand{\barre}{\rule[5mm]{\textwidth}{0.5pt}\\\vspace*{-10mm}~~}%

\DeclareMathOperator \vol {vol}

\newtheorem{definition}{Definition}
\begin{document}
%
%
\begin{frontmatter}
  \title{Perturbation of the Eigenvectors of the Graph Laplacian:\\
    Application to Image Denoising}
  \author[add1]{F.G. Meyer\corref{cor1}}
  \ead{E-mail: fmeyer@colorado.edu}
  \ead[url]{http://ecee.colorado.edu/~fmeyer}
  \cortext[cor1]{Corresponding author}
  \author[add2]{X. Shen}
  \address[add1]{Department of Electrical Engineering, University of Colorado at
    Boulder, Boulder, CO}
  \address[add2]{Department of Diagnostic Radiology, Yale University, CT}
  \begin{abstract}
    The original contributions of this paper are twofold: a new
    understanding of the influence of noise on the eigenvectors of the
    graph Laplacian of a set of image patches, and an algorithm to
    estimate a denoised set of patches from a noisy image. The
    algorithm relies on the following two observations: (1) the
    low-index eigenvectors of the diffusion, or graph Laplacian,
    operators are very robust to random perturbations of the weights
    and random changes in the connections of the patch-graph; and (2)
    patches extracted from smooth regions of the image are organized
    along smooth low-dimensional structures in the patch-set, and therefore
    can be reconstructed with few eigenvectors. Experiments
    demonstrate that our denoising algorithm outperforms the denoising
    gold-standards.
  \end{abstract}
\end{frontmatter}
\section{Introduction}
\subsection{Problem statement and motivation}
The first goal of this work is to study experimentally the
perturbation of the eigenvectors of the graph Laplacian constructed
from a set of image patches. The second goal of the paper is to devise
a novel method to jointly denoise an image and estimate the
eigenvectors of the  patch-graph of the clean image.

Recent work in computer vision and image processing indicates that the
elusive quest for the ``universal'' transform has been replaced by a
fresh new perspective.  Indeed, researchers have recently proposed to
represent images as ``collage'' of small patches. The patches can be
shaped into square blocks \citep{Cheung08} or into optimized contours
that mimic the parts of a jigsaw puzzle \citep{Kannan07}. These
patch-based appearance models are generative statistical models that
can be learned from images, and are then used to represent images.
All the patch-based methods \citep[e.g.,][and references
therein]{Buades05,Gilboa08,Dabov09,Szlam08,Bougleux09,Peyre08,Taylor11b,zontak11}
take advantage of the following fact: the dataset of patches, whether
it is aggregated from a single image, or a library of images, is a
smooth set.  A consequence of this observation is that the
eigenvectors of the graph Laplacian \cite{Belkin03,Coifman06b} provide
a natural basis to expand smooth functions defined on the set of
patches. The image intensity -- seen as a function on the patch-set --
is obviously a smooth function. As a result, the eigenvectors of the
graph Laplacian yield an efficient (measured in terms of sparsity, for
instance) representation of the original image from which the patch-set is
computed.

In practice, images are often corrupted by noise, and the eigenvectors
are to be computed from a graph of noisy patches.  The problem
becomes: what is the effect of the noise on the perturbation of the
eigenvectors? Theoretical results \citep{Stewart90} provide upper
bounds on the angle between the original and the perturbed
eigenspaces. Unfortunately, as noted by \citet{Yan09}, in the context
of the perturbation of the first non trivial eigenvector of the graph
Laplacian, these bounds usually overestimate the actual
perturbation. In addition, the bounds depend on the separation of the
eigenvalues, a quantity that is difficult to predict for image
patches. Finally, bounds on the angle between invariant subspaces
cannot be readily translated in terms of the effect of the
perturbations on the geometric features encoded by the eigenvectors,
or the ability of the perturbed eigenvectors to approximate the
original image.
\subsection{Outline of our approach and results}
The original contributions of this paper are twofold: a new
understanding of the influence of noise on the eigenvectors of the
graph Laplacian and a new method to denoise image patches.

Our approach relies on a series of experiments aimed at understanding
the effect of noise on the perturbations of the eigenvectors of the
patch-graph. In particular, we show that the low order eigenvectors of
the graph Laplacian are robust to random variations in the geometry of
the graph and in the edge weights. These results are connected to
similar results in spectral geometry quantifying the perturbations of
the eigenfunctions of the Laplace-Beltrami operator as a function of changes
in the metric of the manifold \cite{Barbatis96,Davies90}. Equipped
with this experimental understanding of the robustness of the low
index eigenvectors of the graph Laplacian, we propose an iterative
procedure to jointly estimate the eigenvectors and the original image.
We evaluate our novel approach with numerous experiments on a variety
of images, and we compare our approach to some of the denoising
gold-standards. Our approach systematically outperforms these
algorithms.
\section{The manifold of patches from a single image}
We consider an image $u(\bx)$, of size $N \times N$.  We extend the
image by even symmetry when the pixel location $\bx=(i,j)$ becomes close to the border of the
image. We first define the notion of a {\em patch}.
\begin{definition}
  Let $\bx_n=(i,j)$ be a pixel with linear index $n=i \times N + j$. We
  extract  an $m\times m$ block, centered about $\bx_n$,
  \begin{equation}
    \begin{bmatrix}
      u (i-m/2,j - m/2) & \cdots & u (i-m/2,j+m/2)\\
      \vdots & & \vdots\\
      u (i+m/2,j - m/2) & \cdots & u (i+m/2,j+m/2)\\
    \end{bmatrix},
    \label{patch0}
  \end{equation}
where, without loss of generality,  we take $m$ to be an odd integer,
and $m/2$ is the result of the Euclidean division of $m$ by 2.
  By concatenating the columns, we identify the $m \times m$ matrix
  (\ref{patch0}) with a vector in $\R^{m^2}$, and we define the {\em
    patch} $\bu(\bx_n)$ by
  \begin{equation}
    \bu(\bx_n) = 
    \begin{bmatrix}
      u_1(\bx_n)\\
      \vdots\\
      u_{m^2}(\bx_n)
    \end{bmatrix} 
    =
    \begin{bmatrix}
      u (i-m/2,j - m/2) \\
      \vdots \\
      u (i+m/2,j+m/2)\\
    \end{bmatrix}.
    \label{patch}
  \end{equation}
\end{definition} 
As we collect all the patches, we form the {\em patch-set} in $\R^{m^2}$.
\begin{definition}
  The  {\em patch-set} is defined as the set of patches extracted from
  the image  $u$,
  \begin{equation}
    {\cal P}=\{\bu(\bx_n), n = 1,2,\ldots, N^2\}.
    \label{patchset}
  \end{equation}
\end{definition} 
Before we start exploring the effect of noise on the patch-set $\cal
P$, let us pause for an instant and ask ourselves if $\cal P$ is a
smooth set. This question can be answered by ignoring for a moment the
spatial sampling of the image that leads to pixelization. We imagine
that the image intensity is a smooth function $u(\bx)$ defined for $
\bx \in [0,1] \times [0,1]$. In this case, $\cal P$ is clearly a
two-dimensional smooth manifold. Even though the image is a function
of a continuous argument $\bx$, the patch $\bu(\bx)$ is formed by
collecting $m^2$ discrete samples on a lattice. Let $\Delta$ be the
horizontal, or vertical, lattice resolution. For any pair of
translation indices $k,l= -m/2,\ldots,m/2$, we consider the function
that maps $\bx$ to the $\left((l+m/2) + m (k+m/2) +
  1\right)^\text{th}$ coordinate of the patch $\bu(\bx)$,
\begin{equation}
  \bx \longmapsto
u\left(\bx + 
  \begin{bmatrix}
    k\Delta\\ 
    l\Delta
  \end{bmatrix}
  \right).
\end{equation}
This map is a smooth function from $[0,1]\times [0,1]$ to $\R$, since $u$ is 
smooth. Therefore, each of the coordinates of $\bu(\bx)$ is a
smooth function of $\bx$, and the map that assigns a pixel to its patch
\begin{equation}
  \bx \longrightarrow \bu(\bx)
\end{equation}
is a smooth map from $[0,1] \times [0,1]$ to $\R^{m^2}$. The
non-discretized version of $\cal P$ is thus a two-dimensional manifold in
$\R^{m^2}$. We note that \citet{Grimmes05} have argued that a set of
binary images obtained by moving a black object on a white background
forms a sub-manifold that is not differentiable. The lack of
differentiability hinges on the fact that binary images can only have
step edges that are not differentiable. In fact, smoothing the images
allows one to recover the differentiability of the sub-manifold. In
this work, we assume that the image $u$ has been blurred by the acquisition
device, and therefore $u(\bx)$ is a smooth function of $\bx$.

\begin{figure}[H]
  \centerline{
    \includegraphics[width=17pc]{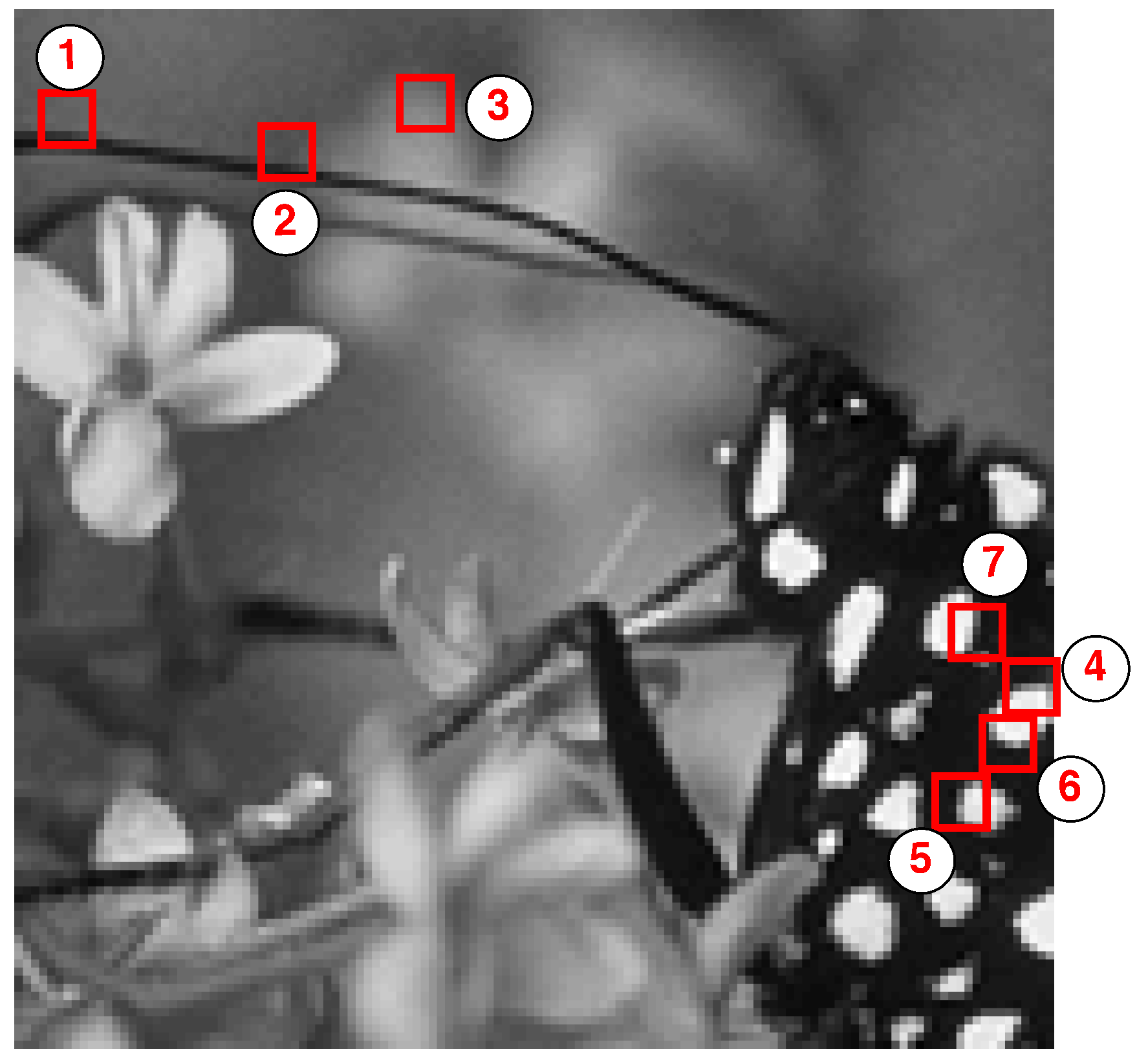}
    \includegraphics[width=20pc]{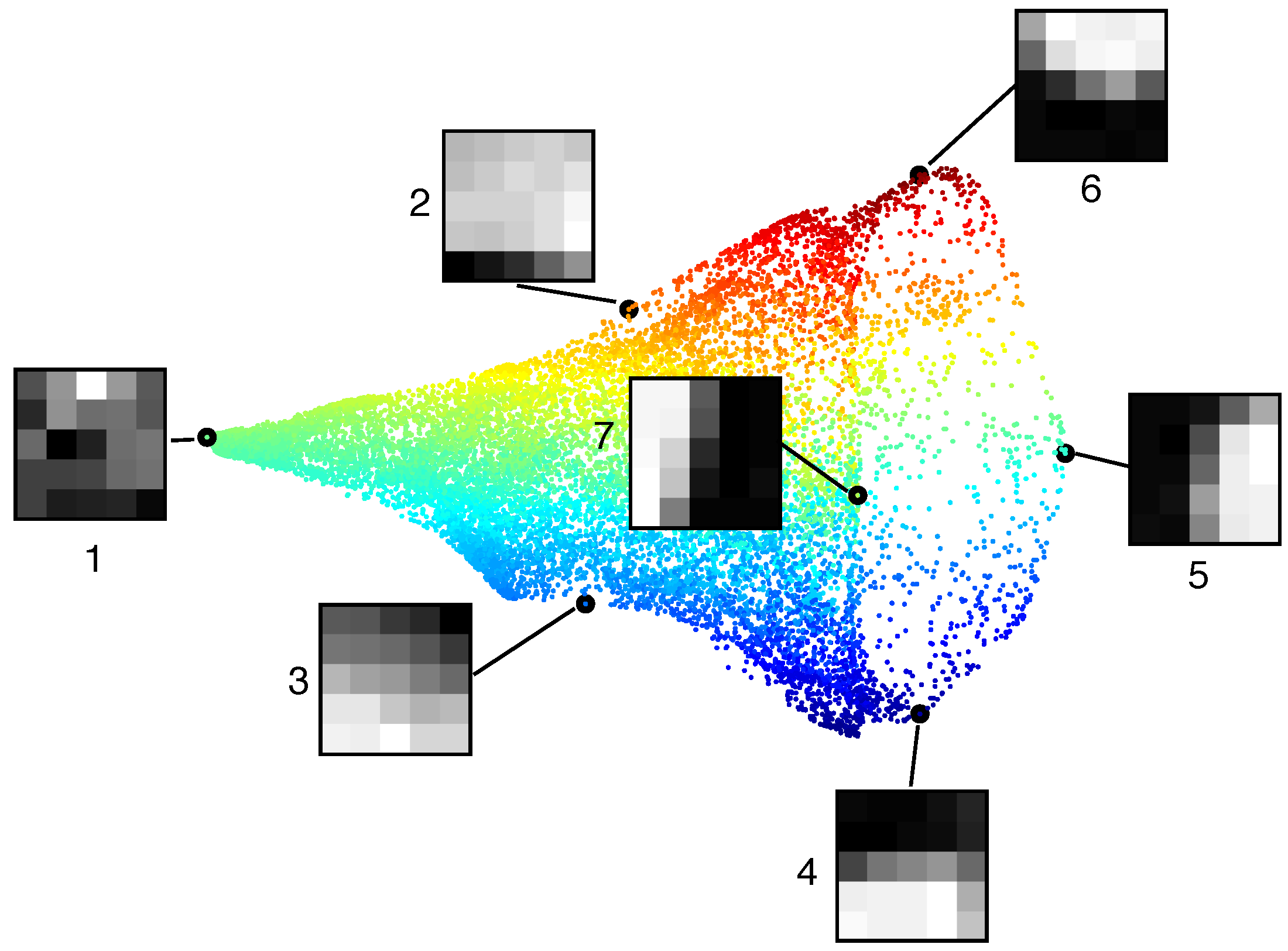}
  } 
  \caption{Manifold of $ 5 \times 5$ patches (right) extracted from the
    $128 \times 128$ butterfly image (left). Each dot represents a
    single patch $\bu(\bx)$. A few patches are labeled in the image and
    on the patch manifold to assist with the interpretation of the manifold
    in terms of the geometric content of the patches.
    \label{patchbutter}}
\end{figure}

We provide an illustration of the patch submanifold $\cal P$ in
Fig.~\ref{patchbutter}, where $5 \times 5$ patches, extracted from the
$128 \times 128$ butterfly image (Fig.~\ref{patchbutter}-left), form a
smooth two-dimensional structure (Fig.~\ref{patchbutter}-right). The
cone-shaped surface encodes the content of the patches: as we move
along the curve formed by the opening of the cone in
Fig.~\ref{patchbutter}-right, we explore patches (numbered 4, 5, 6 and
7) that contain high-contrast edges at different orientations. Each
ruling (generator) of the cone is composed of all the patches that
contain an edge with the same orientation. In effect, the
orientation of the edge inside these patches is encoded by the position of
the ruling on the opening curve. In addition, the white part of the
patch is either always at the top, or always at the bottom (see
e.g. $4\rightarrow 3 \rightarrow 1$ and $6\rightarrow 2 \rightarrow
1$). Finally, as we slide from the opening of the cone to its tip, the
contrast across the edge decreases (see e.g. $4\rightarrow 3
\rightarrow 1$ and $6\rightarrow 2 \rightarrow 1$).
\subsection{Denoising patch datasets}
\begin{figure}[H]
  \centerline{  
    \includegraphics[width= 0.40\textwidth]{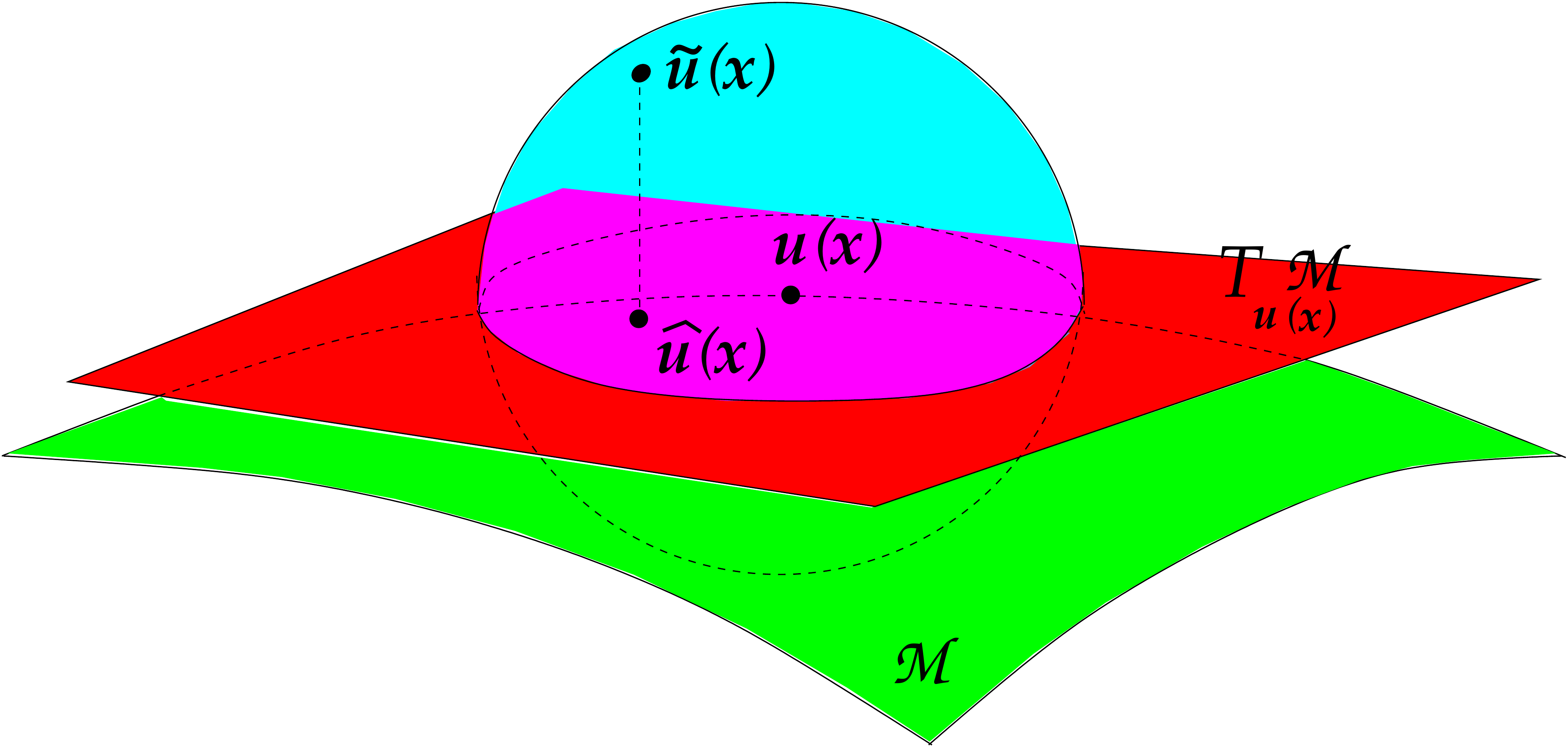}
  }
  \caption{
    Local denoising along the tangent plane of the patch-set. The clean
    patch $\bu(\bx)$ is corrupted by isotropic Gaussian noise. The noisy
    measurement $\tu(\bx)$ is inside a ball centered around $\bu(\bx)$.  An
    estimate of $\bu(\bx)$, $\hu(\bx)$,  is computed by projecting $\tu(\bx)$ on
    the tangent plane to the manifold $T_{\bu(\bx)}{\cal M}$.  
    \label{tangent}}
\end{figure}

We now consider the problem central to this paper: denoising a patch
dataset that lies close to a manifold $\cal M$. We review the two broad
classes of approaches that have been proposed to solve%
\noindent the more general problem of denoising a manifold-valued dataset:
\begin{itemize}
\item local methods that construct a local projector, which varies
  from point to point, and project the noisy measurements on a local
  estimate of the tangent plane $T_{\bu(\bx)}{\cal M}$ (see Fig.\ref{tangent});
\item global methods that denoise the entire dataset. Linear and
  nonlinear methods can be used to reconstruct an estimate of the
  clean manifold.
\end{itemize}
{\noindent \bfseries Local methods: denoising along the tangent
  plane.} The concept of image patches is equivalent to the concept of
{\em time-delay coordinates} in the context of the analysis of a
dynamical system from the time series generated by an observable
\cite{Gilmore98}. Several methods have been proposed to remove the
noise from time-delay embedded data \citep[e.g.,][and references
therein]{Kostelich93}. These methods rely on the estimation of a local
coordinate system formed by the tangent plane to the manifold at
$\bu(\bx)$ (see Fig.~\ref{tangent}). The tangent plane is computed
from the singular value decomposition of a neighborhood of $\bu(\bx)$;
the first singular vectors provide a basis of the local tangent plane
$T_{\bu(\bx)}{\cal M}$ (see Fig.~\ref{tangent}), while the remaining
vectors capture the noise \citep[e.g.,][and references
therein]{Kaslovsky11b}. The noise, which is $m^2$ dimensional, is
squashed by projecting the noisy data onto the tangent plane (see
Fig.~\ref{tangent}). Similar ideas have been proposed recently in the
image processing literature.  The BM3D algorithm \cite{Dabov09} uses
the local patches around a noisy reference patch $\bu(\bx)$ to
construct a local%
\noindent orthonormal basis. A multiscale version of this concept was
recently proposed to denoise curves \cite{feiszli11}. Elad and
co-workers proposed the k-SVD algorithm \cite{Aharon06b} that
assembles multiple singular vectors, computed locally, in order to
construct a global set of functions to represent an image.

{\noindent \bfseries Global methods: diffusion on the manifold.}
Instead of trying to approximate locally the manifold of patches by
the tangent plane $T_{\bu(\bx)}{\cal M}$ around each patch $\bu(\bx)$,
global methods perform a comprehensive denoising on the entire manifold. The
global denoising can be performed by applying a smoothing operator,
defined by a kernel $H$, to the noisy dataset and compute an estimate
$\hu(\bx)$ of $\bu(\bx)$ from the noisy measurement $\tu(\bx)$,
\begin{equation}
  \hu = \int_{\cal M} H(\tu,\bq) \: \bq \: d\mu(\bq).
  \label{diffusion}
\end{equation}
In practice, the Gaussian kernel is usually used, and the geodesic
distance on the manifold is replaced by the Euclidean distance.  The
recent nonlocal means algorithm \cite{Buades05} can be recast as a
discrete implementation of (\ref{diffusion}). \citet{Szlam08} have
explicitly modeled the manifold of patches with a graph and
implemented various diffusions on the graph. \citet{Singer09a}
provided a stochastic interpretation of nonlocal means as a diffusion
process. Similar ideas were proposed in
\cite{Hein06,Bougleux09,Dabov09}. \citet{Gilboa08} proposed a
continuous version of this idea, and defined PDE-based flows that
depend on local neighborhoods.  Early on, the computer graphics
community has been implementing discrete versions of 
(\ref{diffusion}) to smooth and denoise surface meshes
\citep[e.g.,][and references therein]{Taubin95}. Specifically,
\citet{Taubin95} proposed to implement (\ref{diffusion}) using the
diffusion
\begin{equation}
  \frac{\partial \bu(\bx,t)}{\partial t} = \lambda {\cal L} \bu(\bx,t),
\label{flow}
\end{equation}
where ${\cal L}$  is a discrete approximation to the Laplacian operator defined on the
mesh.

{\noindent \bfseries Global methods: from the diffusion operator to
  its eigenvectors.} Instead of applying the diffusion kernel $H$ on
the graph of patches, one can compute the eigenvectors $\bfi_k$ of the
graph Laplacian \cite{Belkin03,Coifman06b}, and use them to expand the
corresponding diffusion kernel $H$. The set of eigenvectors provides a
spectral decomposition, similar to a global Fourier basis. If we keep
only the first few eigenvectors of the expansion, the result is very
similar to applying the diffusion operator $H$ for a long period of
time (see section \ref{diffuse}). In his seminal work,
\citet{Taubin95} proposed to use the eigenvectors of the graph
Laplacian to denoise meshes. \citet{Szlam08} proposed to perform some
nonlinear approximation using the eigenvectors of the graph
Laplacian. \citet{Peyre08} combined thresholding and regularization to
remove noise from images.

In our problem, the diffusion operator $H$ needs to be estimated from
the noisy data. Consequently, we expect that the eigenvectors of the
noisy operator $\widetilde{H}$ may be very different from the
eigenvectors of the true operator $H$. We intend to study the
perturbation of the eigenvectors, and propose a method to denoise
jointly the eigenvectors and the image.
\section{The eigenvectors: a sparse code for images} 
\subsection{The graph of patches: the patch-graph
  \label{graphpatches}}
\begin{figure}[H]
  \centerline{
    \includegraphics[width=26pc]{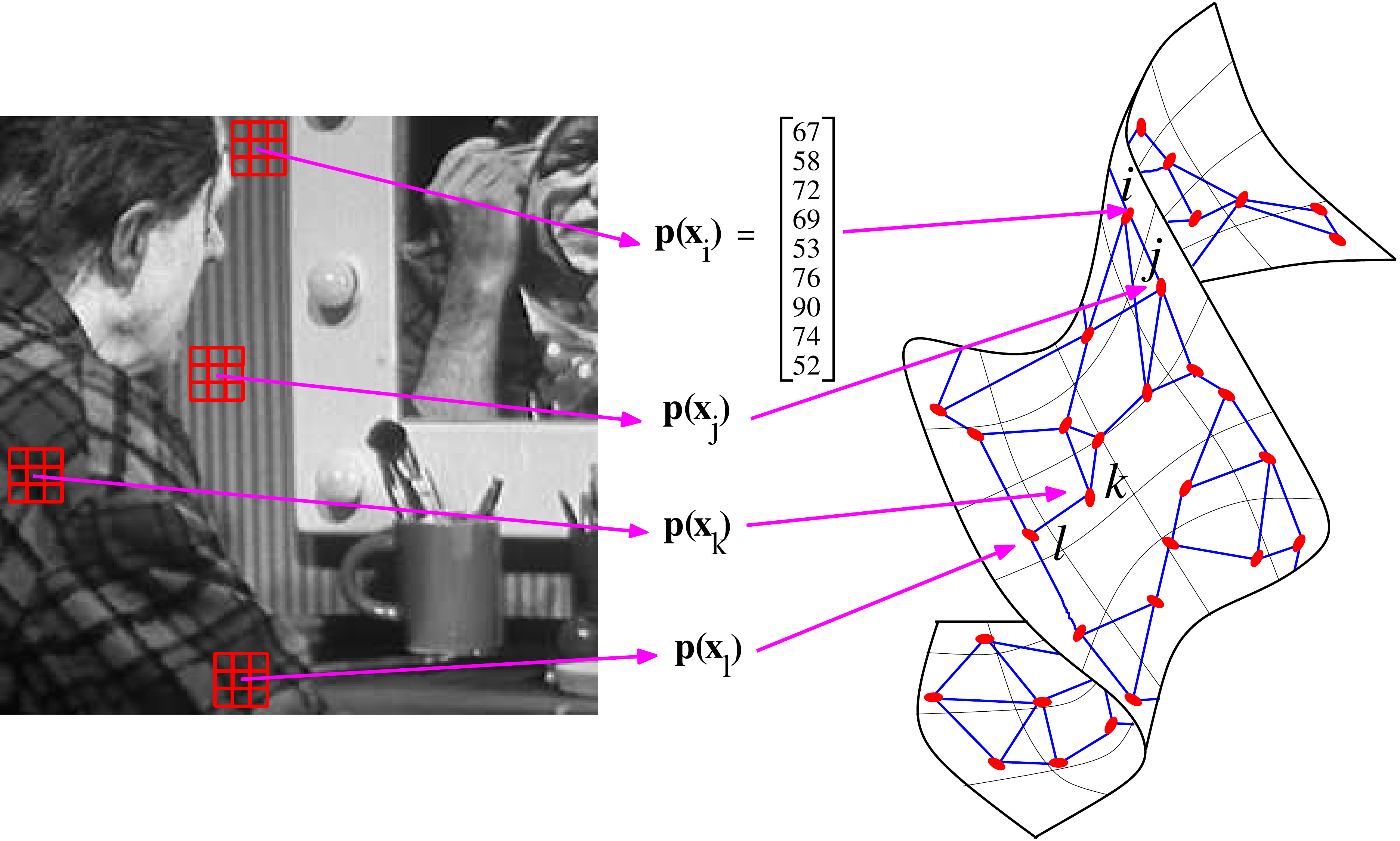}}
  \caption{The $3 \times 3 $ patches $\bu(\bx)$ are identified as  vectors in
    $\R^9$. The patch-graph  encodes the similarities between
    different regions of the images at a given scale. 
    \label{network}}
\end{figure}

In this paper, we think about a patch, $\bu(\bx_n)$, in
several different ways. Originally, $\bu(\bx_n)$ is simply a block of
an image. We also think about $\bu(\bx_n)$ as a point in
$\R^{m^2}$. Finally, we also regard $\bu(\bx)$ as a vertex of a
graph. Throughout this work, we will use these three perspectives. In order to
study the discrete structure formed by the patch-set (\ref{patchset}),
we connect patches to their nearest neighbors, and construct a graph
that we call the {\em patch-graph} (see Fig. \ref{network}).

\begin{definition}
  The {\em patch-graph}, $\Gamma$, is a weighted graph defined as follows.
  \begin{enumerate}
  \item The vertices of $\Gamma$ are the $N^2$ patches $\bu(\bx_n), n
    =1,\ldots, N^2$.
  \item Each vertex $\bu(\bx_n)$ is connected to its $\nu$ nearest
    neighbors using the metric
    \begin{equation}
      d(n,m) = 
      \| \bu(\bx_n) - \bu(\bx_m)\| +
      \beta \| \bx_n -\bx_m\|.
      \label{distance}
    \end{equation}
  \item The weight $w_{n,m}$ along the edge
    $\{\bu(\bx_n),\bu(\bx_m)\}$ is given by
    \begin{equation}
      w_{n,m}= 
      \begin{cases}
        e^{\displaystyle -d^2(n,m)/\delta^2} & \text{ if $\bx_n$ is connected to $\bx_m$,}\\
        0 &\text{ otherwise.}
      \end{cases}
      \label{weight}
    \end{equation}
  \end{enumerate}
\end{definition}
\noindent The parameter $\beta \ge 0$ controls the influence of the penalty term
that measures the distance between the patches in the image
domain. The distance $d(n,m)$ is small if the image intensity is
similar at $\bx_n$ and $\bx_m$, $\bu(\bx_n) \approx \bu(\bx_m)$, and
$\bx_n$ and $\bx_m$ are not too far from one another $\|\bx_n
-\bx_m\|\approx 0$. The parameter $\delta$ controls the scaling of the
similarity $d(n,m)$ between $\bu(\bx_n)$ and $\bu(\bx_m)$ when
defining the edge weight $w_{n,m}$. The particular form of the weight
(\ref{weight}) ensures that $w_{n,m}$ drops rapidly to zero as
$d(n,m)$ becomes larger than $\delta$.  A very large $\delta$
(i.e. $w_{n,m} \approx 1$ for all $n,m$) emphasizes the topology of
the graph and promotes a very fast diffusion of the random walk
through the patch-set. The other alternative: $\delta \approx 0$
accentuates the difference between the patches, but is very sensitive
to noise.  We now define the \textit{weight matrix}, which fully
characterizes the patch-graph.
\begin{definition} 
  The {\em weight matrix} $\bm{W}$ is the $N^2\times N^2$ symmetric matrix with
  entries $\bm{W}_{n,m} = w_{n,m}$. The {\em degree matrix} is the
  $N^2\times N^2$ diagonal matrix $\bm{D}$ with entries $\bm{D}_{n,n} =
  \sum_{m = 1}^{N^2} w_{n,m}$.
\end{definition}
Finally, we define the normalized Laplacian matrix.
\begin{definition} The {\em normalized Laplacian matrix} $\bL$ is the
  $N^2 \times N^2$ symmetric matrix defined by
  \begin{equation}
    \bL = \bI - \bD^{-\frac{1}{2}} \bW \bD^{-\frac{1}{2}}.
    \label{Laplacian}
  \end{equation}
\end{definition}
We note that the sign of the Laplacian is the opposite of the sign of
the Laplacian on a manifold ${\cal M}$ of dimension $d$, defined by
\begin{equation}
  \Delta u (x) = 
-  \lim_{r\rightarrow 0} \; \; \frac{2d}{r^2}
  \left(
    u(x) - \frac{1}{\vol(B(x,r))} \int_{B(x,r)} \!\! \!u(y)dy 
  \right),
  \label{laplace}
\end{equation}
where $B(x,r)$ is the ball of radius $r$ centered around $x$, and
$\vol (B(x,r))$ is the volume of the ball. Despite the fact that the
discrete Laplacian has the wrong sign, it shares the structure of
(\ref{laplace}). Indeed, the entry $n$ of the vector
$\bD^{-\frac{1}{2}} \bW \bD^{-\frac{1}{2}} u$ is an average of the
function $u$ defined on the graph, computed in a small neighborhood
around the node $\bu(\bx_n)$. Similarly,
$\frac{1}{\vol(B(x,r))}\int_{B(x,r)} u(y)dy$ computes the average of
$u$ within a ball centered around $x\in {\cal M}$. In both cases, the
Laplacian measures the difference between $u(\bx)$ and its local
average.

The matrix $\bL$ is symmetric and positive definite, and has $N^2$
eigenvectors $\bfi_1, \cdots,\bfi_{N^2}$ with corresponding
eigenvalues $\lambda_1=0 < \lambda_1 \leq \cdots \leq \lambda_{N^2}$.
Each eigenvector $\bfi_k$ is a vector with $N^2$ components, one for
each vertex of the graph. Hence, we write
\begin{equation*}
  \bfi_k = 
  \begin{bmatrix}
    \bfi_k(\bx_1) &\ldots &\bfi_k(\bx_{N^2})
  \end{bmatrix}
  ^T,
\end{equation*}
to emphasize the fact that we consider $\bfi_k$ to be a function
sampled on the vertices of $\Gamma$. The eigenvectors
$\bfi_1,\ldots,\bfi_{N^2}$ form an orthonormal basis for functions
defined on the graph, where the inner product on the graph is defined by
\begin{equation*}
  \langle f,g \rangle =  \sum_{j=1}^{N^2}  f(j) g (j).
\end{equation*}
\subsection{The eigenvectors encode the geometry of the image}
\begin{figure}[H]
  \centerline{
    \includegraphics[width= 0.33\textwidth]{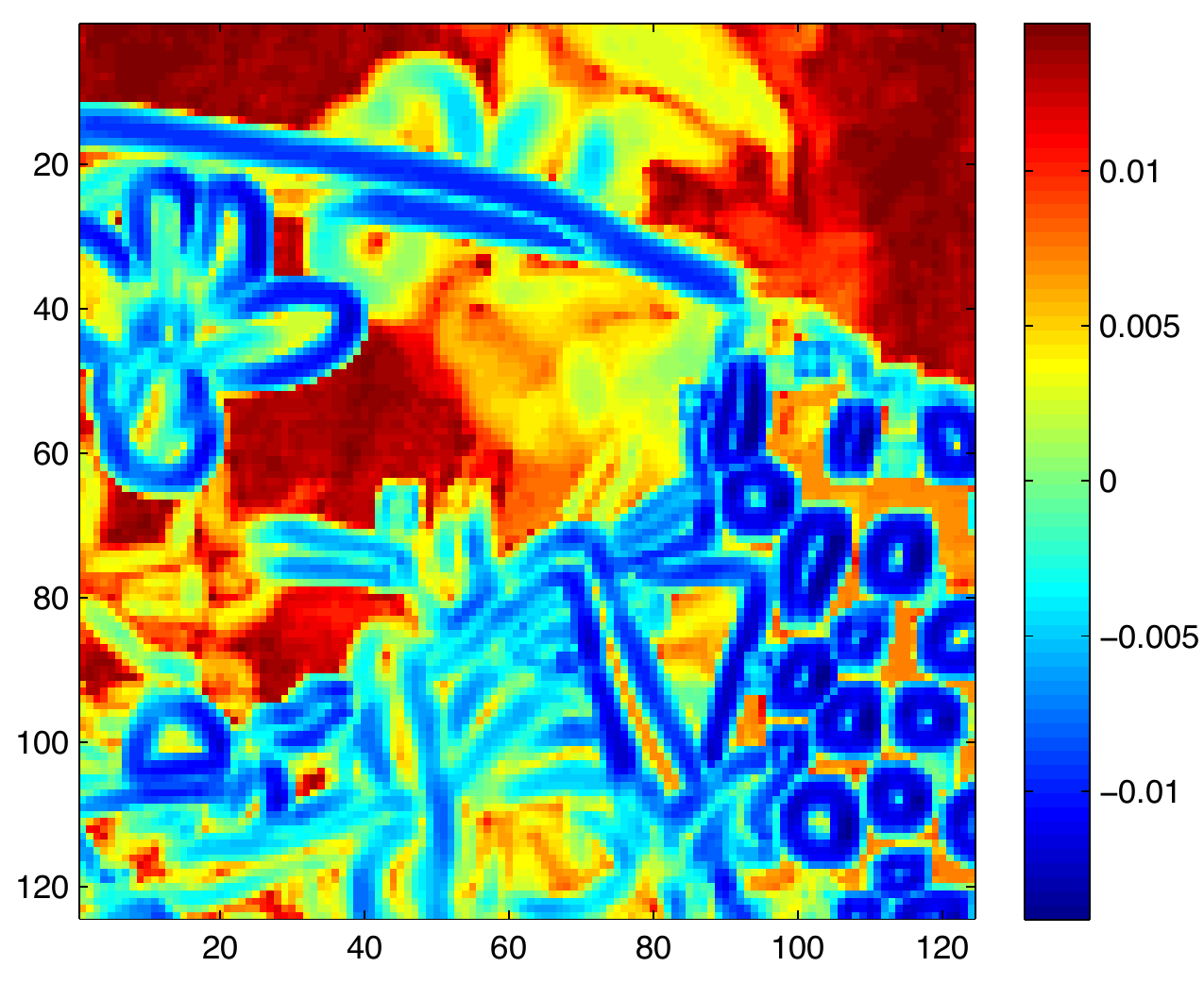}  
    \includegraphics[width= 0.33\textwidth]{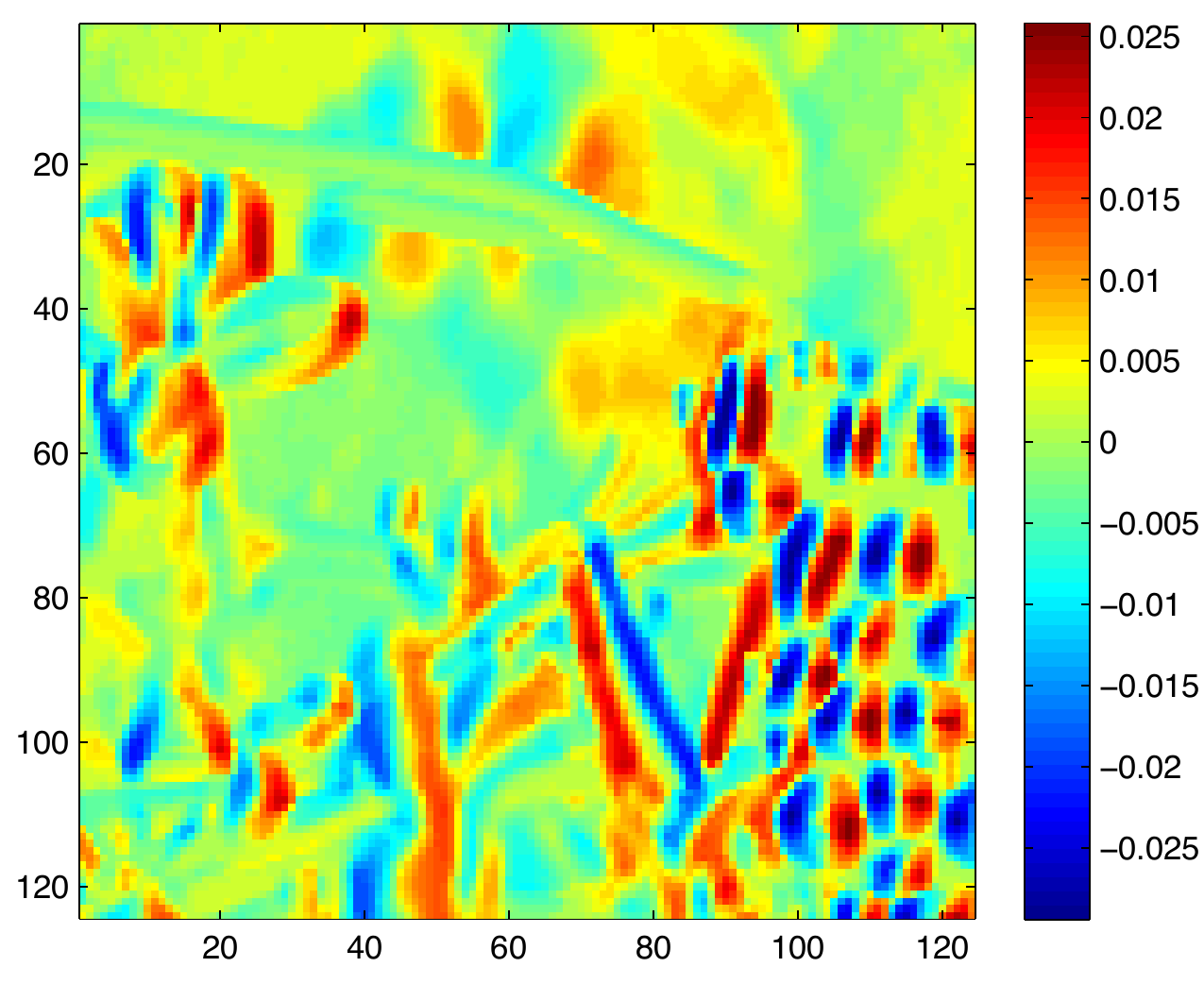}
    \includegraphics[width= 0.33\textwidth]{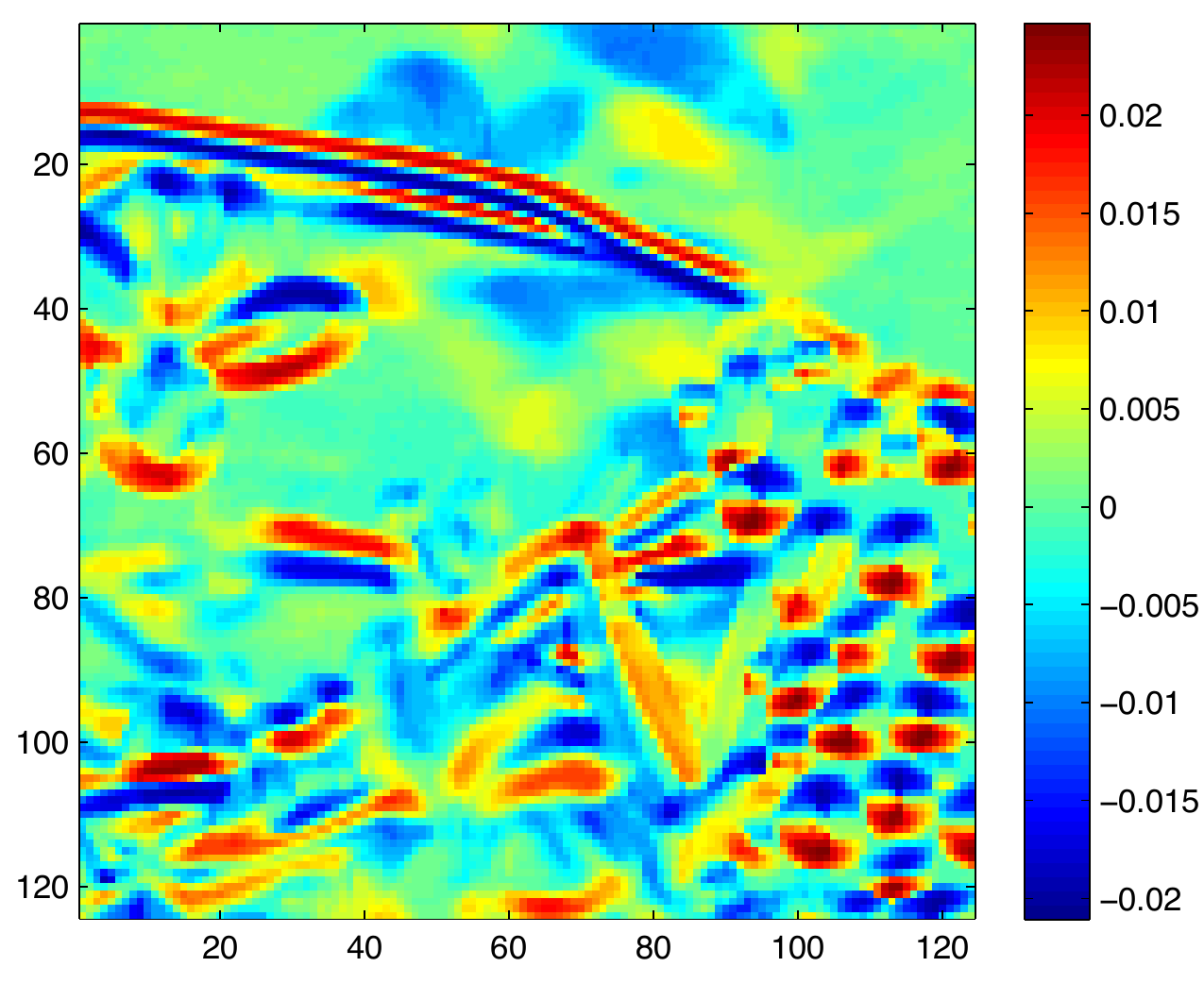}
  }
  \caption{From left to right: eigenvectors  $\bfi_2$,  $\bfi_3$, and $\bfi_4$.
    \label{butter}
  }
\end{figure}

We can represent $\bfi_k=   \begin{bmatrix} \bfi_k(\bx_1) &\ldots &\bfi_k(\bx_{N^2})
\end{bmatrix}^T$ as an image: each pixel $\bx_i$ is color-coded
according to the value of $\bfi_k(\bx_i)$. Figure~\ref{butter} shows the
three vectors $\bfi_2$, $\bfi_3$ and $\bfi_4$ of the butterfly
image (see Fig.~\ref{patchbutter}-left). The first non
trivial eigenvector $\bfi_2$ encodes the gradient of the image
intensity: $\bfi_2$ takes large negative\footnote{We note that the
  sign of $\bfi_k$ is arbitrary.} values at pixels with large gradient
(irrespective of its direction) and takes large
positive values at pixels with small gradient. $\bfi_3$ and $\bfi_4$
clearly encode horizontal and vertical partial derivatives of $u$. The
frequency content of each eigenvector $\bfi_k$ is loosely related to
the eigenvalue $\lambda_k$.  Unlike Fourier analysis, the basis
functions $\bfi_k$ have an%
\noindent intrinsic scale given by the patch size. In
this work we propose to use the eigenvectors $\bfi_k,k=1,\ldots$ as a
basis to denoise the image intensity function $u$.

\begin{figure}[H]
  \centerline{
    \includegraphics[width = 0.50\textwidth]{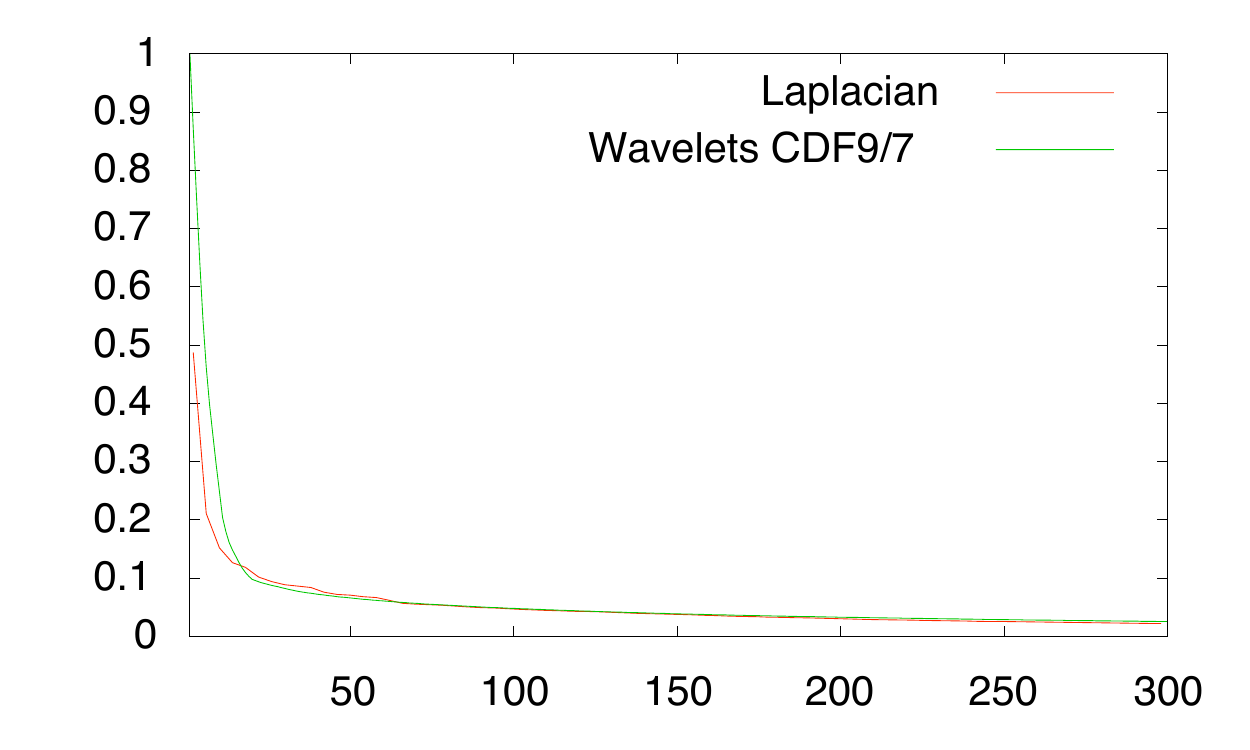}
  }
  \caption{Relative residual energy after reconstructing the
    clown image using the $K$-best eigenvectors (red), and using the $K$-best
    wavelets (green). 
    \label{clean_energy_decay}
  }  
\end{figure}

We conclude this section with experimental evidence that indicates that
the image $u$ has a sparse representation in the basis $\left\{
  \bfi_k\right\}$. We will use this property in the next section to
remove the noise from the image. Figure \ref{clean_energy_decay}
displays the relative residual energy, $\tau_K$, after reconstructing
the clown image (Fig \ref{network}-left) using the first $K$
eigenvectors with the largest coordinates $\langle u, \bfi_k\rangle$,
\begin{equation*}
  \tau_K = \frac{\|u - \sum_{k=1}^K \langle u,\bfi_k \rangle \bfi_k\|^2}{\|u\|^2}.
\end{equation*}
For comparison purposes, the same quantity is computed using a wavelet
decomposition. We see that the basis functions $\bfi_k$ give rise to a
very fast decay of the residual energy, even faster than with the 9/7
wavelet basis.  We performed similar comparisons with several other
images (not shown) that lead us to conclude that we could use this
basis to denoise natural images.
\section{Denoising}
\subsection{Denoising the patch-set using the noise-free eigenvectors
  $\bfi_k$
\label{denoise-algo}}
We describe in this section a denoising algorithm that constructs an
estimate of the patch-set from the knowledge of the noisy image
$\widetilde{u}$. The denoising of the patches is not performed locally
on the patch-set, but rather relies on the global eigenvectors
$\bfi_k$. This allows us to denoise the entire patch-set as a whole.
We assume for the moment that we have access to the eigenvectors
$\bfi_k$ of the noise-free image. While this is clearly unrealistic,
this allows us to explain the principle of the algorithm. In section
\ref{recovery} we study the effect of noise on the
eigenvectors. Finally, in section \ref{twostage}, we describe a
procedure to iteratively reconstruct the patch-set and the
eigenvectors.

The original image $u$ has been corrupted by additive white
Gaussian noise with variance $\sigma^2$, and we measure at every pixel
$\bx$ the noisy image
$\widetilde{u}(\bx)$, given by
\begin{equation*}
  \widetilde{u}(\bx) = u(\bx) + n (\bx).
\end{equation*}
Let $\tu(\bx_n)$ be the patch centered at $\bx_n$ associated with the
noisy image $\widetilde{u}$, and let $\widetilde{\cal P}$ be the
patch-set formed by the collection of noisy patches.  We propose to
construct an estimate $\widehat{\cal P}$ of the clean patch-set
$\cal P$ from the noisy patch-set $\widetilde{\cal P}$.  Lastly, we
will combine several denoised patches $\hu(\bx_m)$ to
reconstruct an estimate $\widehat{u}(\bx_n)$ of the clean image at the
pixel $\bx_n$.

Let $u_1(\bx_n),\ldots,u_{m^2}(\bx_n)$ denote the $m^2$ coordinates of
the patch $\bu(\bx_n)$. We define similarly,
$\widetilde{u}_1(\bx_n),\ldots,\widetilde{u}_{m^2}(\bx_n)$ to be the
$m^2$ coordinates of the patch $\tu(\bx_n)$. We make the following
trivial observation: the coordinate  $\widetilde{u}_j$ is a
real-valued function defined on the set of vertices
\begin{align*}
  V(\Gamma) & \longrightarrow \R\\
  \widetilde{u}_j: \bx_n  & \longmapsto \widetilde{u}_j(\bx_n) 
\end{align*}
In order to denoise simultaneously all the patches, we denoise each
coordinate function $\widetilde{u}_j$ independently. This is achieved
by expanding the function $\widetilde{u}_j$ into the basis formed by
the $\bfi_k$, and performing a nonlinear thresholding. The result is
the denoised function $\widehat{u}_j$, given by
\begin{equation}
  \widehat{u}_j = \sum_{k=1}^{N^2}
  \kappa \left (
    \langle \widetilde{u}_j, \bfi_k \rangle 
  \right )
  \bfi_k,
  \label{denoise}
\end{equation}
\begin{figure}[H]
  \centerline{  
    \includegraphics[width= 0.2\textwidth]{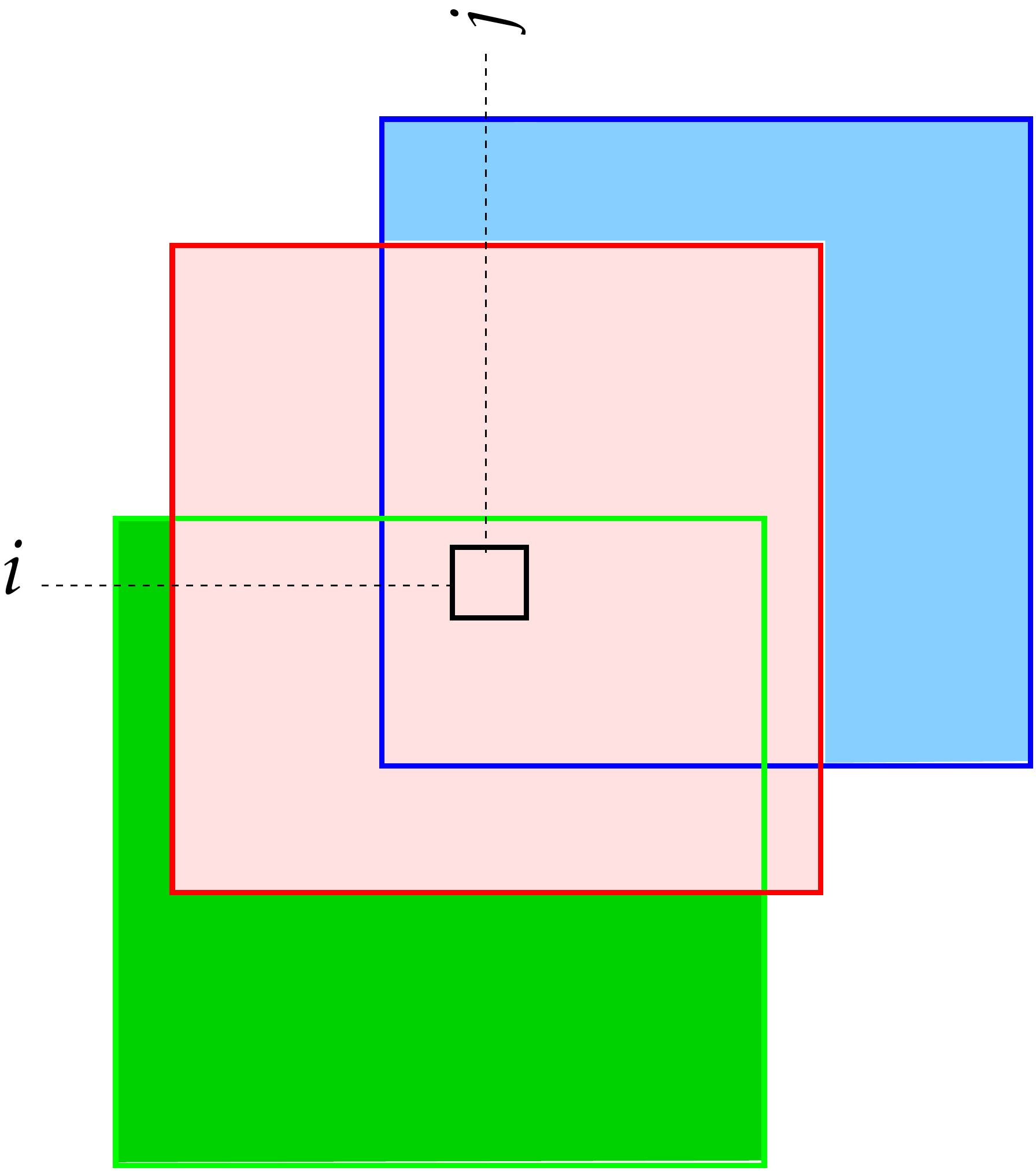}
  }  
  \caption{The image value  $u(i,j)$ (in red) is reconstructed using
    the coordinate of each of the $m^2$ overlapping patches
    $\hu(\bx_l)$ that corresponds to the location $\bx_n = (i,j)$. 
    \label{recons}
  }
\end{figure} 
\noindent where the function $\kappa$ performs a nonlinear thresholding of the
coefficients $\langle \widetilde{u}_j,\bfi_k \rangle$. In practice, we use
a hard thresholding.  After denoising all the coordinate functions
$\widehat{u}_1,\ldots,\widehat{u}_{m^2}$, we have access to an estimate $\left
  \{\hu(\bx_n), n =1,\ldots,N^2\right\}$  of the clean patch-set.

Because the patch-set has  a maximum overlap (i.e. we have
a patch for each pixel), there are $m^2$ patches that overlap (in
the image domain) with the pixel $\bx_n$ (see Fig.~\ref{recons}). Let
$N(\bx_n)$ be the ball for the $l_0$ norm of radius $m/2$ pixels
centered at $\bx_n$,
\begin{equation}
  N(\bx_n) = \left \{\bx_l, \| \bx_n - \bx_l\|_0 \leq m/2\right\}.
\end{equation}
The $m^2$ patches $\bu(\bx_l)$ that overlap with $\bx_n$ all have
their center $\bx_l$ in $N(\bx_n)$ (see Fig.~\ref{recons}). For each
of the pixels $\bx_l \in N(\bx_n)$, there exists an index i$_n$ such
that the coordinate $i_n$ of the patch $\hu(\bx_l)$ corresponds to the
pixel location $\bx_n$. In other words, $[\hu(\bx_l)]_{i_n}$ is an
estimate of the image intensity at the location $\bx_n$ obtained from
patch $\hu(\bx_l)$.  We combine all these estimates and define the
denoised image at pixel $\bx_n$ as the following weighted sum
\begin{equation}
  \hat{u}(\bx_n) =
  \sum_{\bx_l \in N(\bx_n)}
  \alpha (\bx_n,\bx_l)
  \; [\hu(\bx_l)]_{i_n},
  \label{average}
\end{equation}
where the exponential weight $\alpha (\bx_n, \bx_l)$ is given by
\begin{equation}
  \alpha (\bx_n, \bx_l) = 
  \frac{\exp{\left(-\| \bx_n-\bx_l\|^2\right)}}
  {\sum_{\bx_m \in N(\bx_n)}
    \exp{\left(-\| \bx_n -\bx_m\|^2\right)}}\;.
\end{equation}
This choice of weights favors the patch $\bu(\bx_n)$ centered at
$\bx_n$, and disregard exponentially fast the neighboring patches.  We
have also experimented with weights that discourage the mixing of
patches with very different variances (results not shown).

{\noindent \bfseries Connection to translation-invariant denoising.}
We note that we can interpret the $j$th coordinate function $u_j$ as
the original image $u$ shifted by $(p,q)$, where $j = m(p + m/2) + q +
m/2 +1$. For instance, if $j=1$, then the two-dimensional shift is
$(-m/2,-m/2)$ and if $j=m^2$, the shift is $(m/2,/m2)$. The
reconstruction formula (\ref{average}) can then be interpreted as a
translation-invariant denoising procedure \cite{Coifman95} of the
form,
\begin{equation*}
  \widehat{u} (\bx) = 
  \underset{(p,q) \in [-m,m]\times [-m,m]}{\text{Average}}\;\;
  \left\{\underset{(p,q)} {\text{Unshift}}\;\;
    \text{Denoise} \;\;
    \underset{(p,q)}{\text{Shift}} \;\;\; \widetilde{u}(\bx)\right\}.
\end{equation*}
\subsection{Geometric interpretation of the eigenvectors
\label{diffuse}}
We explain in this section the connection between our approach and
denoising methods based on defining a diffusion process on the graph
\citep{Szlam08,Bougleux09}.  The solution to the denoising problem is
related to a diffusion process defined on the patch-set. Indeed, let
us consider the image $v(\bx,t)$ solution to the diffusion defined on
the patch-graph (notice that we use the wrong sign for equation
(\ref{flow}) as explained in section \ref{graphpatches}) defined by,
\begin{equation}
  \frac{\partial v(\bx,t)}{\partial t} =
  -\bL v(\bx,t),
  \label{heat}
\end{equation}
where the initial condition is provided by the noisy image $\widetilde{u}(\bx)$,
\begin{equation}
  v(\bx,0) = \widetilde{u}(\bx).
\end{equation}
We consider the heat kernel formed by the $N^2 \times N^2$ matrix 
\begin{equation}
  \bH_t(n,m) = \sum_{k=1}^{N^2} e^{-\lambda_k t}
  \bfi_k(\bx_n)\bfi_k(\bx_m),
\end{equation}
and we write the solution to the diffusion  as 
\begin{equation}
  v(\bx_n,t) = 
  \sum_{m=1}^{N^2} \bH_t(n,m)\widetilde{u}(m) =
  \sum_{k=1}^{N^2}  e^{-\lambda_k t} \;\langle \widetilde{u}, \bfi_k\rangle \; \bfi_k(\bx_n).
  \label{heatkernel}
\end{equation}
We notice the similarity between the expansions (\ref{denoise}) and
(\ref{heatkernel}). As $t$ becomes large, only a small number of terms
in (\ref{heatkernel}) will be non zero. The attenuation of $\langle
\widetilde{u}, \bfi_k\rangle$ is controlled by the size of the
corresponding eigenvalue $\lambda_k$. This is in contrast to
(\ref{denoise}), where the coefficients used to reconstruct the
denoised signal are chosen based on the magnitude of $\langle
\widetilde{u},\bfi_k\rangle$. The similarity between the two
expansions allows us to understand what is the information encoded by
each $\bfi_k$. In (\ref{heatkernel}), each function $e^{-\lambda_kt}
\bfi_k(\bx)$ is a stationary solution to the diffusion
(\ref{heat}). The amplitude of this stationary solution decreases
exponentially fast with time, but the geometry, encoded by $\bfi_k$
remains unchanged. The eigenvectors $\bfi_k$ therefore encode the
geometric features present in the image: edges, texture, etc. as
suggested by Fig.~\ref{butter}.  Finally, we note that if the graph of
patches is replaced with a regular lattice formed by the image
sampling grid, and if each patch is reduced to a single pixel, then
the diffusion (\ref{heat}) models one of the standard diffusion-based
denoising algorithms \cite{Morel95}.
\subsection{Estimating the eigenvectors $\bfi_k$ from the noisy images
  \label{recovery}}
We described in the previous section a denoising procedure that relies
on the knowledge of the eigenvectors $\bfi_k$ associated with the
clean patch-set. Of course, if we had access to the clean patch-set,
then we would have the clean image. While our approach appears to be
circular, we claim that it is possible to bootstrap an estimate of the
$\bfi_k$ and iteratively improve this estimate. This approach relies
on the stability of the low-frequency eigenspace of the graph
Laplacian $\bL$. In the next section we study experimentally the
perturbation of the eigenvectors $\bfi_k$ when a significant amount of
noise is added to an image. We demonstrate that changes in the
topology of the graph, and not changes in the weights, is the source
of the perturbation of the $\bfi_k$. This result leads to an iterative
method that recovers the clean eigenvectors $\bfi_k$ that is
described in section \ref{twostage}.
\subsubsection{The eigenvectors of the perturbed Laplacian}
Each patch $\bu(\bx_n)$ of the clean image is corrupted by a multivariate Gaussian noise
$\bn(\bx_n)$ with diagonal covariance matrix, $\sigma^2 I_{m^2}$,
\begin{equation}
  \tu(\bx_n) = \bu(\bx_n) + \bn(\bx_n).
\end{equation}
The distance between any two noisy patches
$\tu(\bx_n)$ and $\tu(\bx_m)$ is given by
\begin{equation}
  \| \tu(\bx_n) - \tu(\bx_m)\|^2 = \|\bu(\bx_n) - \bu(\bx_m)\|^2 + 2
  \langle 
  \bu(\bx_n) - \bu(\bx_m),
  \bn(\bx_n) - \bn(\bx_m)
  \rangle
  + 
  \|\bn(\bx_n) - \bn(\bx_m)\|^2.
\label{noisy-patch}
\end{equation}
with the corresponding expected value,
\begin{equation}
  \E \| \tu(\bx_n) - \tu(\bx_m)\|^2 = \|\bu(\bx_n) - \bu(\bx_m)\|^2 
  + 
  2 \sigma^2 m^2.
  \label{xi2}
\end{equation}
The weights associated with the noisy image are given by
\begin{equation}
  \widetilde{w}_{n,m}  = w_{n,m} 
  \;\;
  e^{-2 \langle 
    \bu(\bx_n) - \bu(\bx_m),
    \bn(\bx_n) - \bn(\bx_m)
    \rangle/\delta^2}
  \;e^{-\|\bn(\bx_n) - \bn(\bx_m)\|^2/\delta^2}.
  \label{noisyweight}
\end{equation}
While $e^{-\|\bn(\bx_n) - \bn(\bx_m)\|^2/\delta^2} < 1$, the term 
$e^{-2 \langle \bu(\bx_n) - \bu(\bx_m),\bn(\bx_n) -
  \bn(\bx_m)\rangle/\delta^2}$ may be greater than 1, thereby
increasing $w_{n,m}$.
Overall, the presence of noise has  two effects:
\begin{enumerate}
\item the topology of the graph is modified since points that
  were previously neighbors may become far away and vice-versa
  (caused by changes in the distance (\ref{xi2})),
\item the weights along the edges are modified according to (\ref{noisyweight}).
\end{enumerate} 

Let $\widetilde{\Gamma}$ be the graph associated with the noisy
patch-set $\widetilde{\cal P}$. $\widetilde{\Gamma}$ and $\Gamma$
have the same vertices, $V(\Gamma) = V(\widetilde{\Gamma})$, and the
same total number of edges (since both are based a nearest neighbor
topology). The weight matrix $\bW$ is perturbed according to
(\ref{noisyweight}). We define $\widetilde{\bW}$ and $\widetilde{\bL}$
to be the weight and Laplacian matrices computed from the noisy
image, respectively. We expect that the eigenvectors $\tfi_k$ of $\widetilde{\bL}$
will be different from the $\bfi_k$.  As explained in the next
section, the low-frequency eigenvectors remain stable, but the higher
frequency eigenvectors rapidly degrade. As a result, $\tfi_k$ cannot
be used to reconstruct a clean image.

Theoretical results that provide upper bounds on the angle between an
eigenspace of $\bL$ and the corresponding eigenspace of
$\widetilde{\bL}$ exist \cite{Stewart90}. Unfortunately, as noted in
\cite{Yan09}, these bounds usually overestimate the actual
perturbation of the eigenvectors. Furthermore, these results depend on
the separation of the eigenvalues, a quantity that is difficult to
predict in our problem. In fact the separation can be very small
leading to very large bounds. Another limitations of the usual
theoretical bounds is that the angle between two invariant subspaces
cannot be readily translated in terms of the ability of the perturbed
eigenvectors $\tfi_k$ to approximate the original image using a small number of
terms. We propose therefore to study experimentally the effect of the
noise on the ability of $\tfi_k$ to encode geometric information about
the image. We employ standard methodology used in vision
research. Specifically, let $\ftt_k$ be the Fourier transform of
$\tfi_k$, we study the energy distribution of $\ftt_k$ using polar
coordinates $(\rho,\theta)$ in the Fourier plane and we compute the
average energy of $\ftt_k$ over all orientation $\theta$ for a given
radius $\rho$.

\begin{figure}[H]
  \centerline{\hspace*{6pc}\hfill original \hfill noisy\hfill\hfill}
  \centerline{
    \includegraphics[width= 0.20\textwidth]{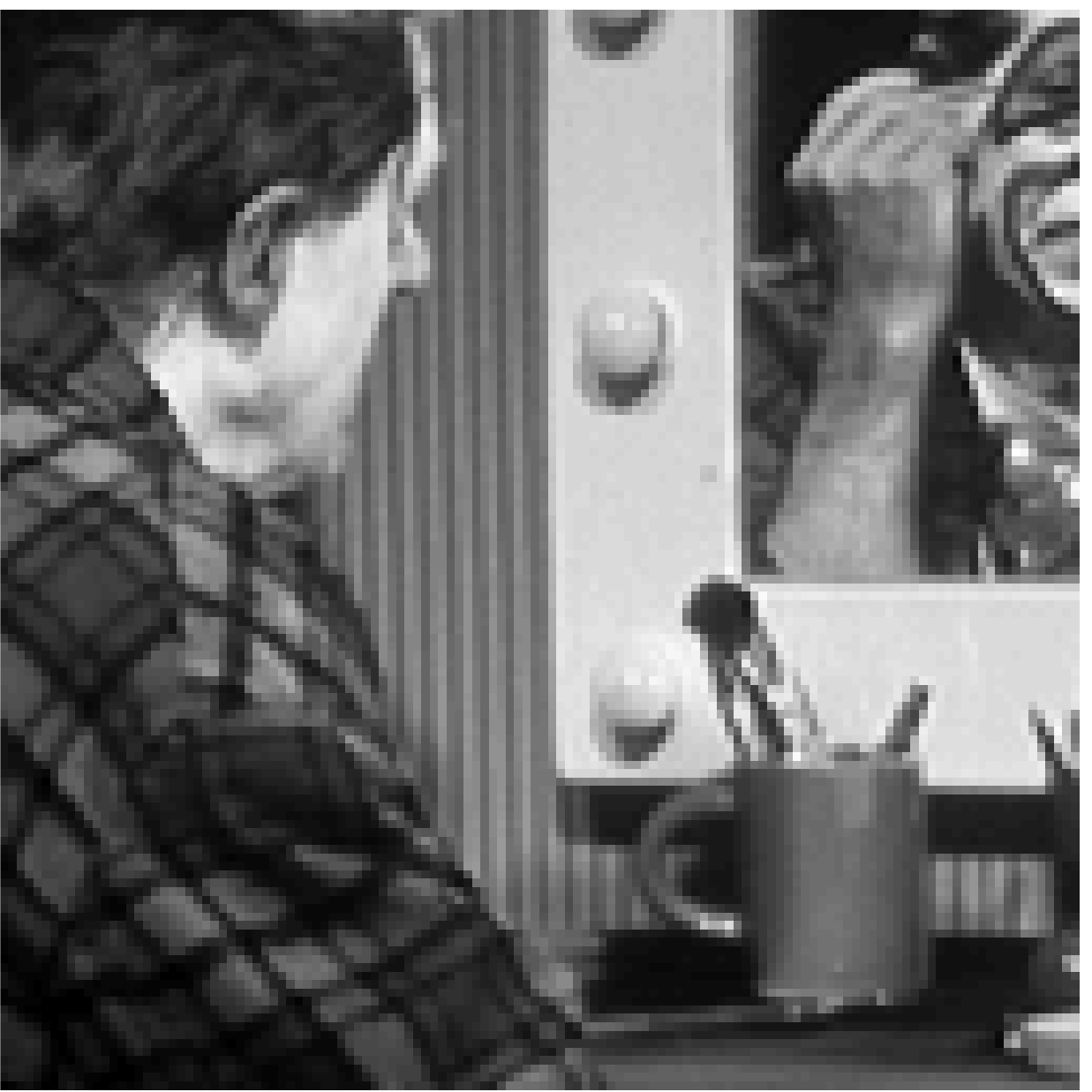}  
    \includegraphics[width= 0.20\textwidth]{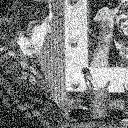}
}
  \caption{Original $128 \times 128$  ``clown'' image (left);  clown
    image with white Gaussian noise, $\sigma =  40$ (right).
\label{clown}}
\end{figure}

\begin{figure}[H]
  \centerline{\hspace*{4pc} $\bfi_2$ \hfill\hspace*{2pc}   $\bfi_{32}$ \hfill $\bfi_{128}$\hfill $\bfi_{256}$\hfill} 
  \centerline{
    \includegraphics[width= 0.20\textwidth]{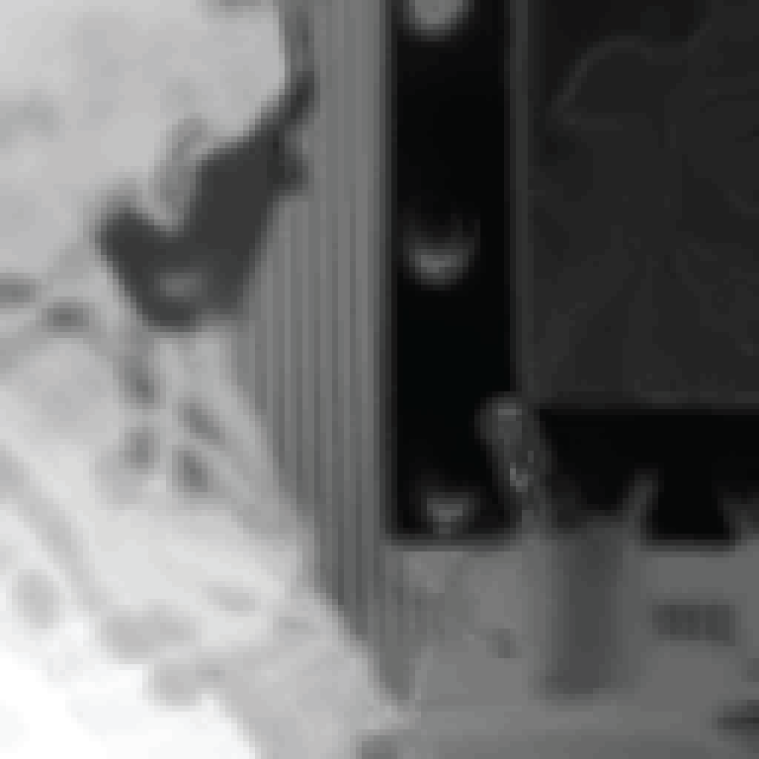}\hspace*{-0.25pc}  
    \includegraphics[width= 0.20\textwidth]{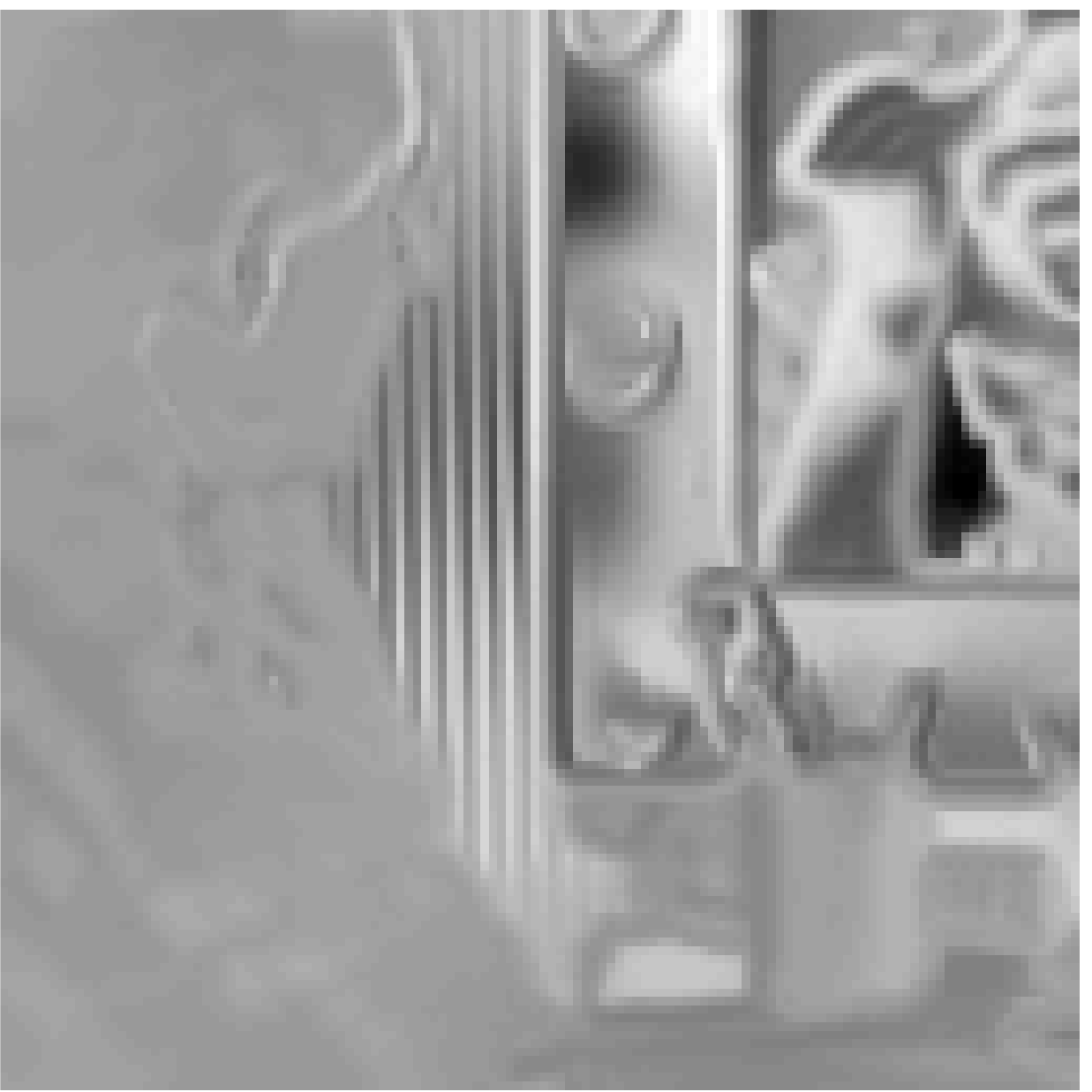}\hspace*{-0.25pc}  
    \includegraphics[width= 0.20\textwidth]{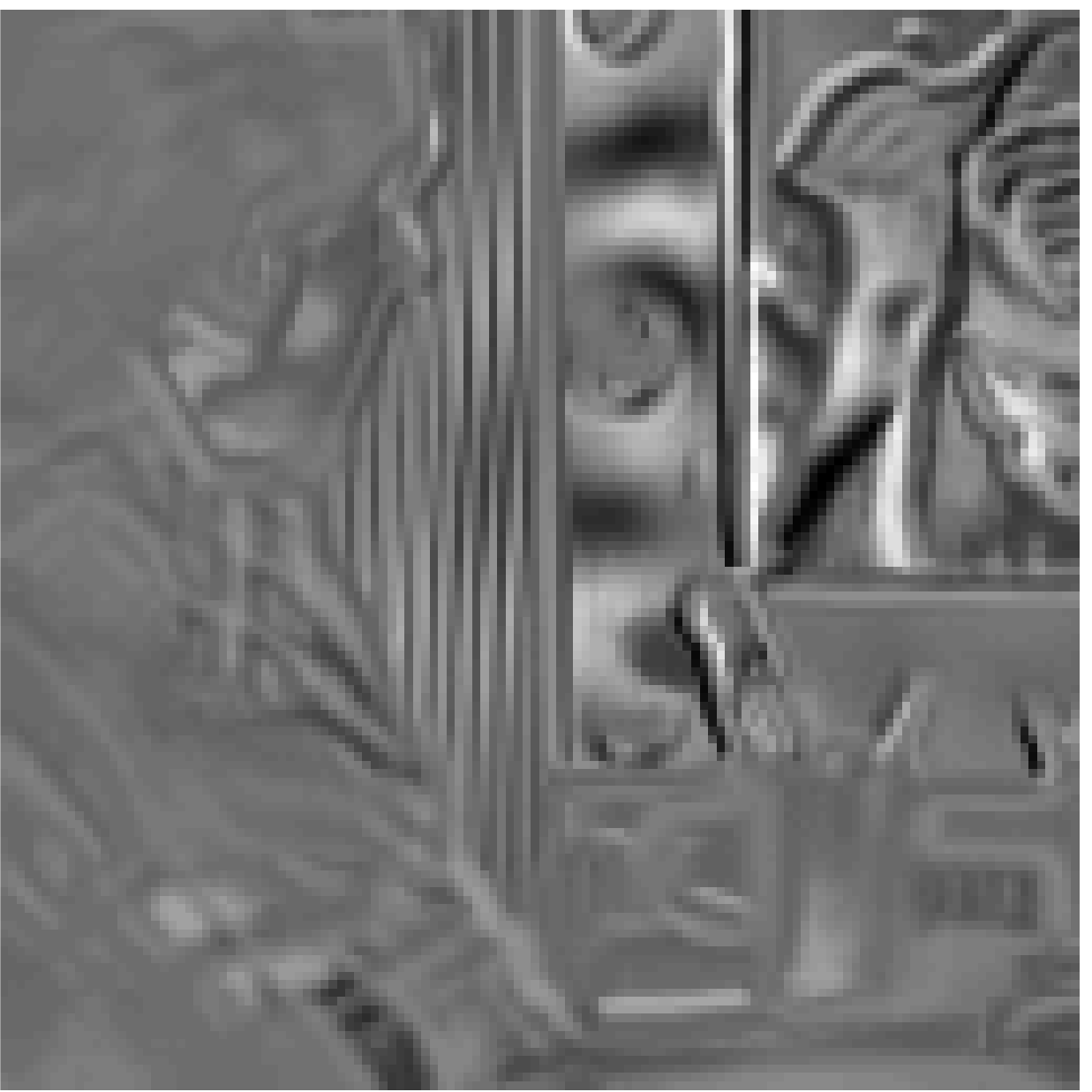}\hspace*{-0.25pc}  
    \includegraphics[width= 0.20\textwidth]{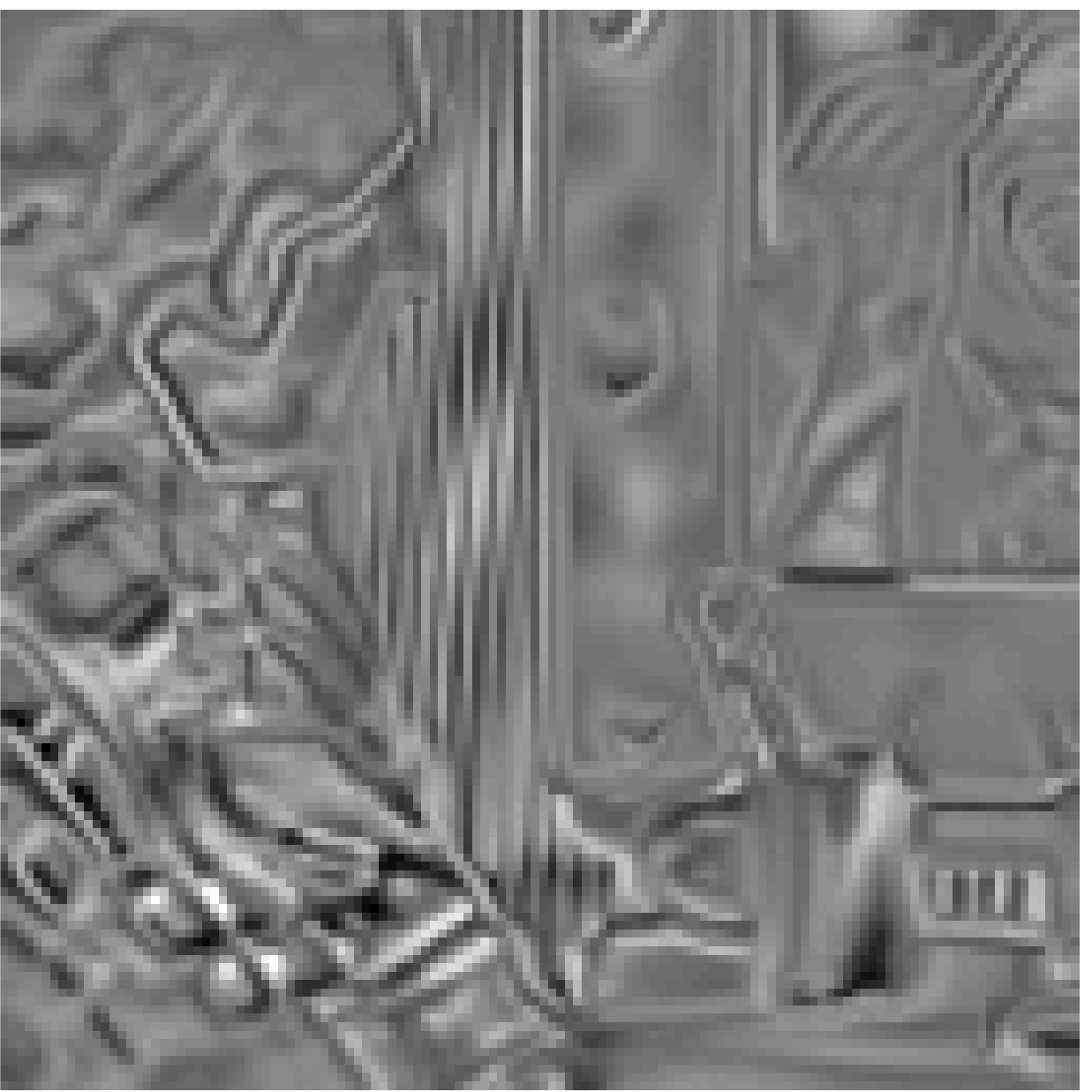}  
  }
  \centerline{  
    \includegraphics[width= 0.20\textwidth]{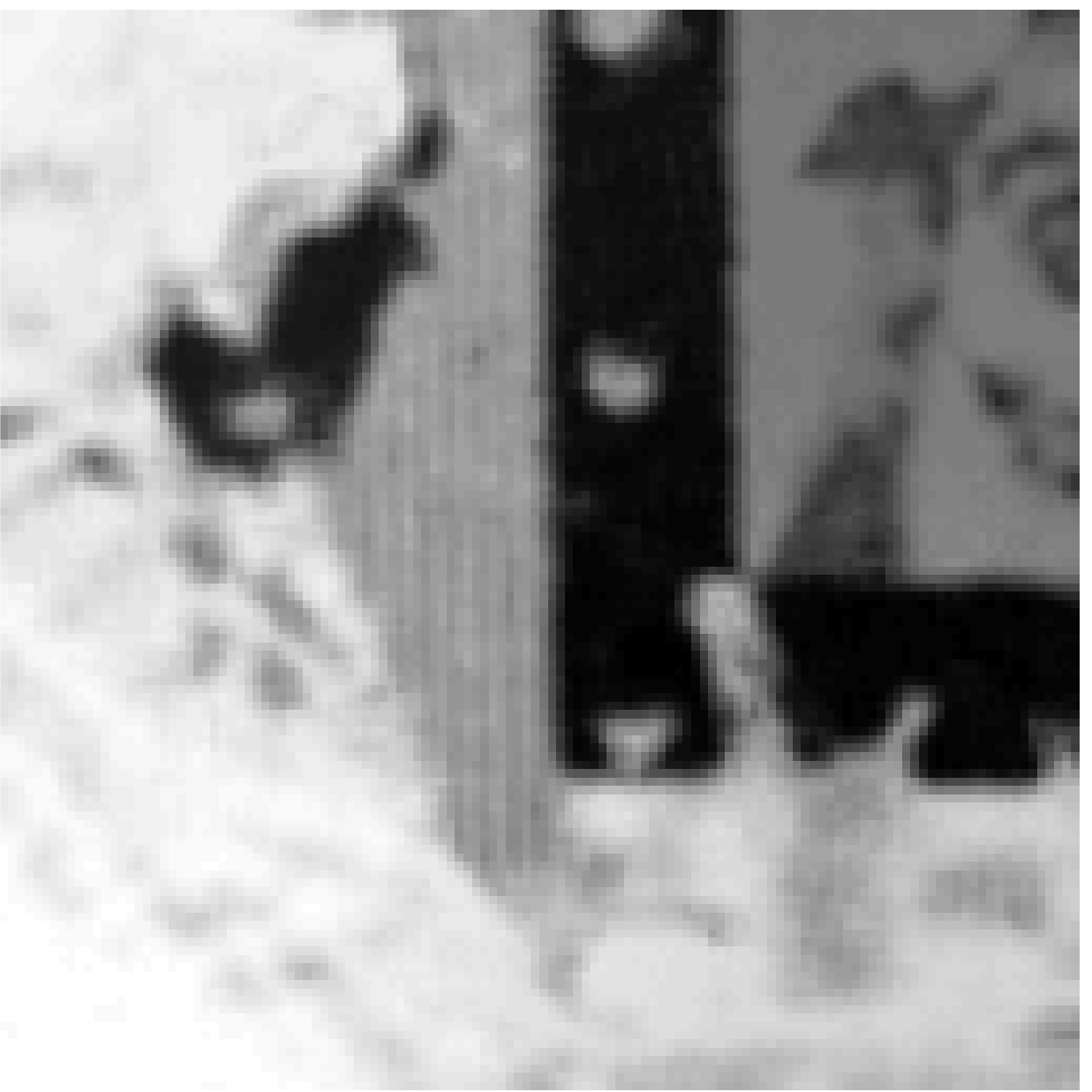}\hspace*{-0.25pc}  
    \includegraphics[width= 0.20\textwidth]{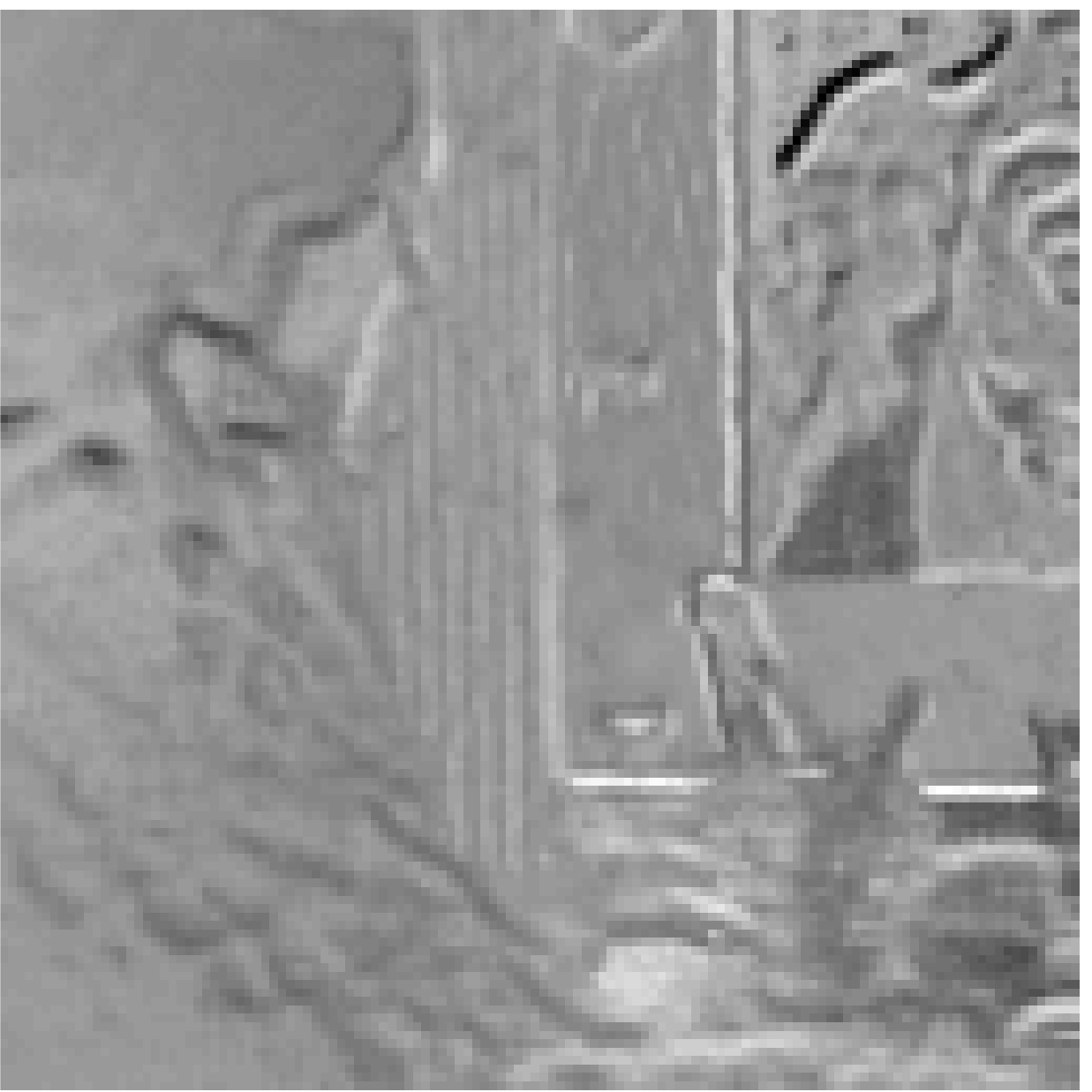}\hspace*{-0.25pc}  
    \includegraphics[width= 0.20\textwidth]{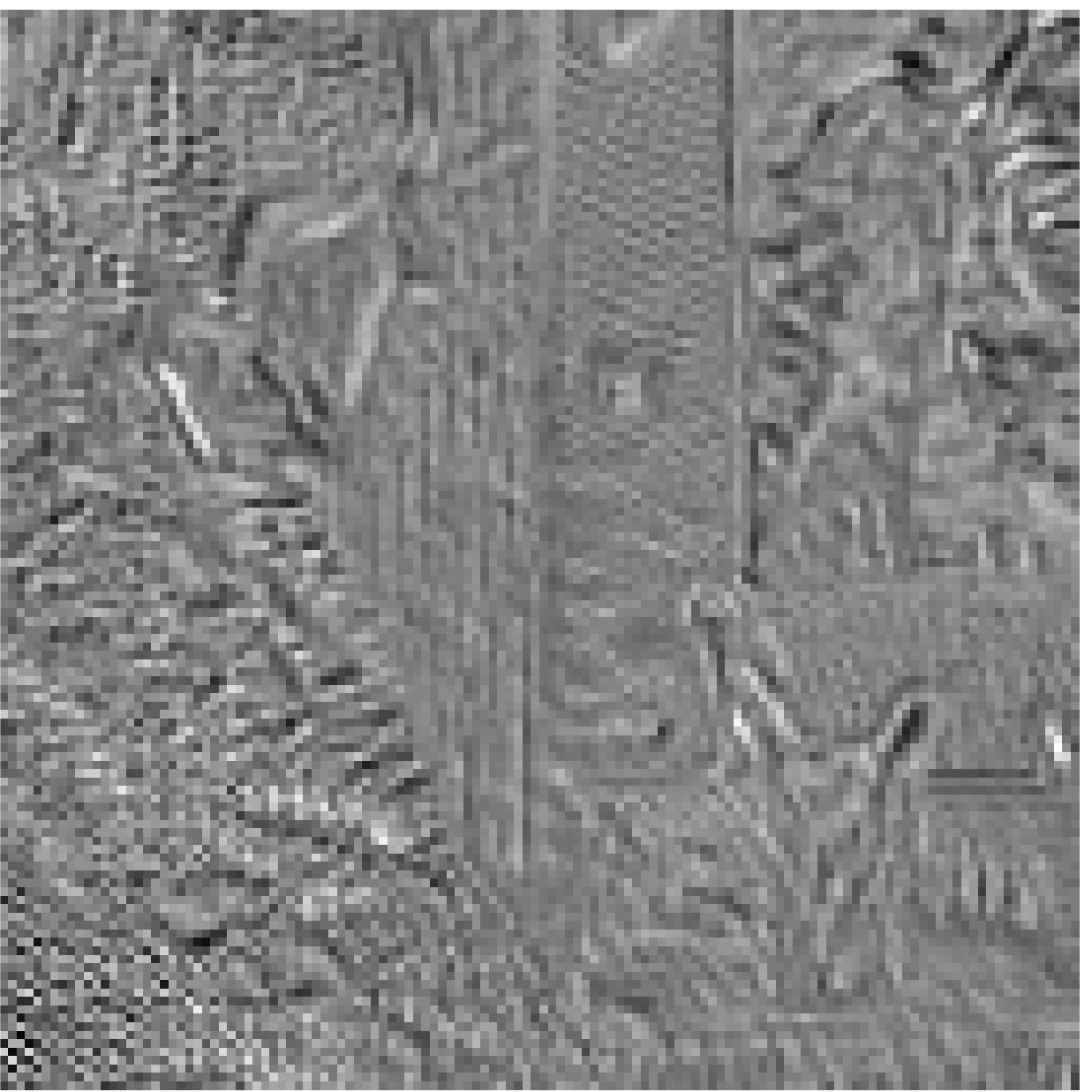}\hspace*{-0.25pc}  
    \includegraphics[width= 0.20\textwidth]{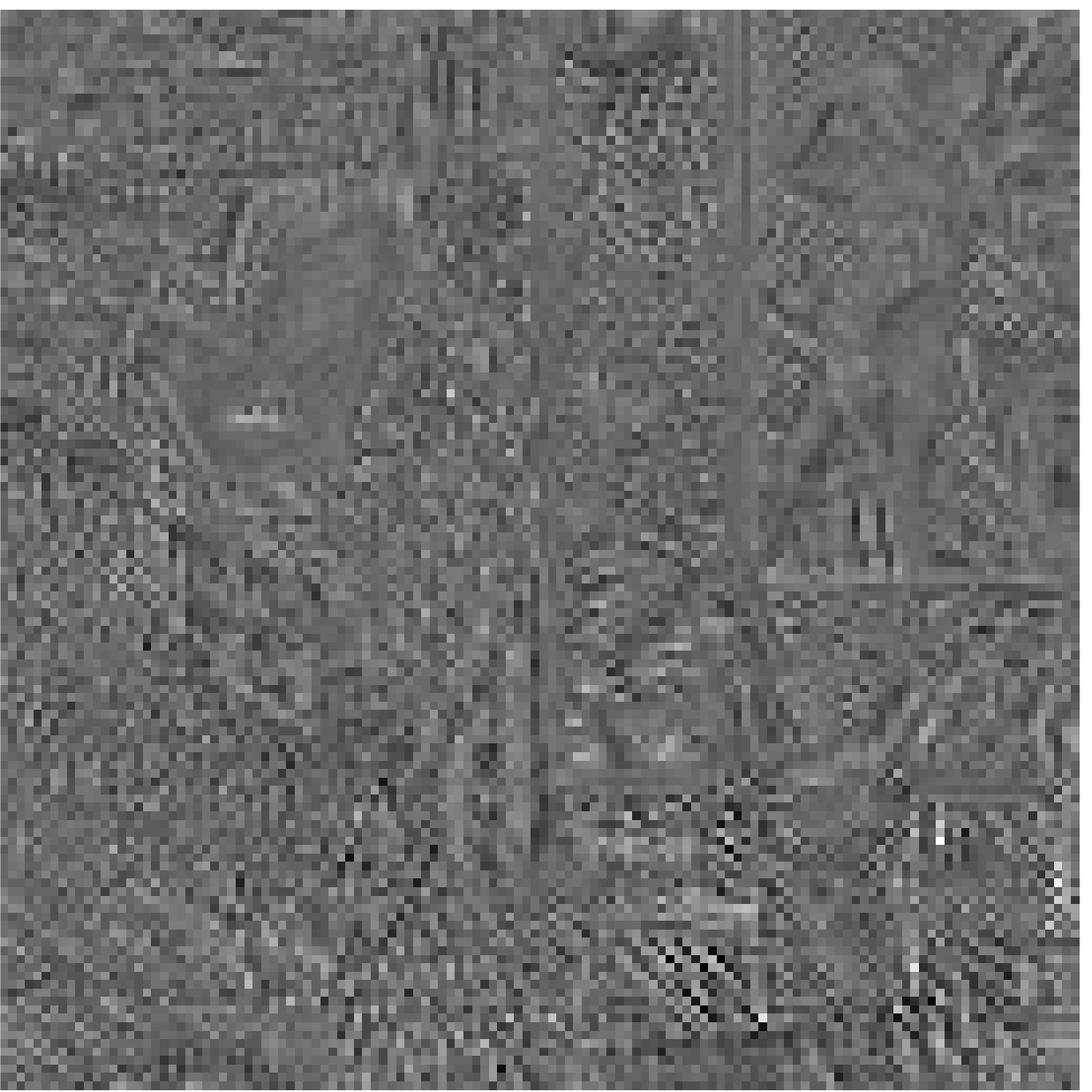}  
  }
 \centerline{\hspace*{4pc} $\tfi_2$ \hfill\hspace*{2pc}   $\tfi_{32}$ \hfill $\tfi_{128}$\hfill $\tfi_{256}$\hfill} 

  \caption{Top row: eigenvectors of the clean image. Bottom row:
    eigenvectors of the noisy image. 
    \label{noisy-eig}
  }
\end{figure}

Our experiment was performed as follows. We add white Gaussian noise
($\sigma=40$) to the clean clown image (see Fig.~\ref{clown}), and we
compute the eigenvectors $\tfi_k$ of the graph Laplacian. This image
provides a good balance between texture and smooth regions; we
performed similar experiences with other images, and obtained
results that were quantitatively similar. The perturbed eigenvectors
$\tfi_2$, $\tfi_{32}, \tfi_{128}$, and $\tfi_{256}$ are shown in the
bottom row of Fig.~\ref{noisy-eig}. The eigenvectors of the clean
image are shown in the top row of Fig.~\ref{noisy-eig} for comparison
purposes. We can make the following visual observations, which will be
confirmed by a quantitative analysis in the next paragraph. For $k\leq 32$,
the eigenvectors $\tfi_k$ appear to be reasonably similar to
the original eigenvectors $\bfi_k$. For $k\ge 128$,%

\noindent $\tfi_k$ appear to mostly capture the random noise instead
of the edges and texture present in the original image. For instance,
the texture on the wallpaper and on the shirt of the clown, which is
present in $\bfi_{256}$, is replaced by noise in $\tfi_{256}$ (see
Fig.~\ref{noisy-eig}).

\begin{figure}[H]
  \centerline{\hfill $i = 0$\hfill \hfill $i=1$\hfill\hfill$i =2$\hfill}
  \centerline{\small
    \begin{rotate}{90}~~~~~~~~~~$\widetilde{\cal E}^i(l)$\end{rotate}
    \includegraphics[width= 12pc]{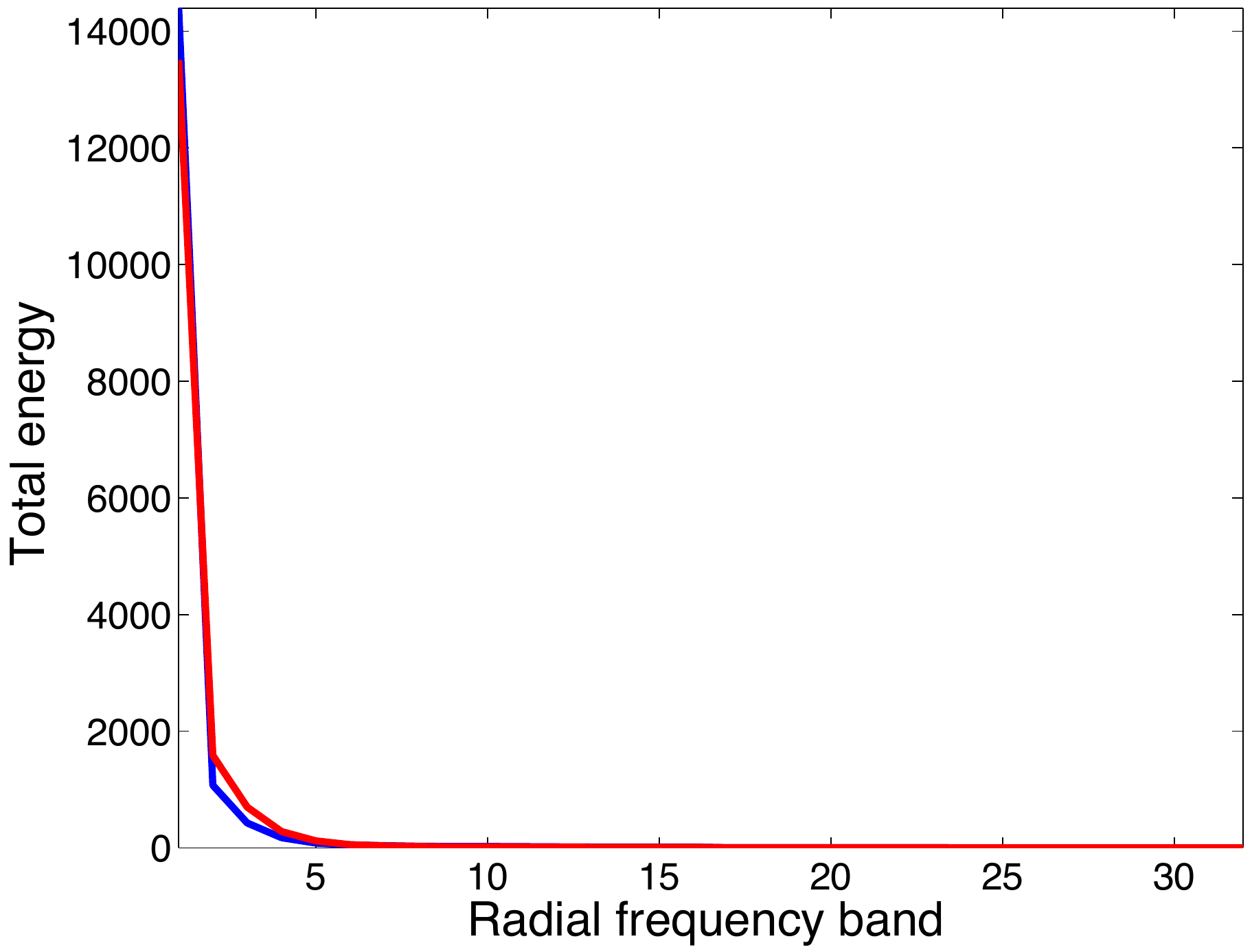}
    \includegraphics[width= 12pc]{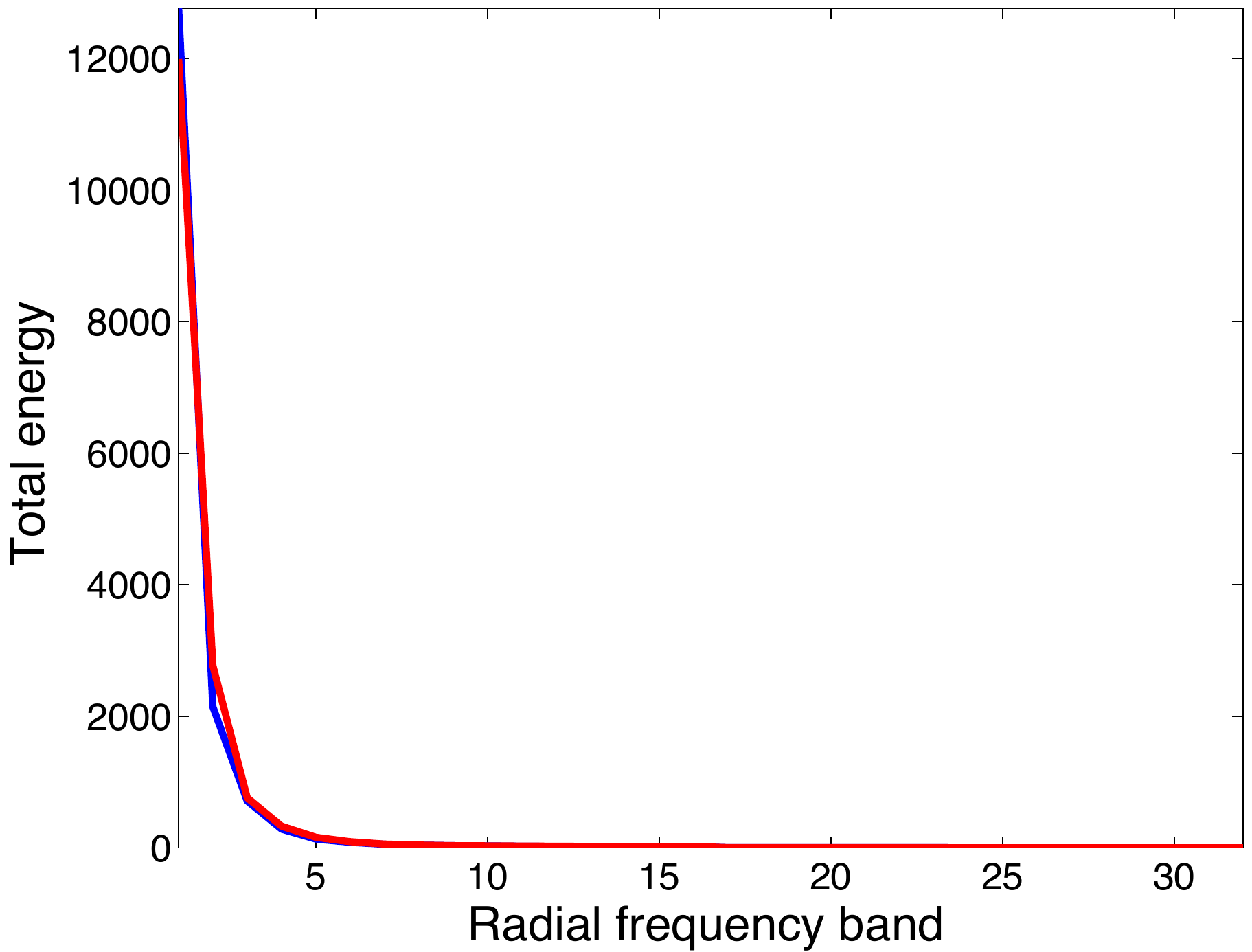}
    \includegraphics[width= 12pc]{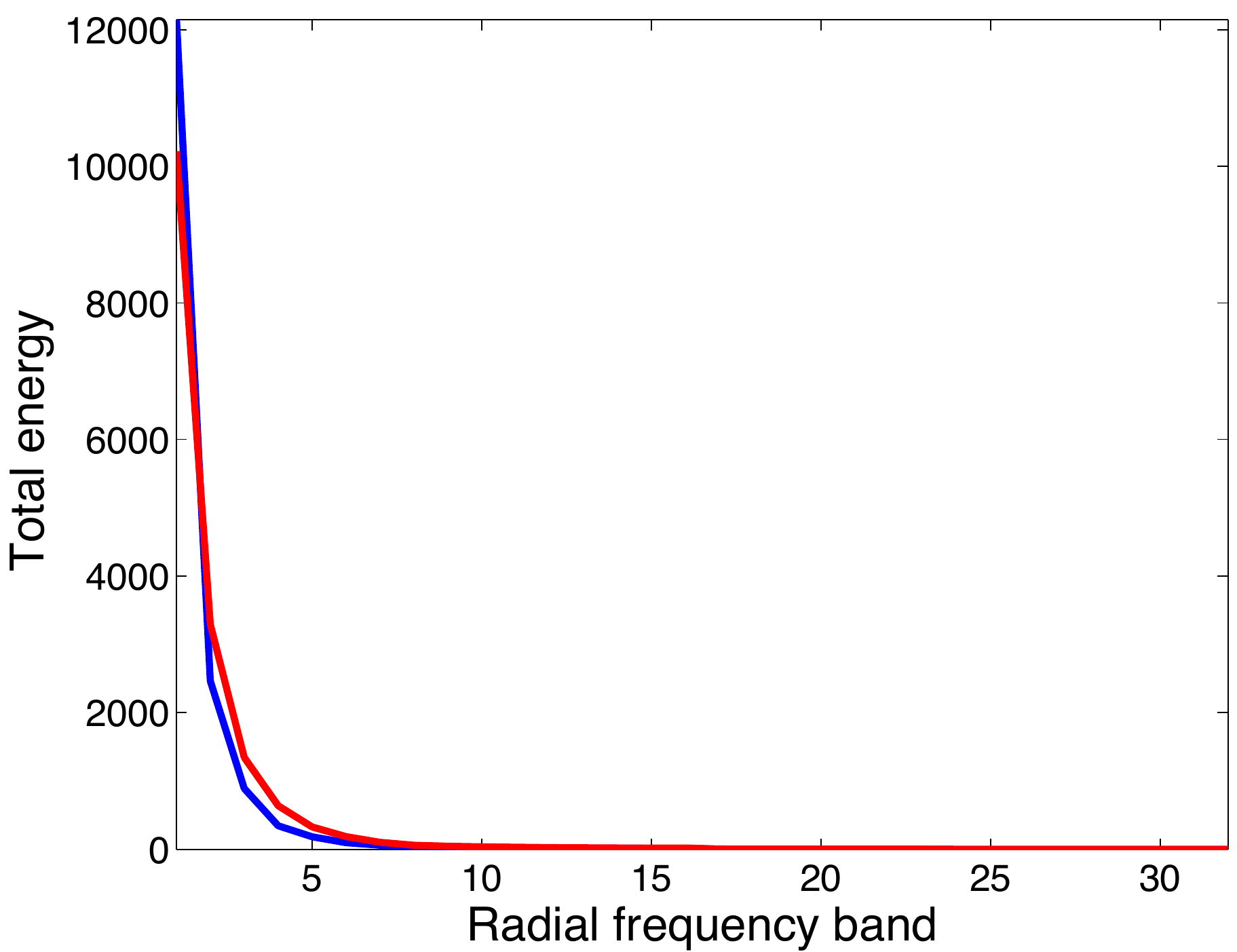}
  }
  \centerline{\small \hfill $i = 3$\hfill \hfill $i=4$\hfill \hfill$i =5$\hfill}
  \centerline{\small   
    \begin{rotate}{90}~~~~~~~~~~$\widetilde{\cal E}^i(l)$\end{rotate}
    \includegraphics[width= 12pc]{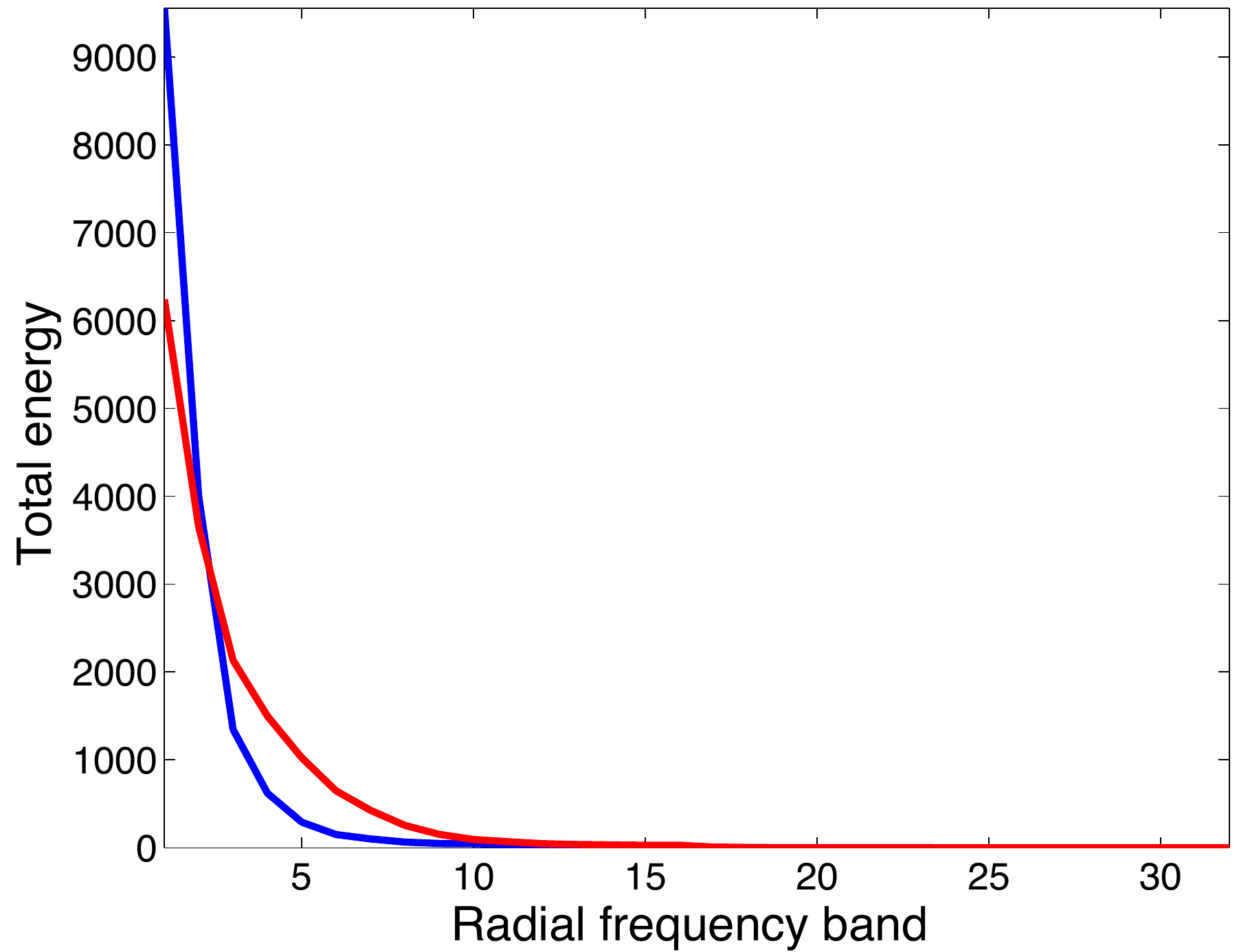}
    \includegraphics[width= 12pc]{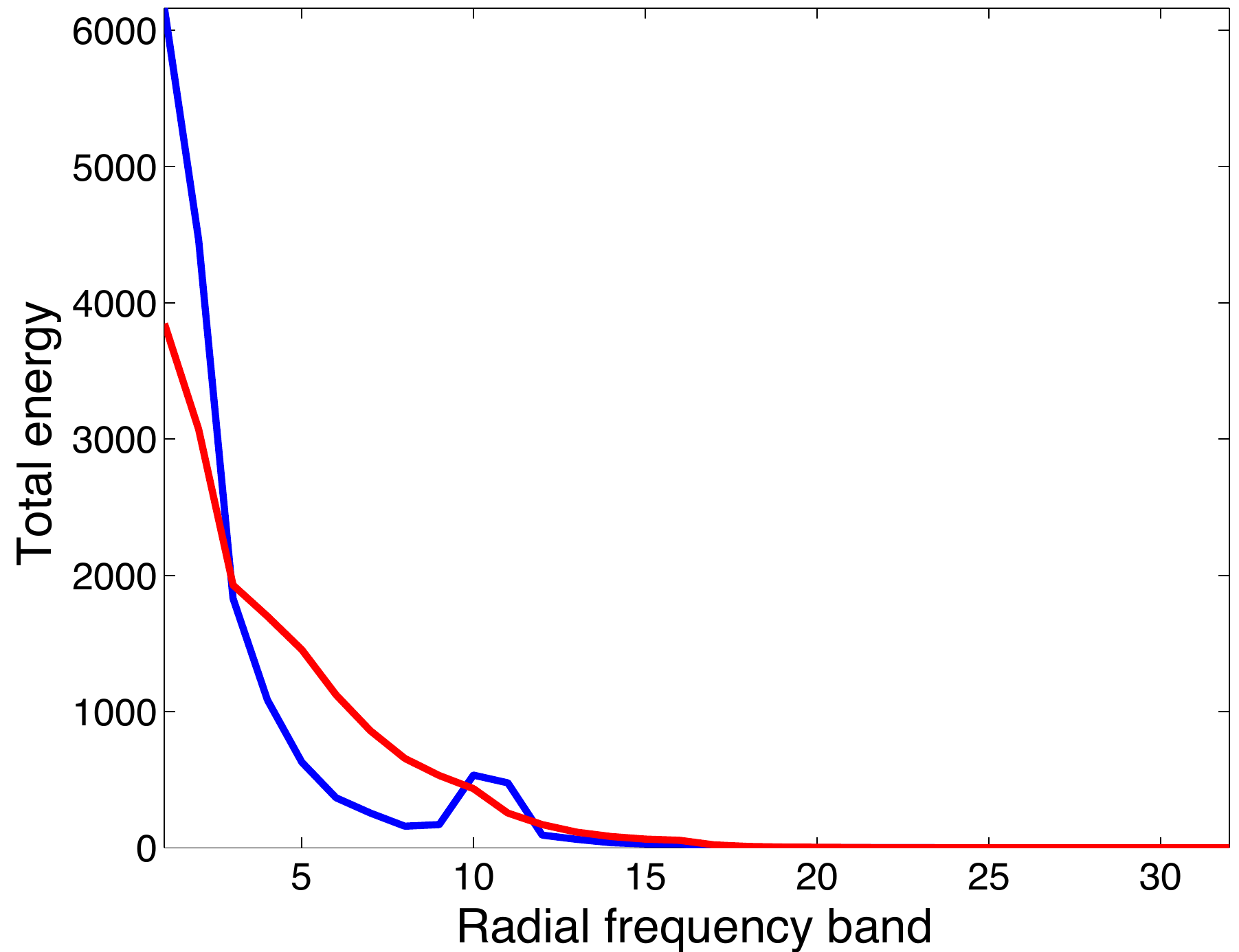}
    \includegraphics[width= 12pc]{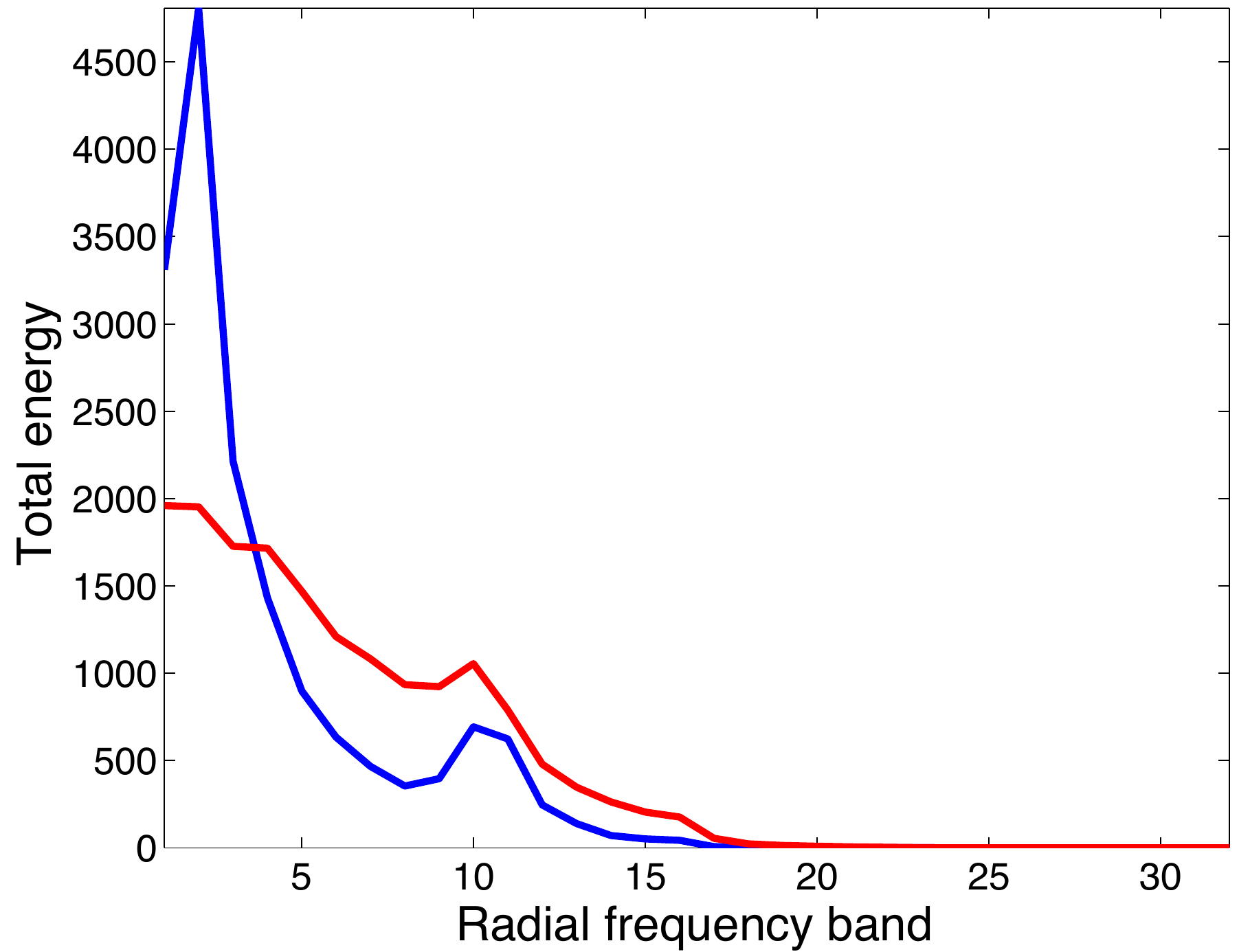}
  }
  \centerline{\small  \hfill $i = 6$\hfill $i=7$\hfill \hfill}
  \centerline{\small 
    \begin{rotate}{90}~~~~~~~~~~$\widetilde{\cal E}^i(l)$\end{rotate}
    \includegraphics[width= 12pc]{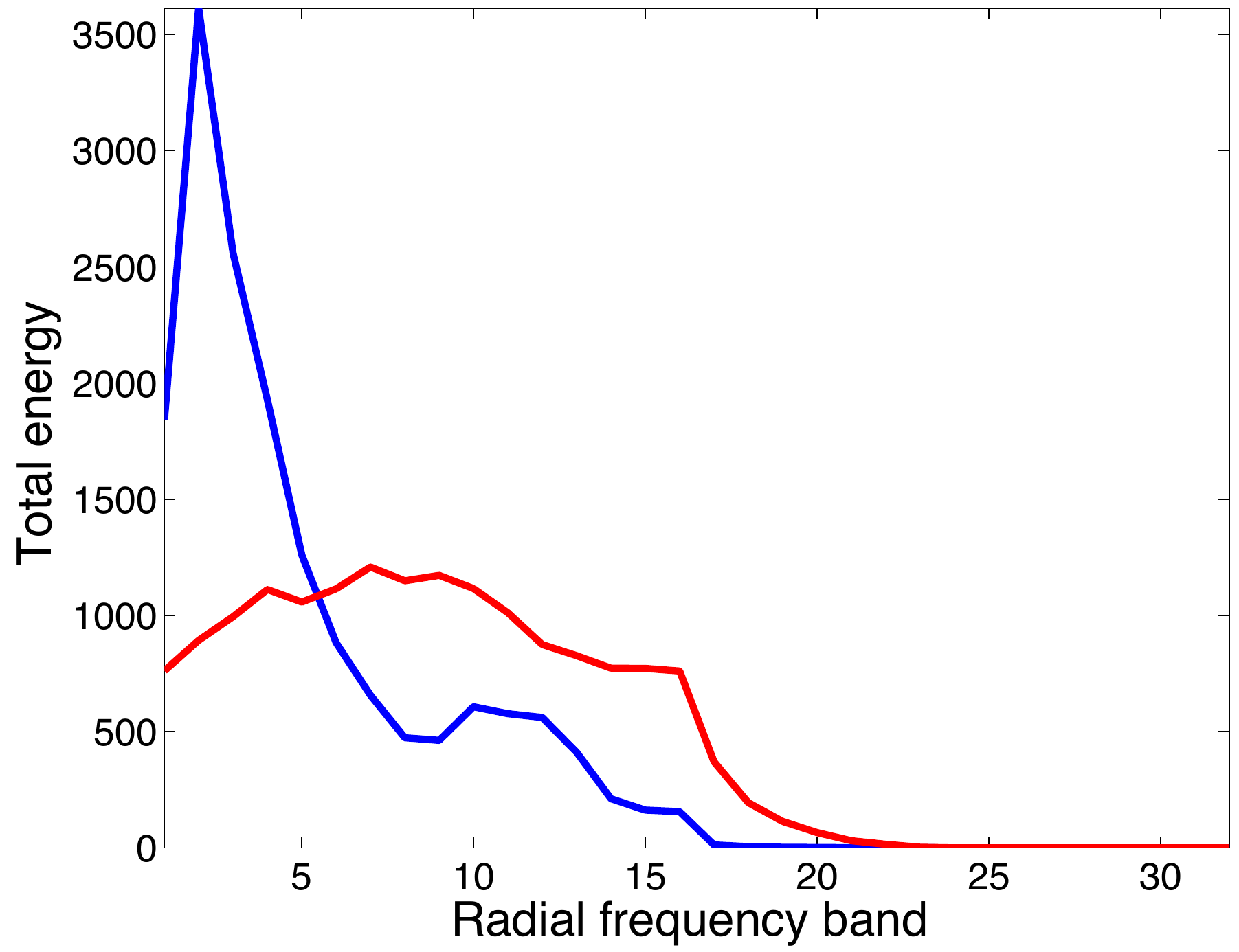}
    \includegraphics[width= 12pc]{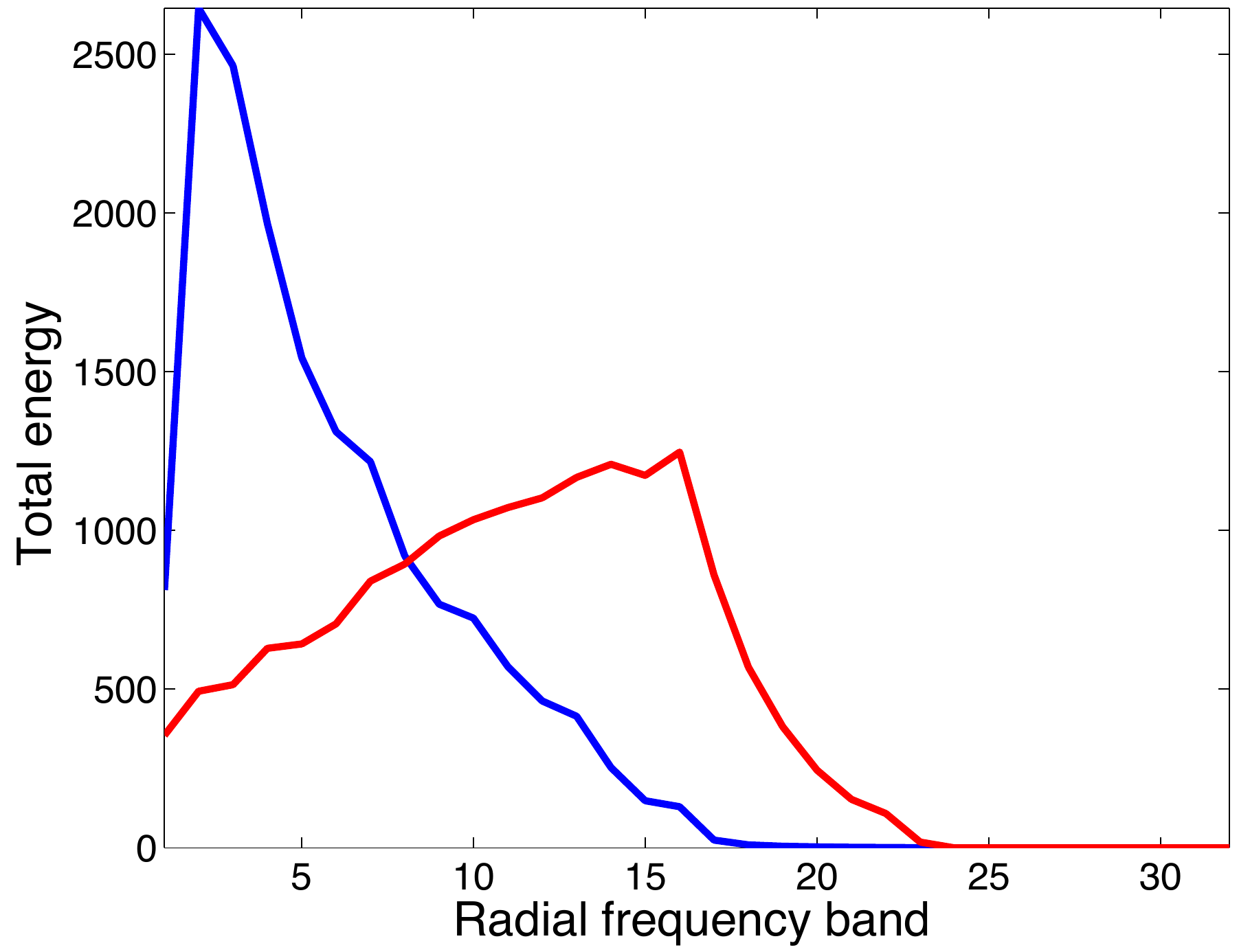}
    ~~~~\raisebox{4pc}{
      \begin{minipage}[b]{7pc}
        $\bfi_k$ in {\blue blue} \\
        $\widetilde{\bfi}_k$ in {\red red}
      \end{minipage}
    }}
  \caption{Energy $\widetilde{\cal E}^i(l)$ and ${\cal E}^i(l)$ of the
    eigenvectors $\bfi_k$ and $\tfi_k$, respectively, as a function of
    the radial frequency index $l$ for the ``clown'' image
    ($i=0,\ldots 7$ from top to bottom, and left to right). The
    $\tfi_k$ are computed using the noisy image.
 \label{noisy_edp} }
\end{figure}

{\noindent \bfseries A quantitative analysis of the perturbation of
  the $\bfi_k$}. We provide here a quantitative comparison between
$\bfi_k$ and $\tfi_k$. Our evaluation is based on the comparison
between the energy distribution of the Fourier transforms $\ftt_k$ and
$\ftf_k$.  Following a practice standard in vision research, we
analyse $\ftt_k$ as a function of the radial and angular frequencies,
$(\rho, \theta)$.  Because we assume that the visual features (edges,
textures, etc) can have any random orientation, we integrate $|\ftt_k|^2$
over all possible orientation $\theta$ and compute the energy
of $\ftt_k$ within a thin  ring $R_l$,
\begin{equation}
  \widetilde{\cal E}_k(l) = \int_{R_l} |\ftt_k|^2(\rho,\theta) \,d\rho \,d\theta,
\end{equation}
where the ring $R_l$ is defined by
\begin{equation*}
  R_l = \left \{(\rho,\theta) \in [l \Delta_\rho , (l+1)\Delta_\rho] \times [0,2\pi]  
    \right\}, \qquad \qquad l = 0, \ldots, L-1.
\end{equation*}
The width of each ring ($\Delta_\rho$) depends on the sampling
frequency and the size of the image. In the experiments we use a total
of $L=32$ rings, and $\Delta_\rho =2$ for an image of size $128 \times
128$. In other words, there were two discrete frequencies within each
ring $R_l$.  We do not study the perturbation of each eigenvector
individually. Instead, we group the eigenvectors $\tfi_k$ according to
a dyadic scale of the index $k$: the scale $i$ corresponds to the
eigenvectors
\begin{equation*}
  \left\{\tfi_{2^i+1},\cdots, \tfi_{2^{i+1}}\right\}, \quad i=0,\ldots.
\end{equation*} 
This dyadic repacking of the eigenvectors was motivated by the
experimental observation that the eigenvectors $\tfi_{k's}$ with the
same dyadic scale $i$ had a similar Fourier transform. Finally, we
compute at each scale $i=0,\ldots$ the total energy contribution,
integrated inside the ring $R_l$, from the group of eigenvectors at
that scale $i$,
\begin{equation*}
  \widetilde{\cal E}^i(l) = \sum_{k=2^i+1}^{2^{i+1}} \widetilde{\cal E}_k(l).
\end{equation*}
Figure \ref{noisy_edp} shows the energy distribution $\widetilde{\cal
  E}^i(l)$ as a function of the radial frequency index $l$, for the
scales $i=0, \ldots 7$. For each plot, the $x$-axis is the index $l$
of the ring $R_l$, and therefore provides a quantization of the radial
frequency. The $y$-axis is the total energy contribution, measured
inside the ring $R_l$, from the group of eigenvectors at scale $i$. We
plot the energy $\widetilde{\cal E}^i(l)$ of the perturbed
eigenvectors (in blue), as well as the energy ${\cal E}^i(l)$ of the
clean eigenvectors (in red).

For the first three scales $i=0,1,2$, corresponding to $k=2,\ldots,8$,
there is hardly any difference between the energy distribution of the
clean and perturbed eigenvectors (see Fig. \ref{noisy_edp}). Starting
at scale $i=3$, the energy of the perturbed eigenvectors becomes
different from the energy of the clean eigenvectors: $\widetilde{\cal
  E}^i_l$ has a flatter distribution with more energy in the higher
frequencies. The energy leak into the high radial frequencies is
created by the noise and confirms the visual impression triggered by
$\tfi_{256}$ in Fig.~\ref{noisy-eig}. The high-index eigenvectors
$\tfi_k$ are trying to capture the noise present in the image
$\widetilde{u}$. As the scale $i$ further increases, the departure of
$\widetilde{\cal E}^i_l$ from ${\cal E}^i_l$ becomes even more%
%
\noindent 
dramatic. At scale $i=7$, the radial energy distribution of the
eigenvectors $\tfi_{129},\ldots,\tfi_{256}$ is almost flat, indicating
that the $\tfi_k$ have adapted to the noise. These eigenvectors are
obviously no longer suitable for denoising.\\

{\noindent \bf Which perturbation matters most: the topology of the graph, or the
  edge weights?}
As demonstrated in some recent studies \cite{mcgraw08,milanese10}
topological changes of a graph (created by the addition or removal of
edges) can significantly perturb the spectrum and the eigenvectors of
the graph Laplacian. Inspired by these studies, we analyze the effect
of the modifications of the topology of the graph on the
eigenvectors~$\widetilde{\bfi_k}$.

Precisely, we verify experimentally the following result: if the
non-zero entries in the matrix $\bW$ (\ref{weight}) are kept at the
same locations (we keep the graph topology), and if we randomly
perturb the weights according to (\ref{noisyweight}), then the
perturbed eigenvectors have almost the same energy distribution in the
Fourier domain as the eigenvectors computed from the original weight
matrix. In the previous paragraph we observed that random fluctuations
of the weights combined  with perturbations of the graph topology
(caused by the random changes of the local distance, and the
corresponding neighbors) significantly perturbed the eigenvectors of
$\bL$.  Combining these two experiments, we conclude that the topology
of the graph is the most important factor, since, if preserved, it
guarantees the stability of the eigenvectors.

Let us now describe the experiment. We first build a patch-graph and a
weight matrix $\bW$ based on the clean image. We now add a random
realization of white Gaussian noise ($\sigma = 40$), and compute the
new weights $\widetilde{w}_{n,m}$ according to (\ref{weight}), where
the neighbors are now defined by the clean patch-graph. We then update
all the non-zero entries of the matrix $\bW$ using the perturbed
weights $\widetilde{w}_{n,m}$. This constitutes the perturbed matrix
$\widecheck{\bW}$. We define $\widecheck{\bL}$ accordingly.

We note that $\widecheck{\bW}$ is not equal to the matrix
$\widetilde{\bW}$, described in the previous paragraph. Indeed, the
nonzero entries of $\widetilde{\bW}$ correspond to the edges of the
graph defined by the perturbed distance $\widetilde{d}_{n,m}$. In
other words, $\widetilde{w}_{n,m} =\widecheck{w}_{n,m}$ only if
$\widetilde{\bu}(\bx_n)$ and $\widetilde{\bu}(\bx_m)$ are neighbors,
and $\bu(\bx_n)$ and $\bu(\bx_m)$ are also neighbors. 

\begin{figure}[H]
  \centerline{  
    \includegraphics[width= 0.20\textwidth]{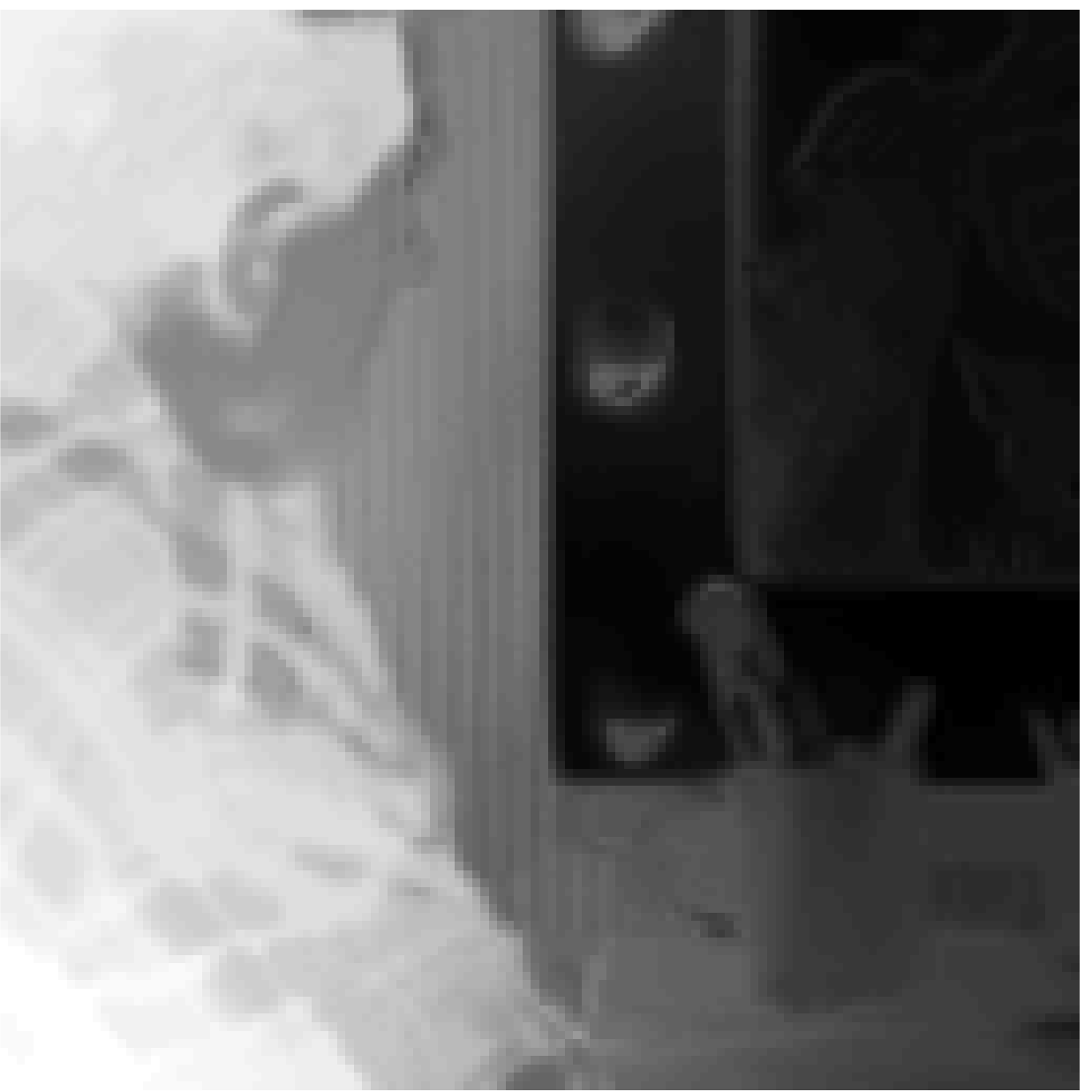}\hspace*{-0.25pc}  
    \includegraphics[width= 0.20\textwidth]{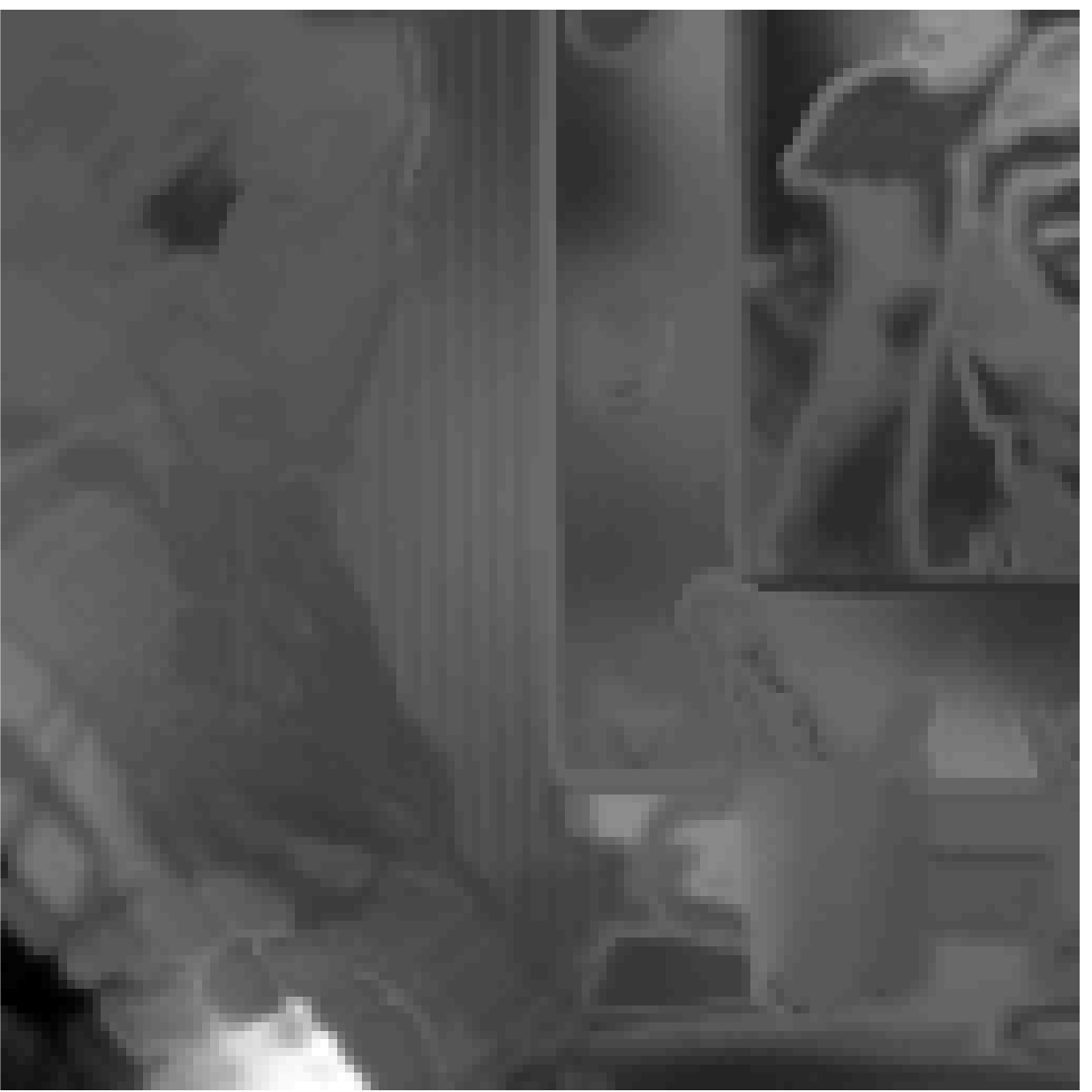}\hspace*{-0.25pc}  
    \includegraphics[width= 0.20\textwidth]{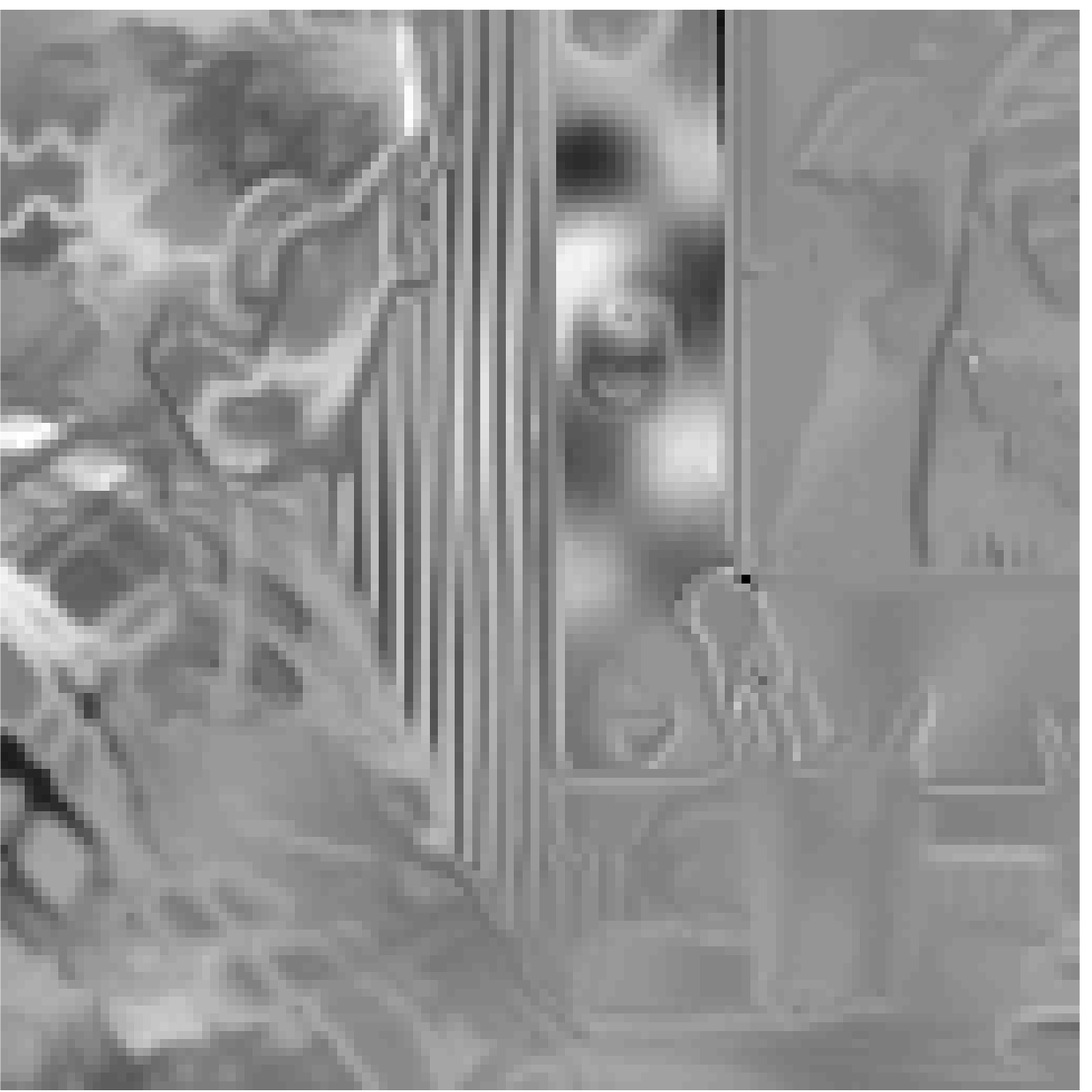}\hspace*{-0.25pc}  
    \includegraphics[width= 0.20\textwidth]{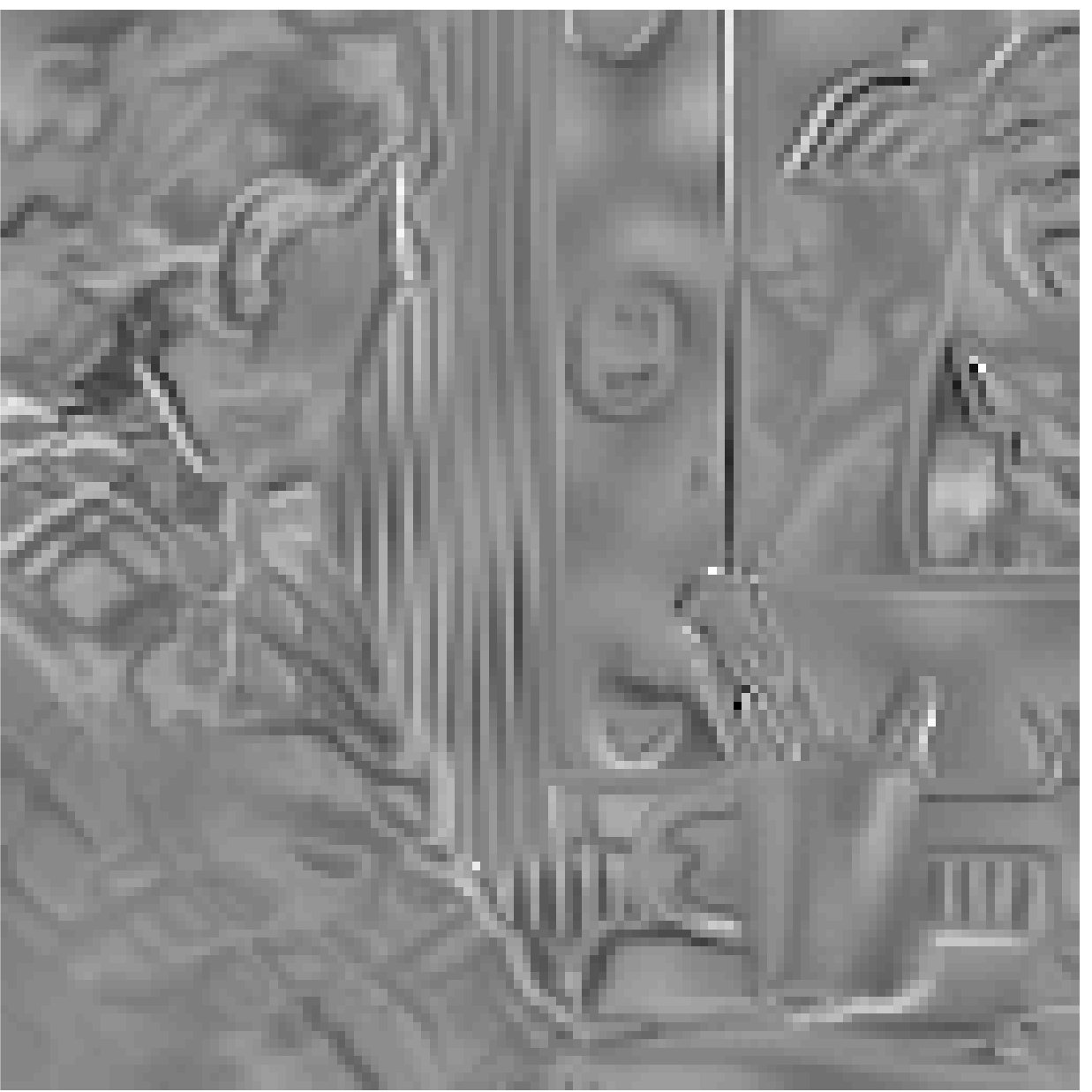}
  }
 \centerline{\hspace*{4pc} $\cfi_2$ \hfill\hspace*{2pc}   $\cfi_{32}$ \hfill $\cfi_{128}$\hfill $\cfi_{256}$\hfill} 
  \caption{The topology of the graph that is used to compute the $\cfi_k$ is
    determined from the clean image. 
    \label{noisy_on_clean_eigs}
  }
\end{figure}

\begin{figure}[H]
  \centerline{\small \hfill $i = 0$\hfill\hfill $i=1$\hfill\hfill$i =2$\hfill}
  \centerline{\small
    \begin{rotate}{90}~~~~~~~~~~$\widecheck{{\cal E}^i}(l)$\end{rotate}
    \includegraphics[width= 12pc]{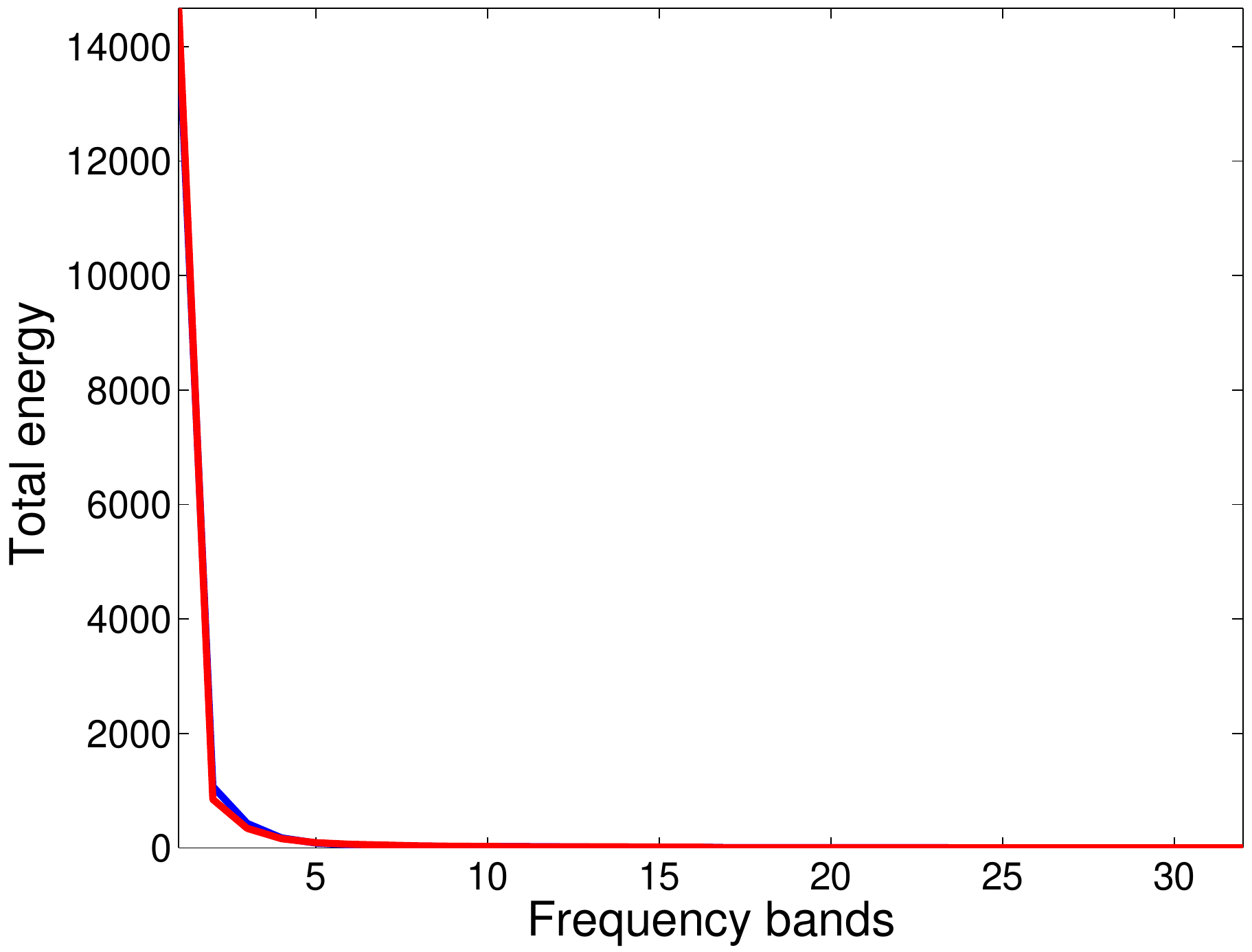}
    \includegraphics[width= 12pc]{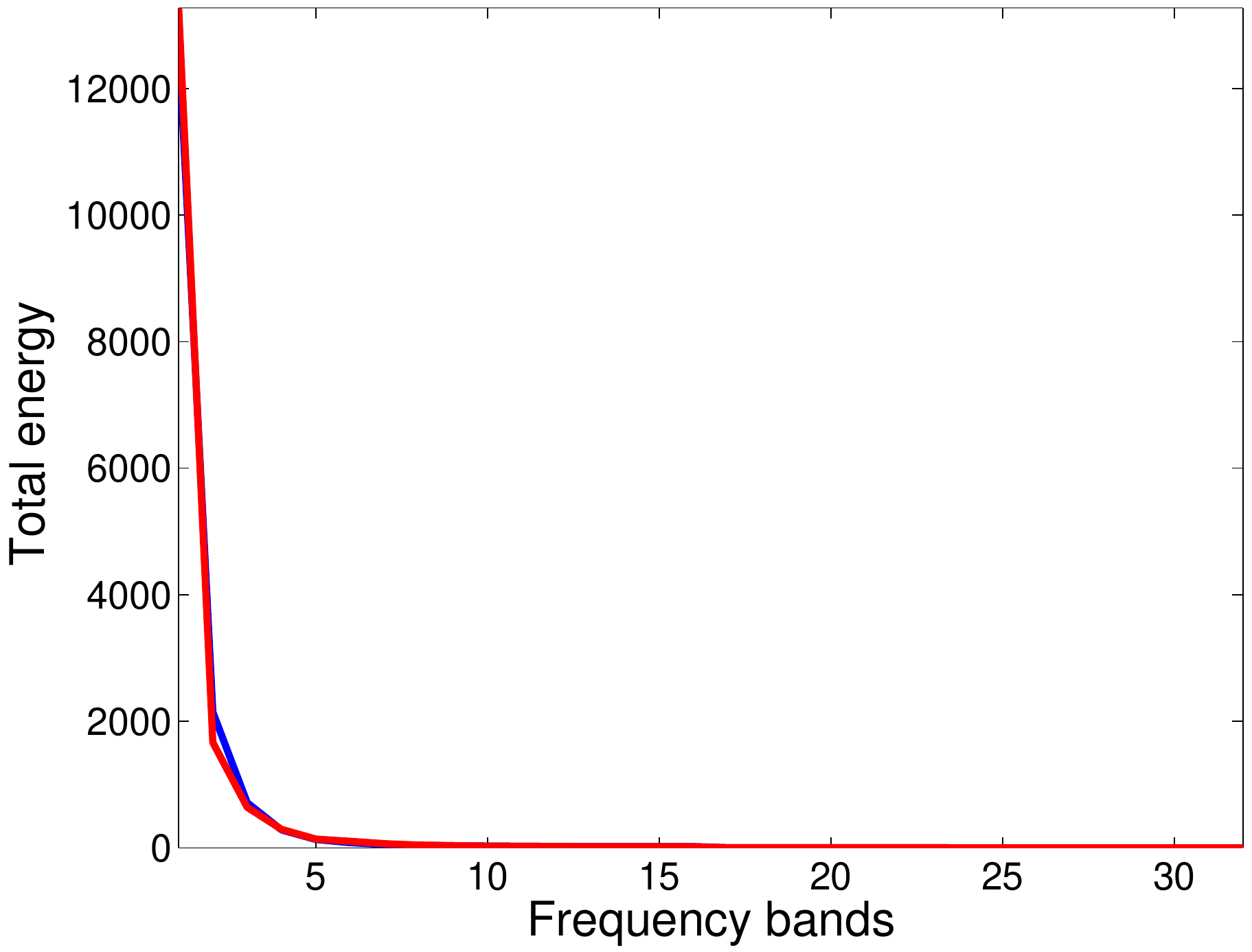}
    \includegraphics[width= 12pc]{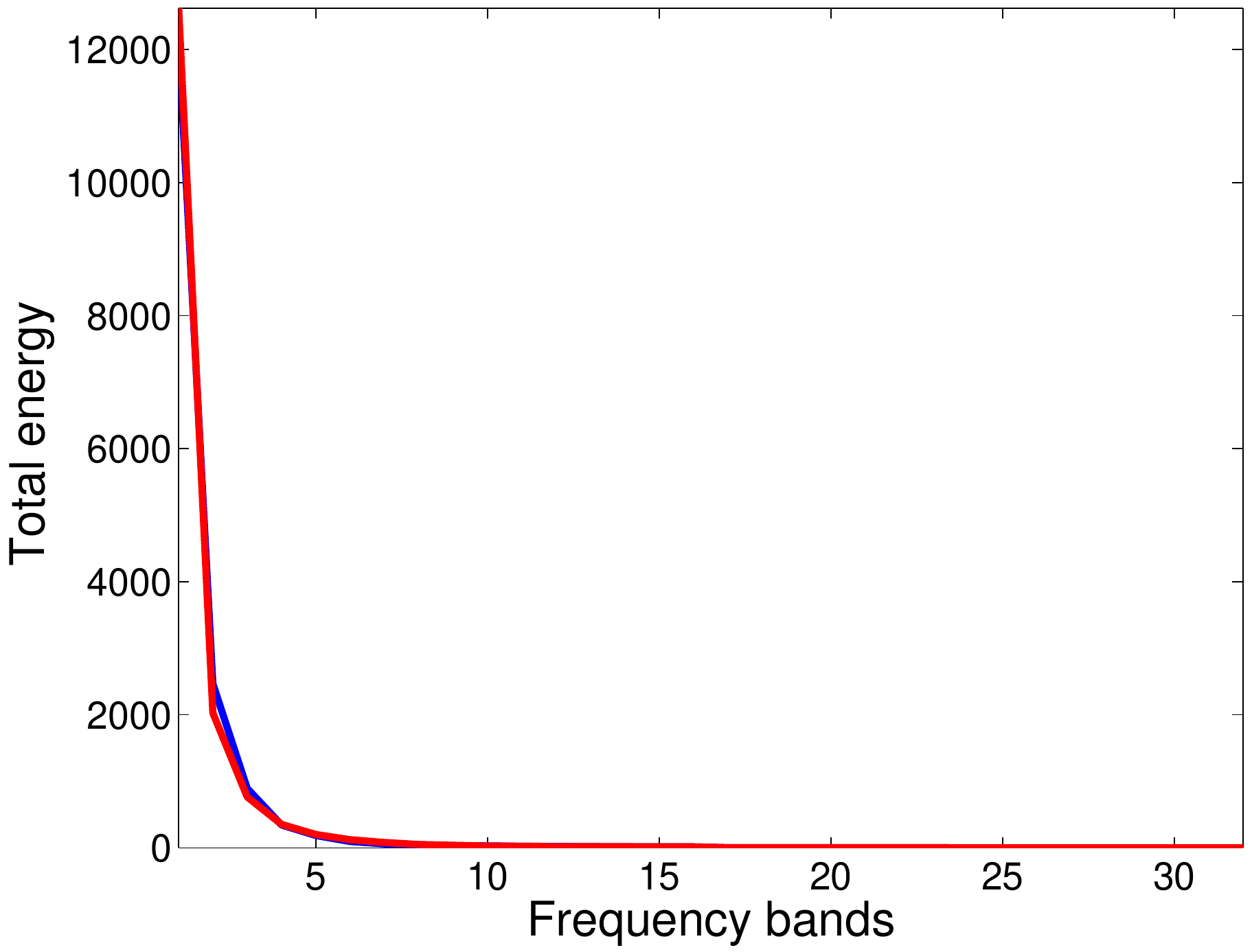}
  }
  \centerline{\small \hfill $i = 3$\hfill \hfill$i=4$\hfill\hfill$i =5$\hfill}
  \centerline{\small   
    \begin{rotate}{90}~~~~~~~~~~$\widecheck{{\cal E}^i}(l)$\end{rotate}
    \includegraphics[width= 12pc]{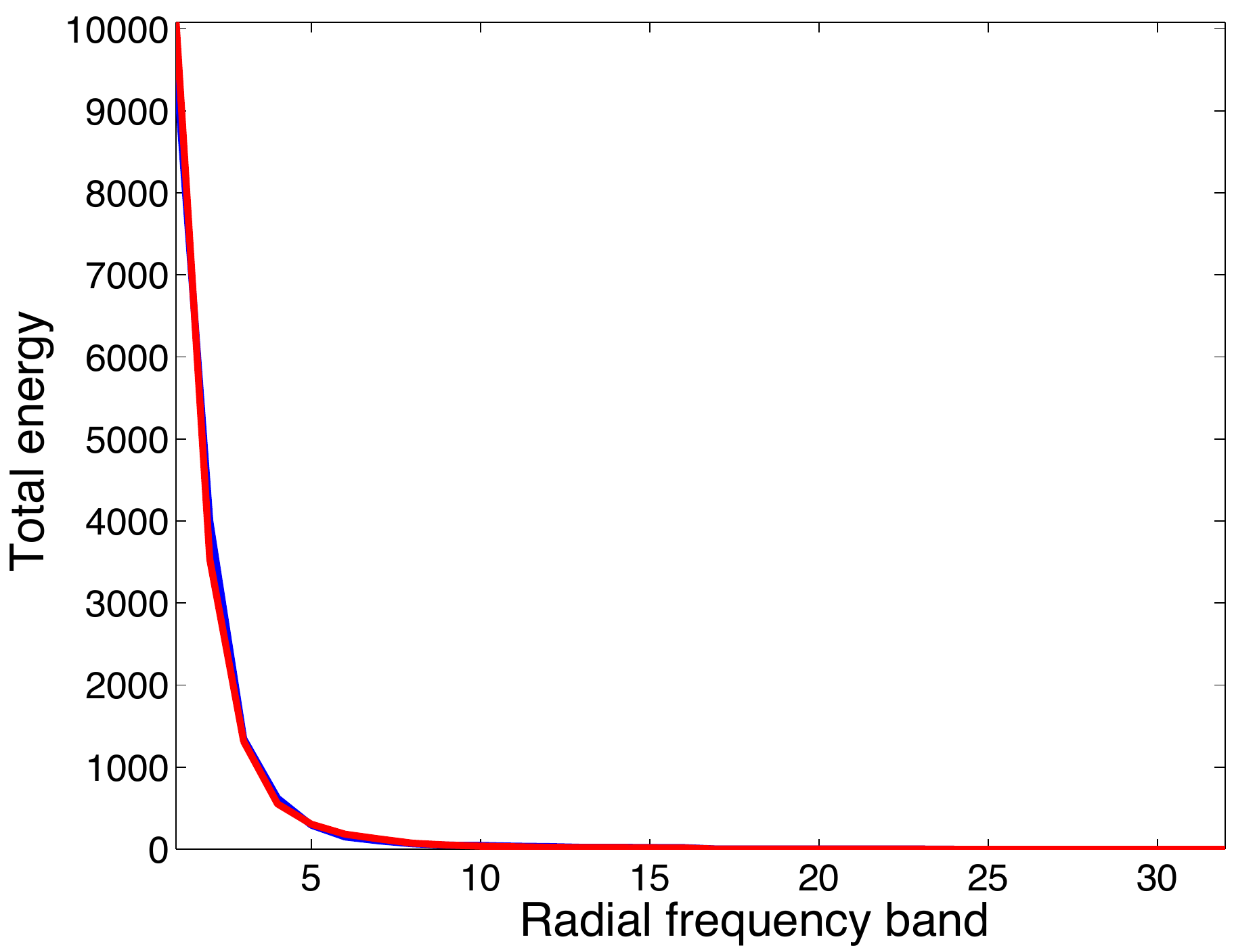}
    \includegraphics[width= 12pc]{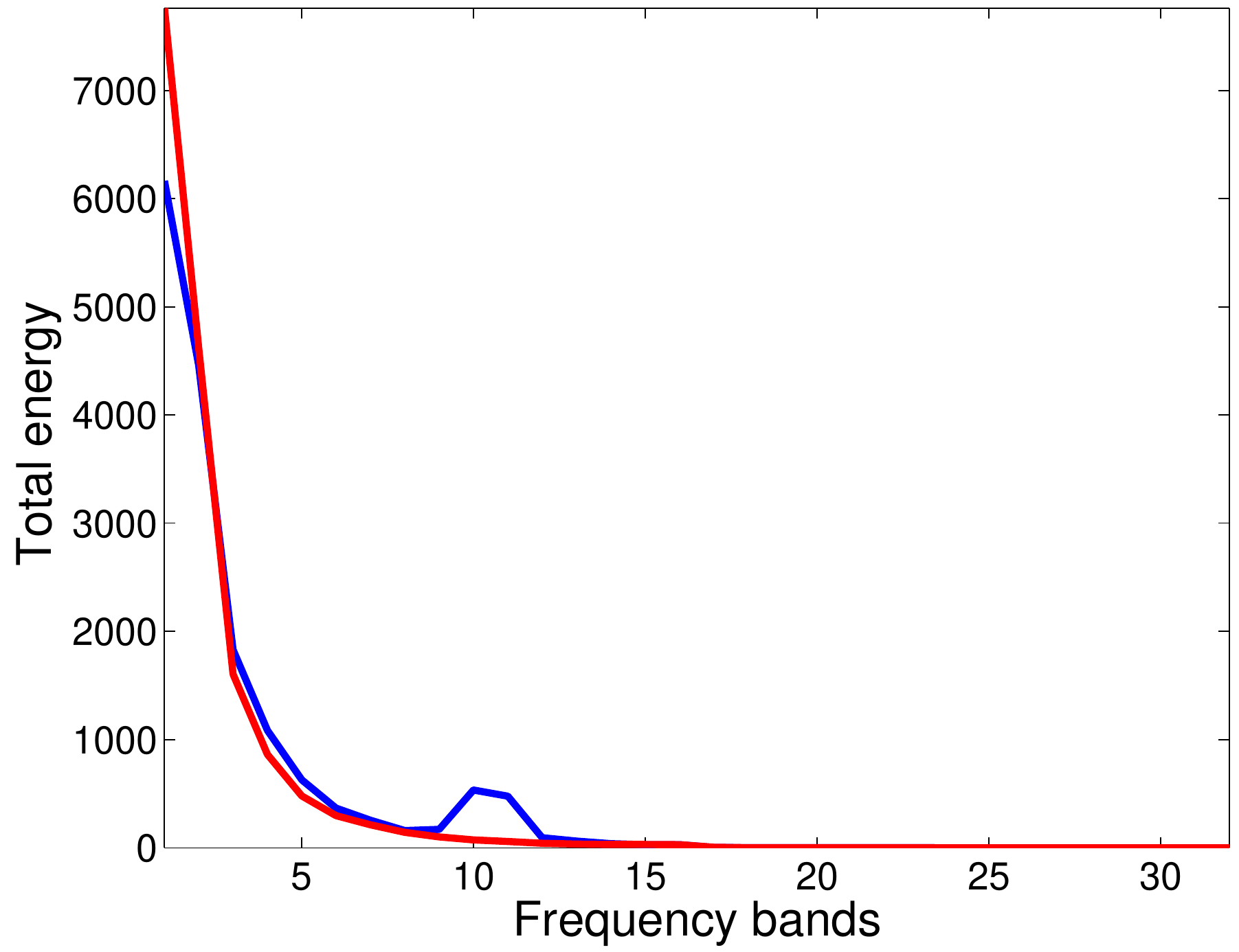}
    \includegraphics[width= 12pc]{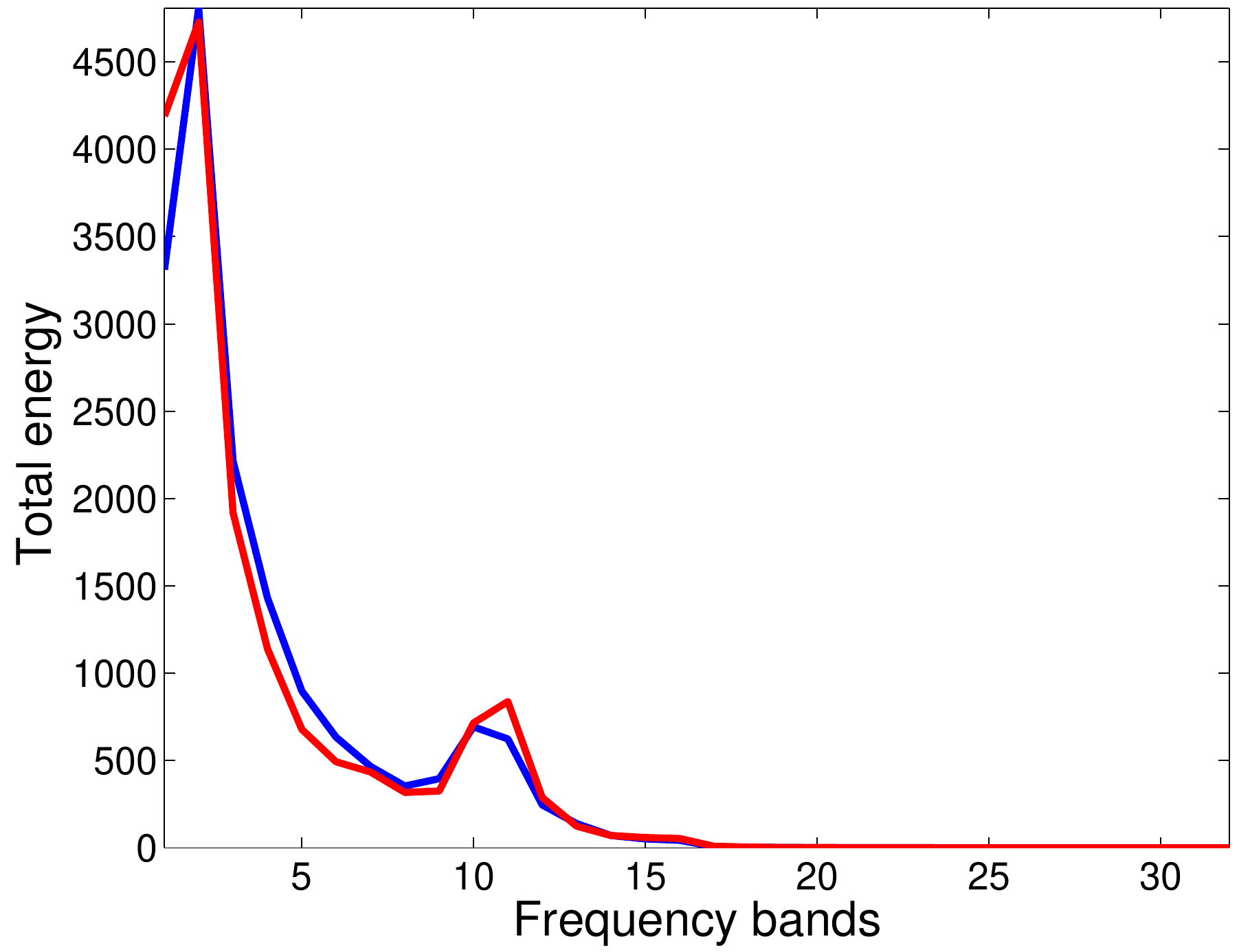}
  }
  \centerline{\small  \hfill $i = 6$\hfill $i=7$\hfill \hfill}
  \centerline{\small 
    \begin{rotate}{90}~~~~~~~~~~$\widecheck{{\cal E}^i}(l)$\end{rotate}
    \includegraphics[width= 12pc]{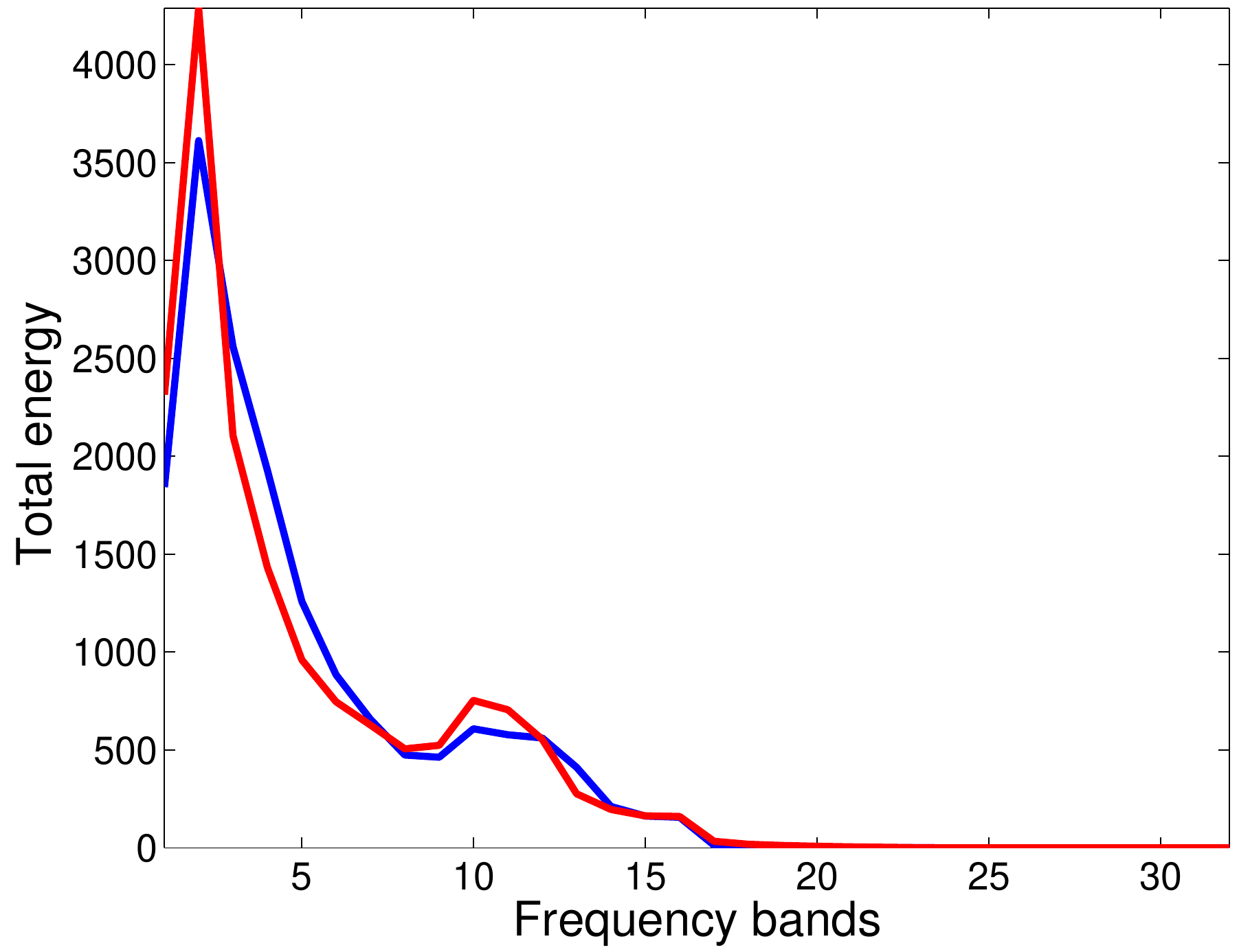}
    \includegraphics[width= 12pc]{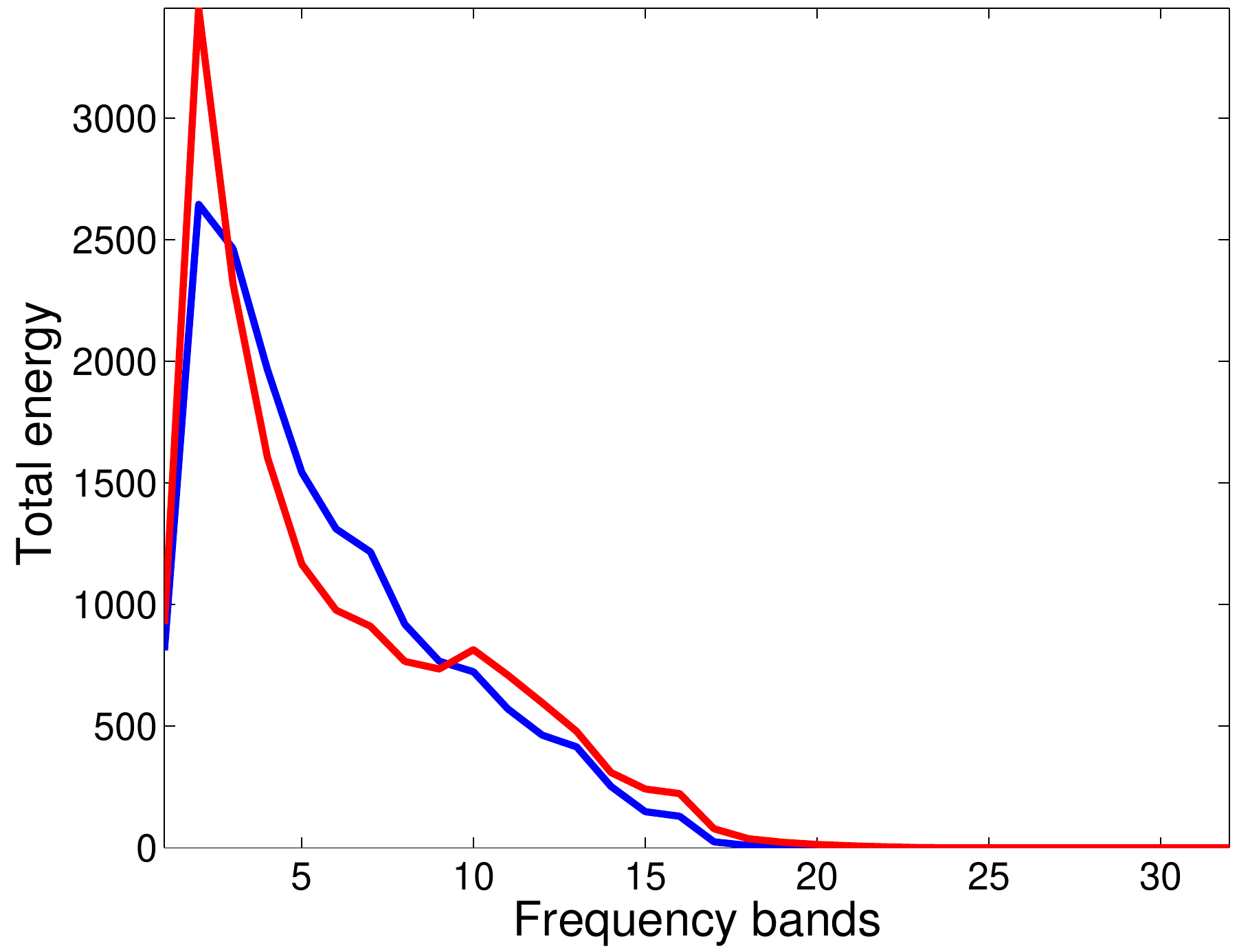}
    ~~~~\raisebox{4pc}{
      \begin{minipage}[b]{7pc}
        $\bfi_k$ in {\blue blue} \\
        $\cfi_k$ in {\red red}
      \end{minipage}
    }}
  \caption{Energy $\widecheck{{\cal E}^i}(l)$ and ${\cal E}^i(l)$  of
    the eigenvectors $\widecheck{\bfi}_k$ (red) and $\bfi_k$ (blue) as
    a function of the radial frequency index $l$ for the ``clown''
    image ($i=0,\ldots 7$ from top to bottom, and left to  right). The
    topology of the graph that is used to compute the $\cfi_k$ is
    determined from the clean image. 
    \label{noisy_on_clean_edp}}
\end{figure}

Let $\cfi_k$ be the eigenvectors of the perturbed matrix
$\widecheck{\bL}$.  Figure \ref{noisy_on_clean_eigs} displays
$\cfi_2$, $\cfi_{32}, \cfi_{128}$, and $\cfi_{256}$. Despite the fact
that the entries of the matrix $\bL$ have been perturbed by adding a
large amount of noise to the image, the eigenvectors of the resulting
matrix, $\widecheck{\bL}$, appear visually very similar to the
original eigenvectors of $\bL$. This observation is confirmed in Fig.
\ref{noisy_on_clean_edp}, which shows the energy distribution ${\cal
  E}^i(l)$ as a function of the radial frequency for the two sets of
eigenvectors: $\bfi_k$ and $\cfi_k$. Remarkably, the distribution of
energy of $\bfi_k$ and $\cfi_k$, measured across the different radial
frequencies in the Fourier plane, are very similar. The eigenvectors
$\cfi_k$ do not suffer from any significant perturbation despite very
significant changes in the image patches.

This experimental result is related to similar properties in the
context of spectral geometry. If we replace the graph $\Gamma$ by a
manifold, and we consider that the weight matrix encodes the metric,
then our experimental results are related to the more general
phenomenon that involves the stability%
\noindent of the eigenspaces of the Laplace-Beltrami operator when the
metric on the manifold is perturbed continuously
\cite{Barbatis96,Davies90}. In our case, the continuous changes of the
metric correspond to changes in the weights $w_{n,m}$.

In practice, we do not have access to $\bW$, and therefore we cannot
construct $\widecheck{\bW}$. However, the results of this section
suggest that we should try and reconstruct an estimate of the graph
topology, associated with $\bW$. We combine the following
two results to construct an estimate of $\bW$.
\begin{enumerate}
\item For most images, many of the entries of  the weight matrix $\bW$
  can be
  estimated from a lowpass version of the image. In other words,
\begin{equation*}
\| \bu(\bx_n) - \bu(\bx_m)\| \approx
\| \bm{h u}(\bx_n) - \bm{h u}(\bx_m)\|,
\end{equation*}
where $h$ is a lowpass filter, and $\bm{h u}$ is a patch from the
image $h u$. Indeed, unless we are studying patches
that contain only high frequencies, most of the energy within the
patch comes from the low frequencies.
\item The eigenvectors from the very first scales, $i=0,\ldots,4$, are
  very stable even when a large amount of noise is added to the
  image. We can therefore use these eigenvectors to estimate a
  lowpass version of the original image.
\end{enumerate}
The strategy for denoising is now clear. We compute the weight matrix
$\widetilde{\bW}$ and calculate the eigenvectors associated with the
first scales $i=0,1,2,3,4$.  We use these eigenvectors to compute a
coarse estimate of the lowpass part of the image, $\widehat{u}^{(1)}$,
using the procedure described in section \ref{denoise-algo}. Because
too few eigenvectors are used in this reconstruction, we add back a
scaled version of the residual error (the difference between the noisy
image $\widetilde{u}$ and the denoised image $\widehat{u}^{(1)}$) to
obtain the intermediate image $\widehat{u}^{(2)}$ defined by
\begin{equation}
  \widehat{u}^{(2)} = \widehat{u}^{(1)} +
  \gamma(\widetilde{u}-\widehat{u}^{(1)}) = (1-\gamma) \widehat{u}^{(1)} +
  \gamma \widetilde{u},  
  \label{addback}
\end{equation}
where $\gamma \in (0,1)$. at this point the image $\widehat{u}^{(2)}$
has been sufficiently denoised, and we construct a new graph. We note
that we have experimented with a different procedure to compute the
image $\widehat{u}^{(2)}$, using standard lowpass filtering, or
wavelet denoising. Using these alternate approaches, we obtain similar
results, albeit of lower quality (results not shown).

\begin{figure}[H]
  \centerline{  
    \includegraphics[width= 0.20\textwidth]{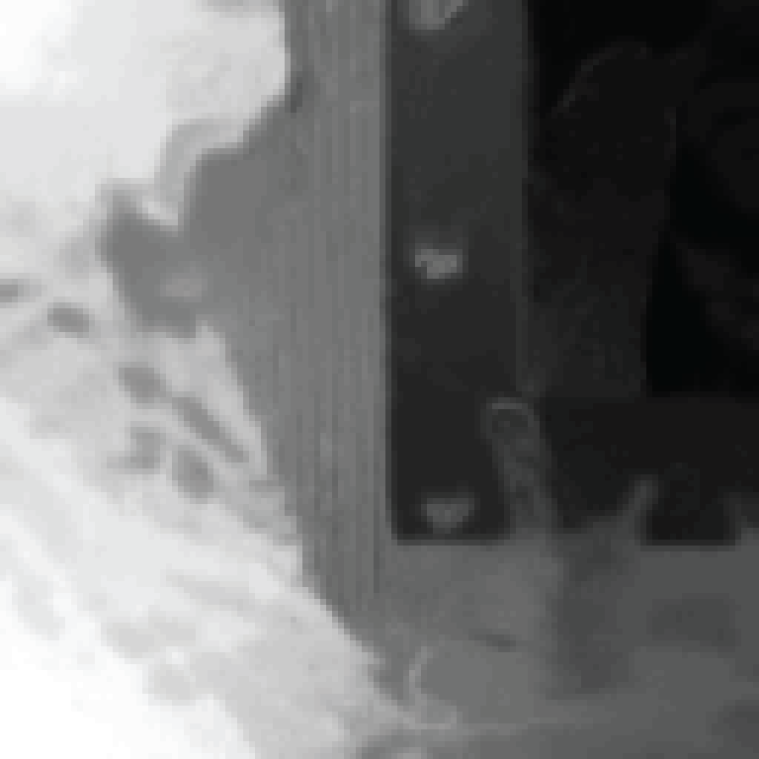}\hspace*{-0.25pc}  
    \includegraphics[width= 0.20\textwidth]{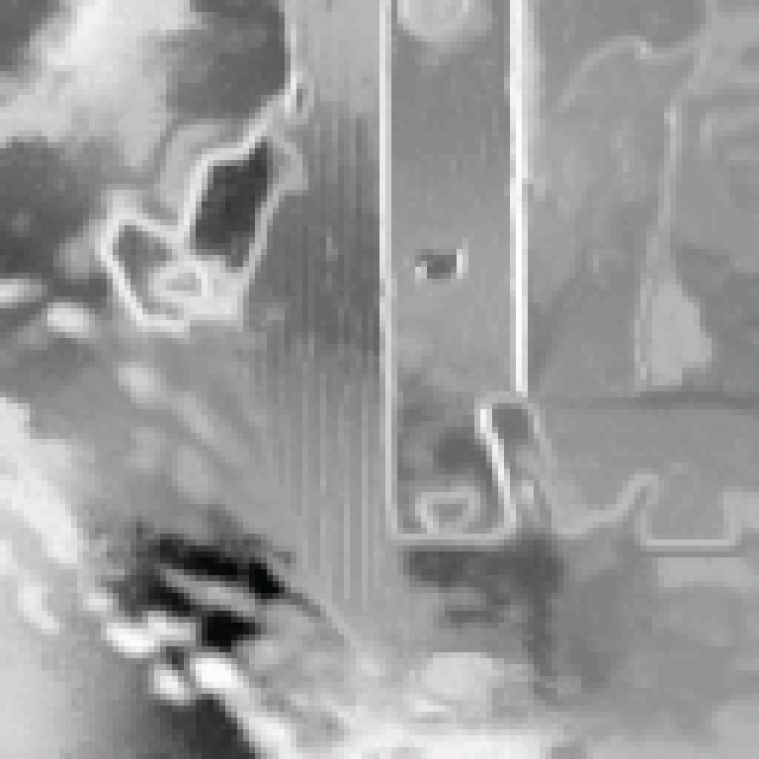}\hspace*{-0.25pc}  
    \includegraphics[width= 0.20\textwidth]{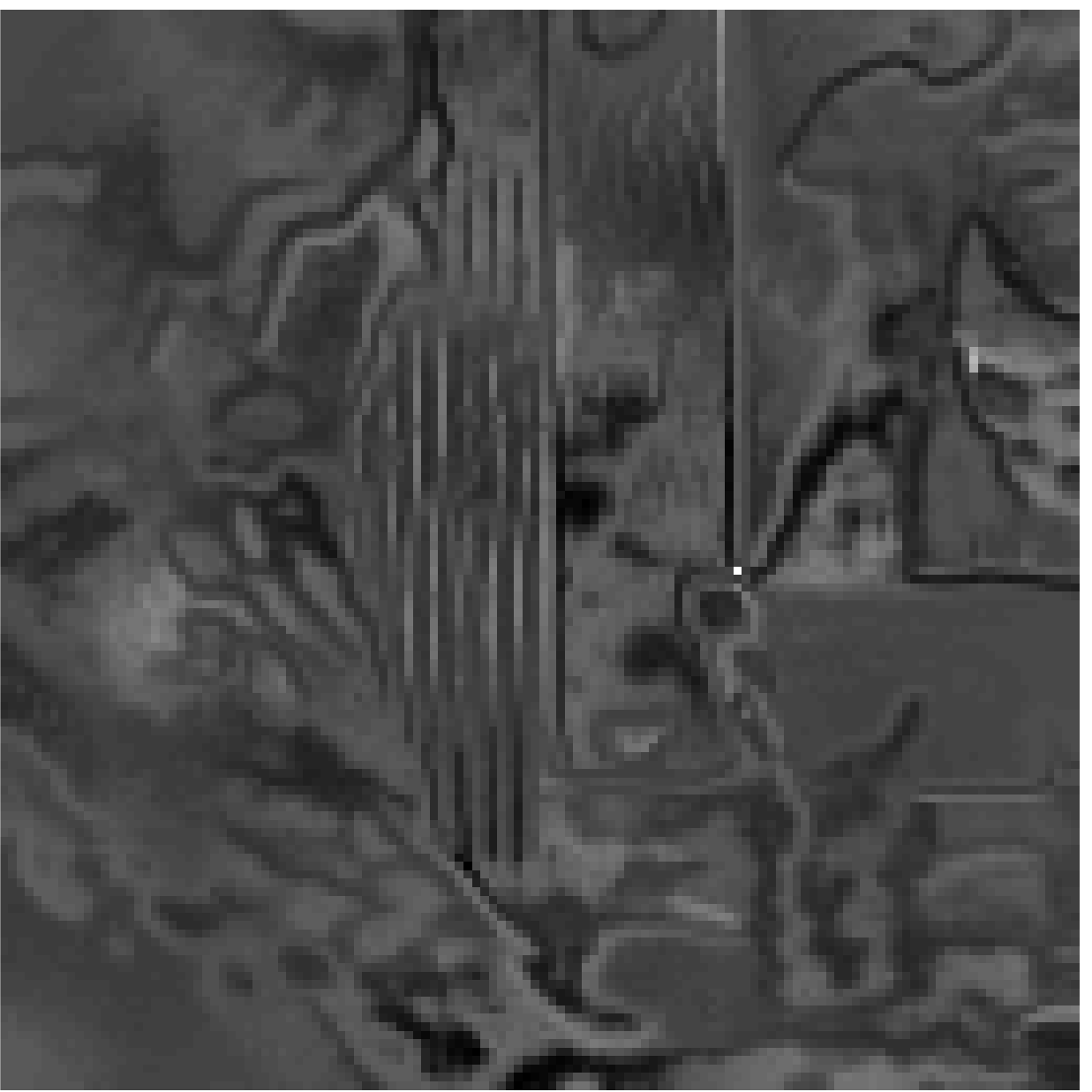}\hspace*{-0.25pc}  
    \includegraphics[width= 0.20\textwidth]{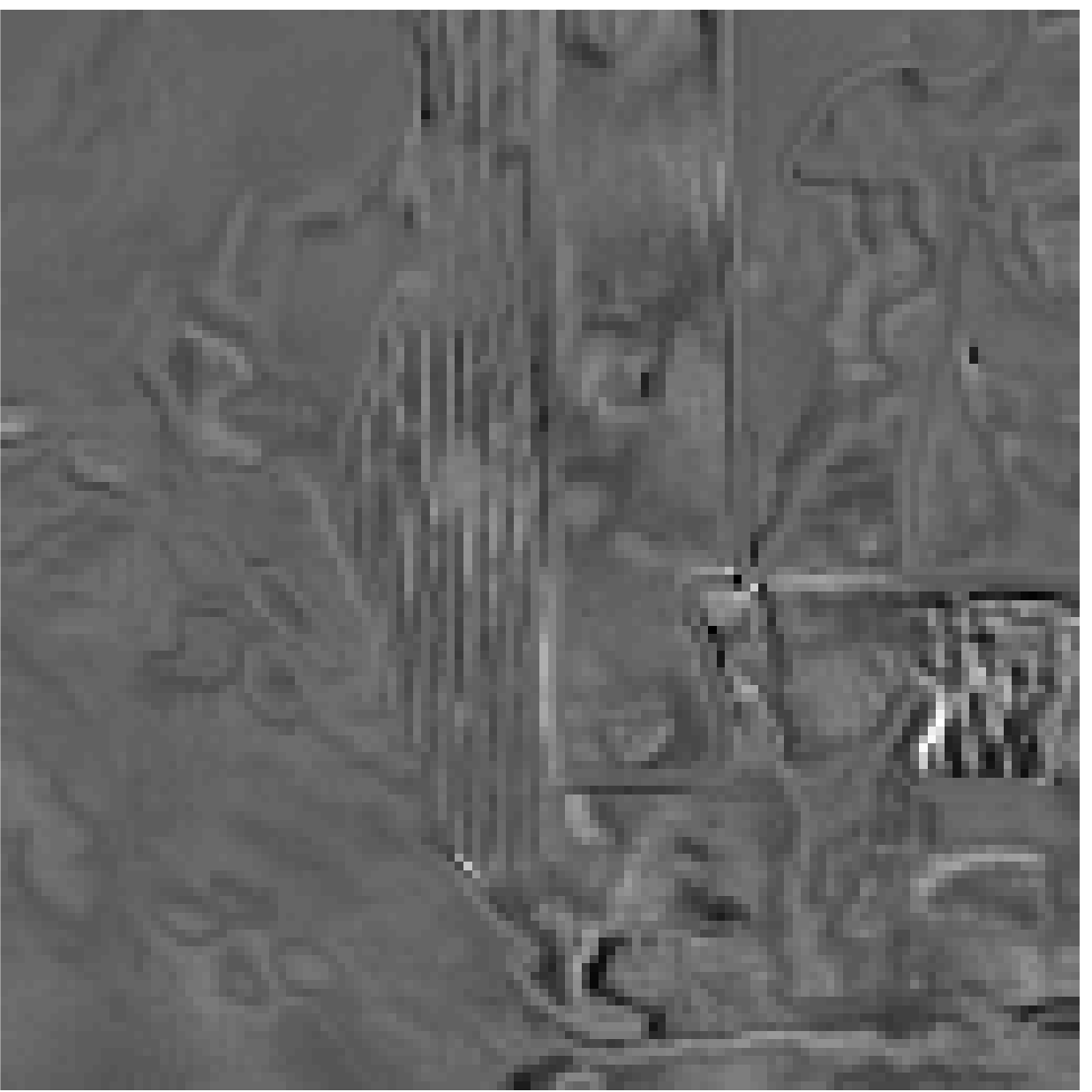}  
  }
 \centerline{\hspace*{4pc} $\hfi_2$ \hfill\hspace*{2pc}   $\hfi_{32}$
   \hfill $\hfi_{128}$\hfill $\hfi_{256}$\hfill}  
  \caption{The eigenvectors $\hfi_k$ are computed using a second graph that is built
    from the denoised image $\widehat{u}^{(2)}$. 
    \label{twostage_eig}
  }
\end{figure}

We finally compute the eigenvectors $\hfi_k$ of 
$\widehat{\bL}^{(2)}$ associated to the image $\widehat{u}^{(2)}$.
Figure \ref{twostage_eig} displays the eigenvectors $\hfi_2$,
$\hfi_{32}$, $\hfi_{128}$, and $\hfi_{256}$. These eigenvectors are
visually similar to those%
\begin{figure}[H]
  \centerline{\small \hfill $i = 0$\hfill\hfill $i=1$\hfill\hfill$i =2$\hfill}
  \centerline{\small
    \begin{rotate}{90}~~~~~~~~~~$\widehat{\cal E}^i(l)$\end{rotate}
    \includegraphics[width= 12pc]{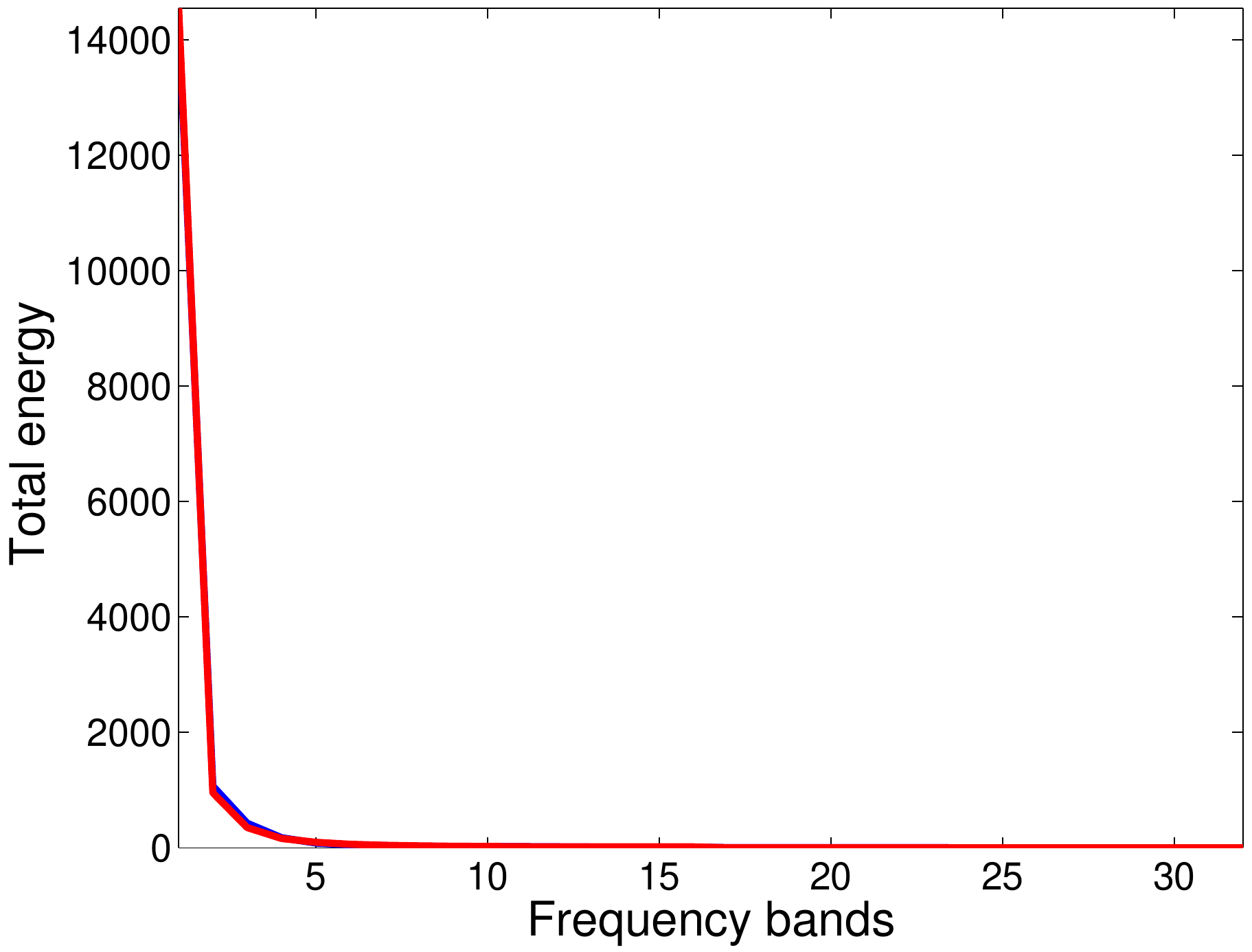}
    \includegraphics[width= 12pc]{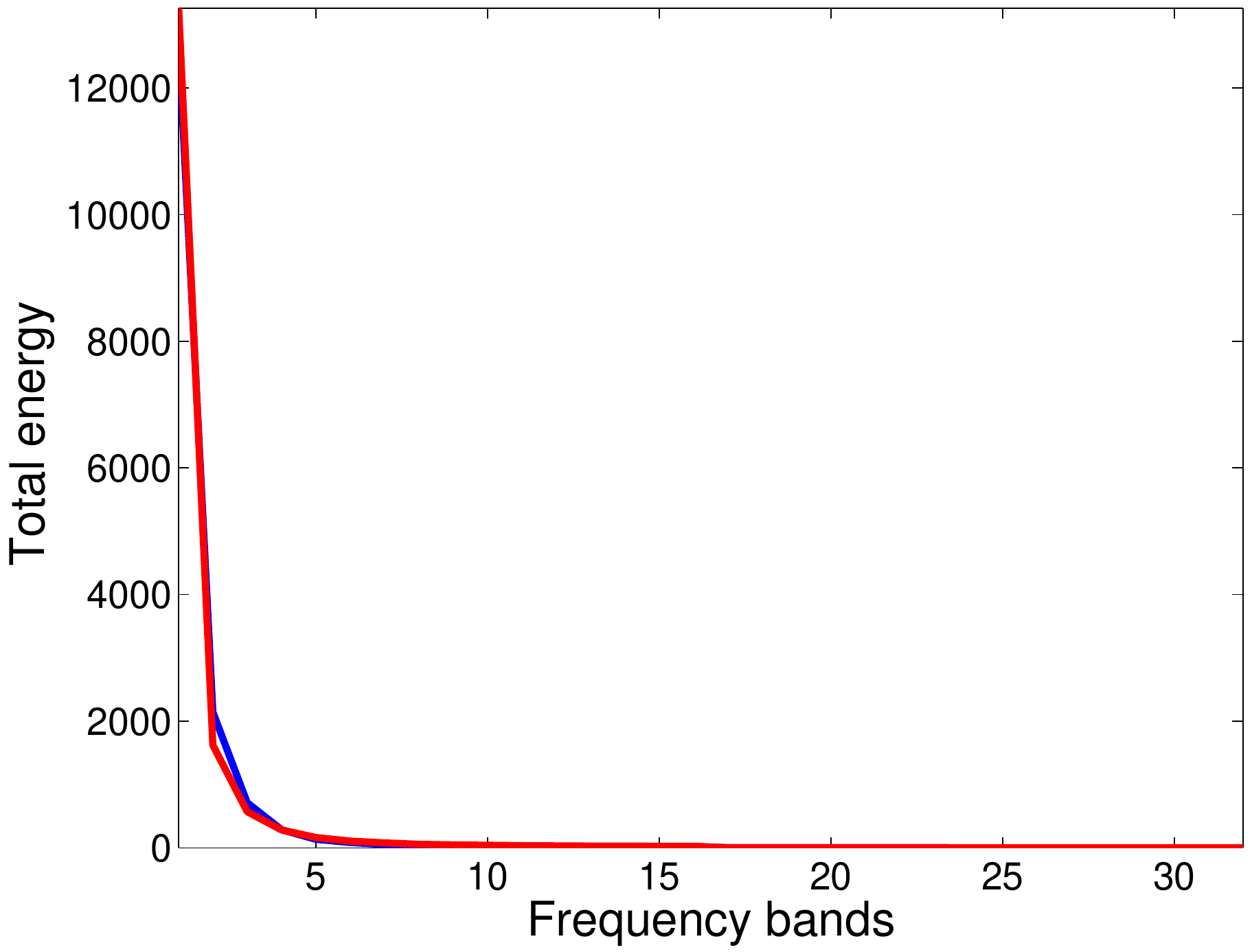}
    \includegraphics[width= 12pc]{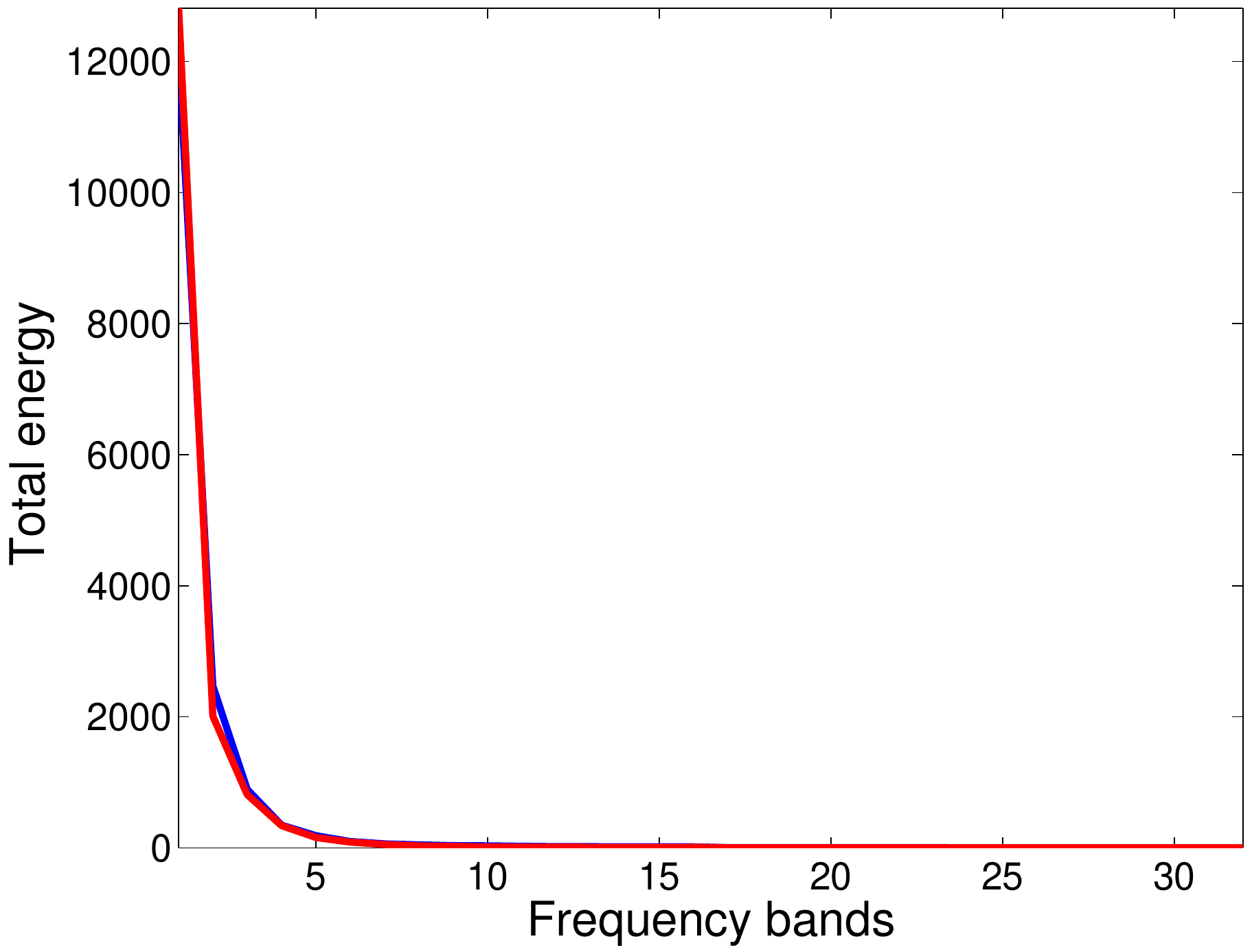}
  }
  \centerline{\small \hfill $i = 3$\hfill \hfill $i=4$\hfill \hfill$i =5$\hfill}
  \centerline{\small   
    \begin{rotate}{90}~~~~~~~~~~$\widehat{\cal E}^i(l)$\end{rotate}
    \includegraphics[width= 12pc]{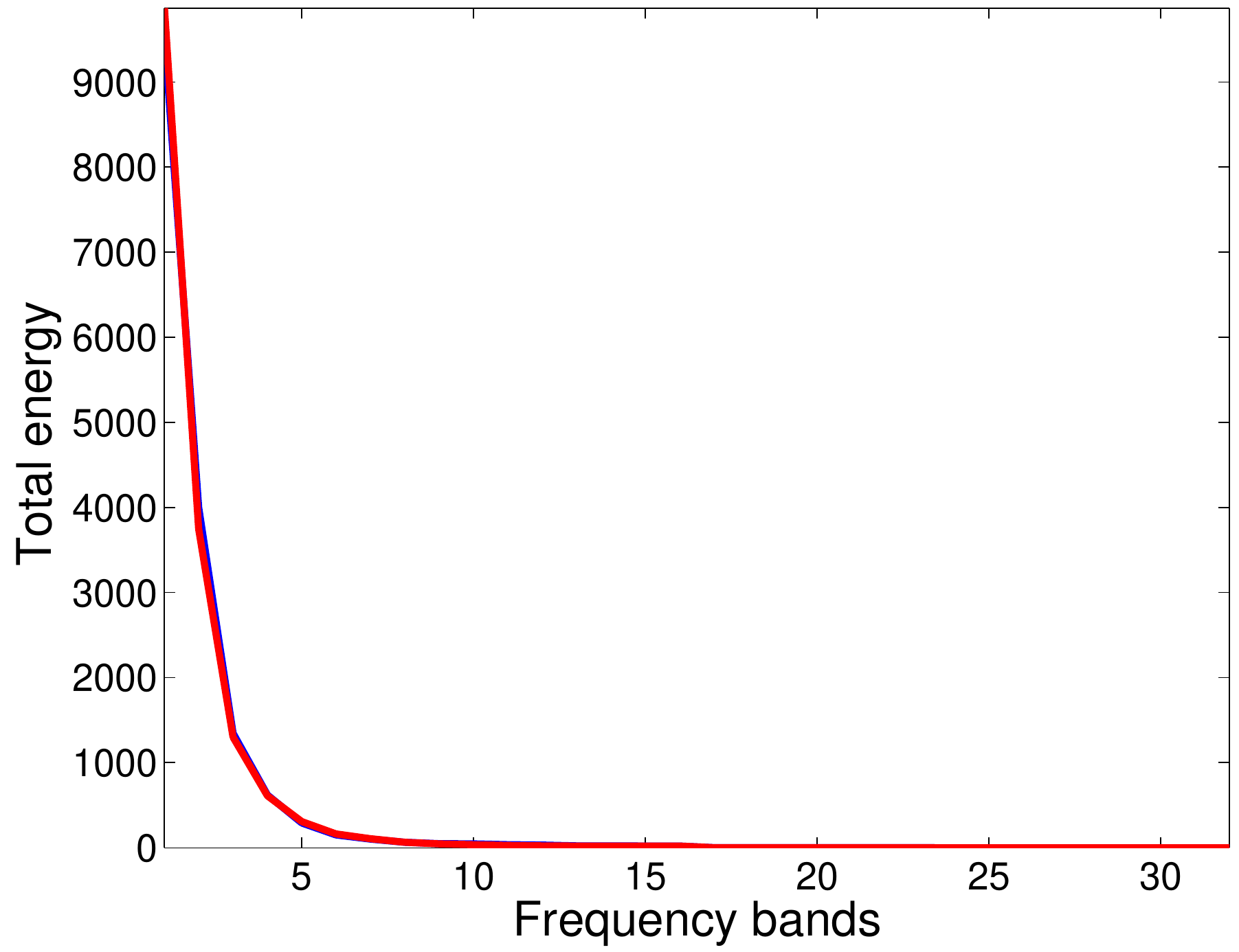}
    \includegraphics[width= 12pc]{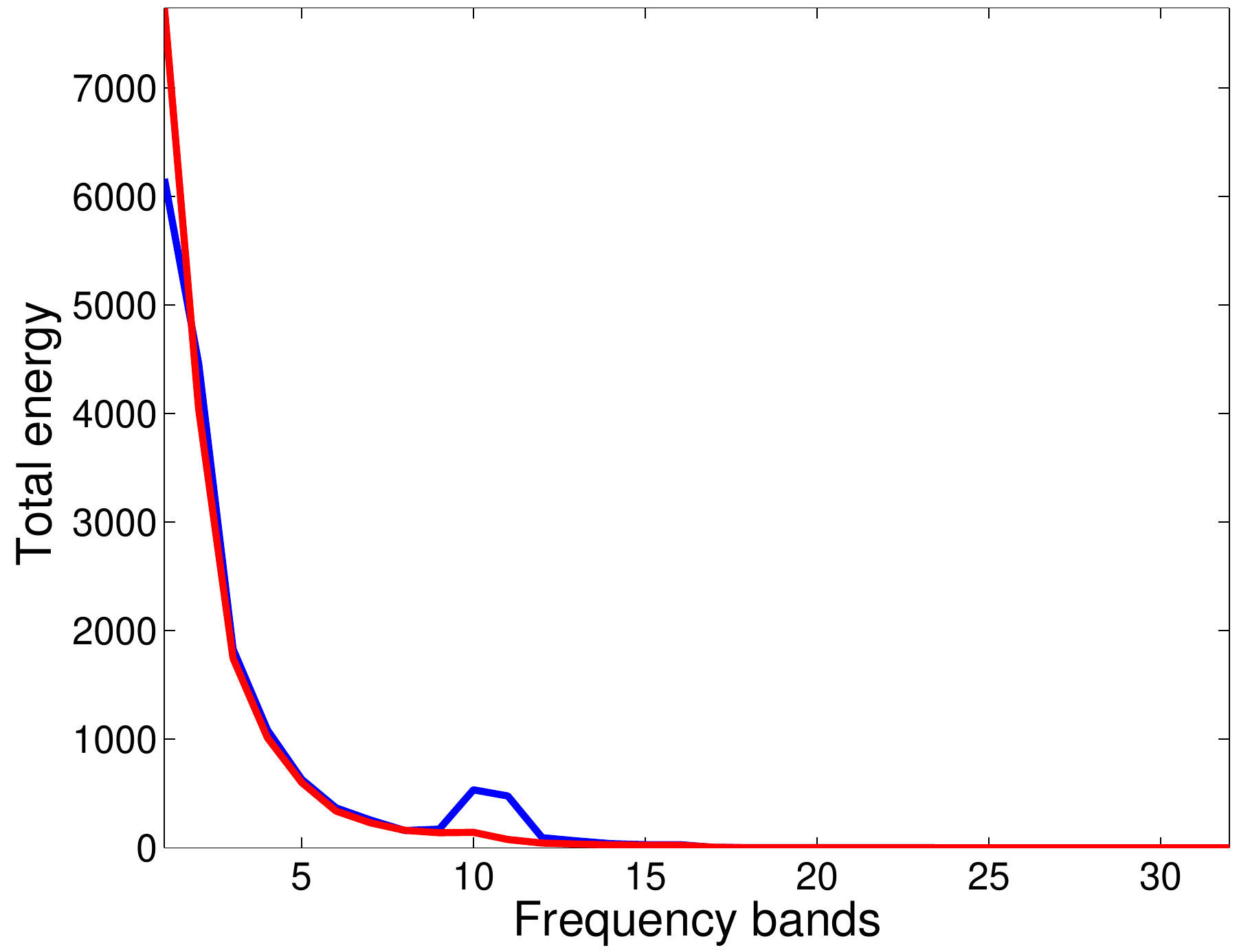}
    \includegraphics[width= 12pc]{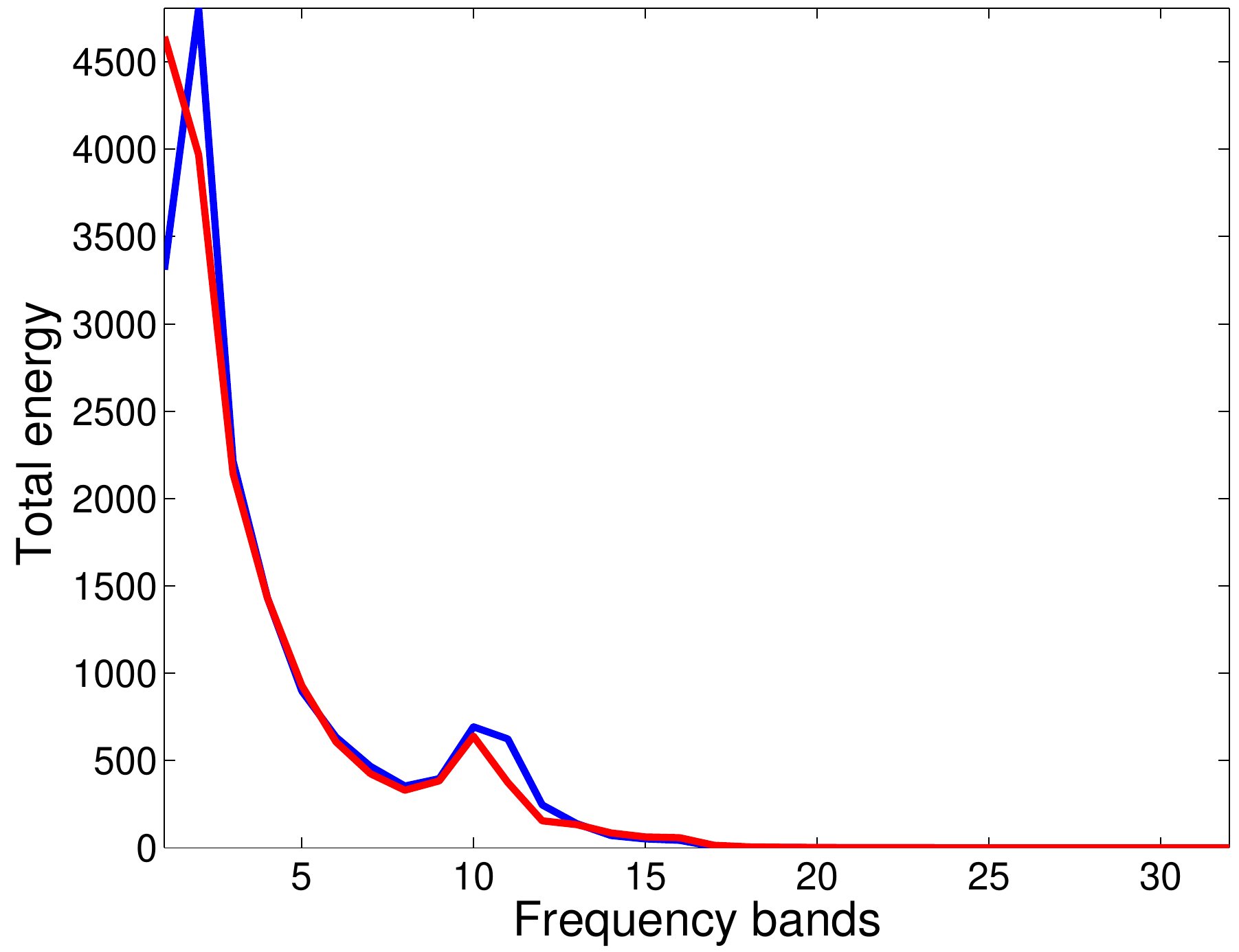}
  }
  \centerline{\small  \hfill $i = 6$\hfill $i=7$\hfill \hfill}
  \centerline{\small 
    \begin{rotate}{90}~~~~~~~~~~$\widehat{\cal E}^i(l)$\end{rotate}
    \includegraphics[width= 12pc]{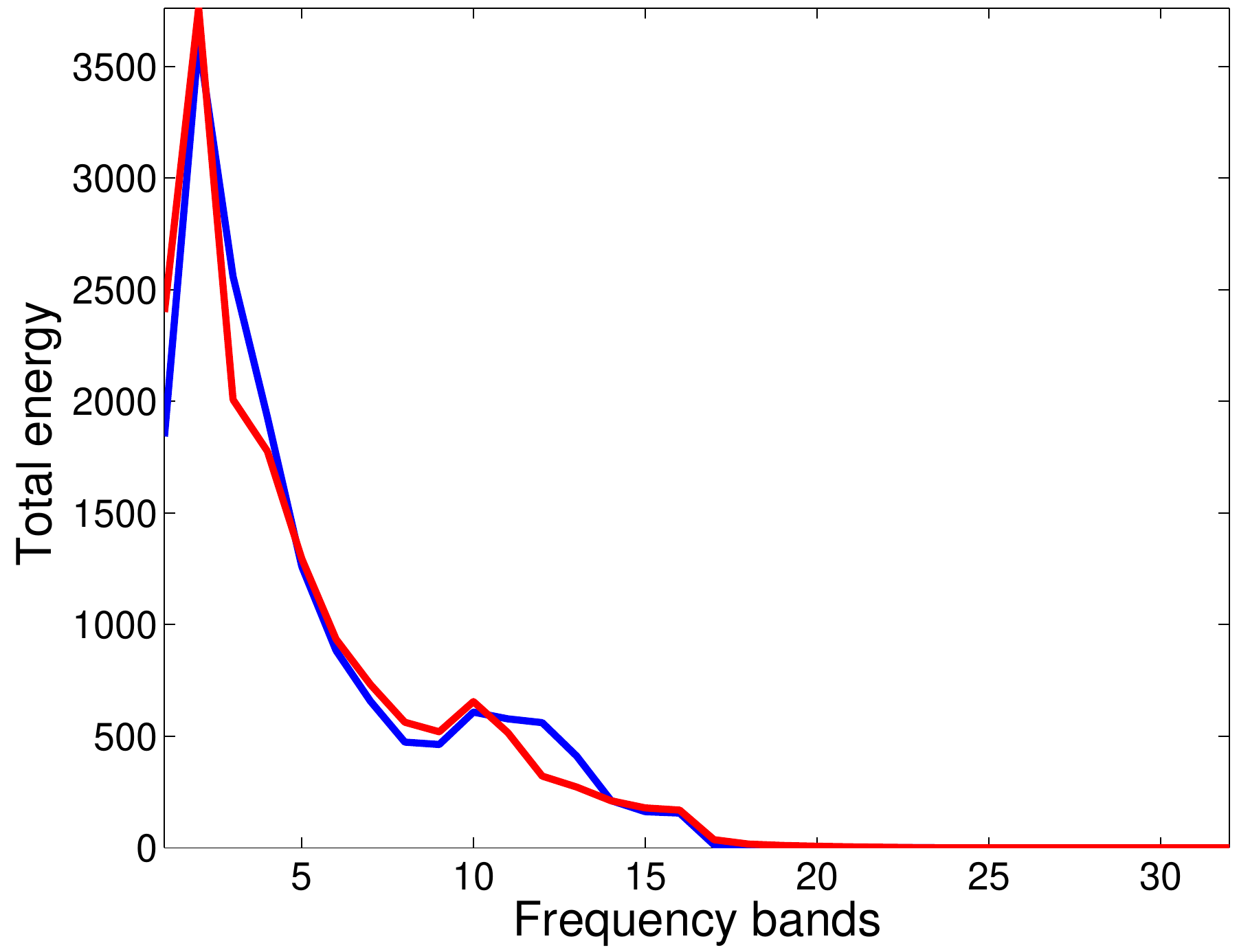}
    \includegraphics[width= 12pc]{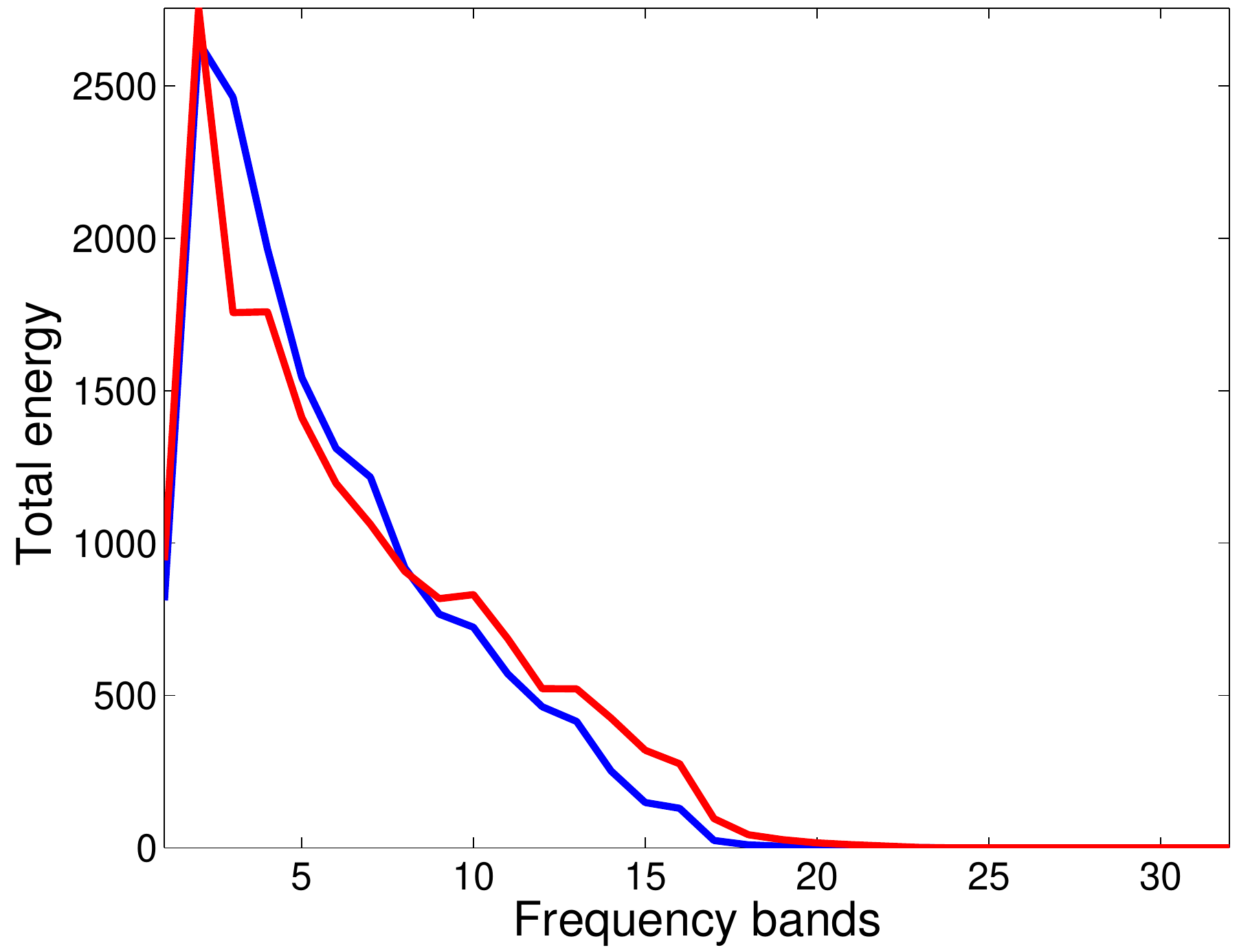}
    ~~~~\raisebox{4pc}{
      \begin{minipage}[b]{7pc}
        $\bfi_k$ in {\blue blue} \\
        $\hfi_k$ in {\red red}
      \end{minipage}
    }}
  \caption{Energy $\widehat{\cal E}^i(l)$ and ${\cal E}^i(l)$ of the
    reconstructed eigenvectors $\hfi_k$ (red) and $\bfi_k$ (blue) as a
    function of the radial frequency index $l$ for the ``clown'' image
    ($i=0,\ldots 7$ from top to bottom, and left to right). The
    eigenvectors $\hfi_k$ are computed using a second graph that is
    built from the denoised image
    $\widehat{u}^{(2)}$.  \label{twostage_edp} }
\end{figure}
\noindent  computed using the graph of the clean image
(see Fig.~\ref{noisy_on_clean_eigs}). Figure \ref{twostage_edp}
displays the distribution of the energy of the eigenvectors $\hfi_k$ as a
function of the radial frequency. Even at large scale $i=6$ and 7, the
distribution of the energy of the $\hfi_k$ across the radial frequency
almost coincides with the distribution of the energy of the $\bfi_k$. 

\citet{Szlam08} proposed a related approach to bootstrap a
semi-supervised classifier: a new graph is constructed after one
iteration of the classifier to improve the classification performance.
\subsection{Iterative denoising
  \label{twostage}}
We propose an iterative procedure to denoise an image based on the
results of the previous section. In practice we use two passes of 
denoising. The algorithm is described in Fig.~\ref{algo}.  In the next
section we study in details the two most important parameters of the
algorithm: the patch sizes ($m_1$ and $m_2$) and the numbers of
eigenvectors ($K_1$ and $K_2$) that are used during the two stages of
the algorithm.
\subsection{Optimization of the parameters of the algorithm}
\subsubsection{The patch size}
While the eigenvectors $\bfi_k$ provide a global basis on the
patch-set, the patch size introduces a notion of local scale in the
image. We notice visually that as the patch size increases, the
eigenvectors become less crisp and more blurred. As explained in
\citep{Singer09a}, \citep{Taylor11b} when the patch size is larger,
patches are at a larger distance of one another in the
patch-set. Consequently, a larger patch size prevents patches that are
initially different from accidentally becoming neighbors after the
image is corrupted by noise, since their mutual distance is relatively
less perturbed by the noise. Because the eigenvectors $\tfi_k$ are
sensitive to the geometry of the patch-set, larger patches help
minimize the perturbation of the eigenvectors.  We note that, as
patches become larger, the eigenvectors become more blurred, therefore
requiring more eigenvectors to describe the texture. Finally,
we need to be aware of the following practical limitation: if the
patch size is large, and the image size is relatively small, then we
have few patches to estimate the geometry of the patch-set; a problem
aggravated by the fact that the patch-set lives in a large ambient
space in that case.

\begin{figure}[H]
  \centerline{  
    \includegraphics[width= 0.40\textwidth]{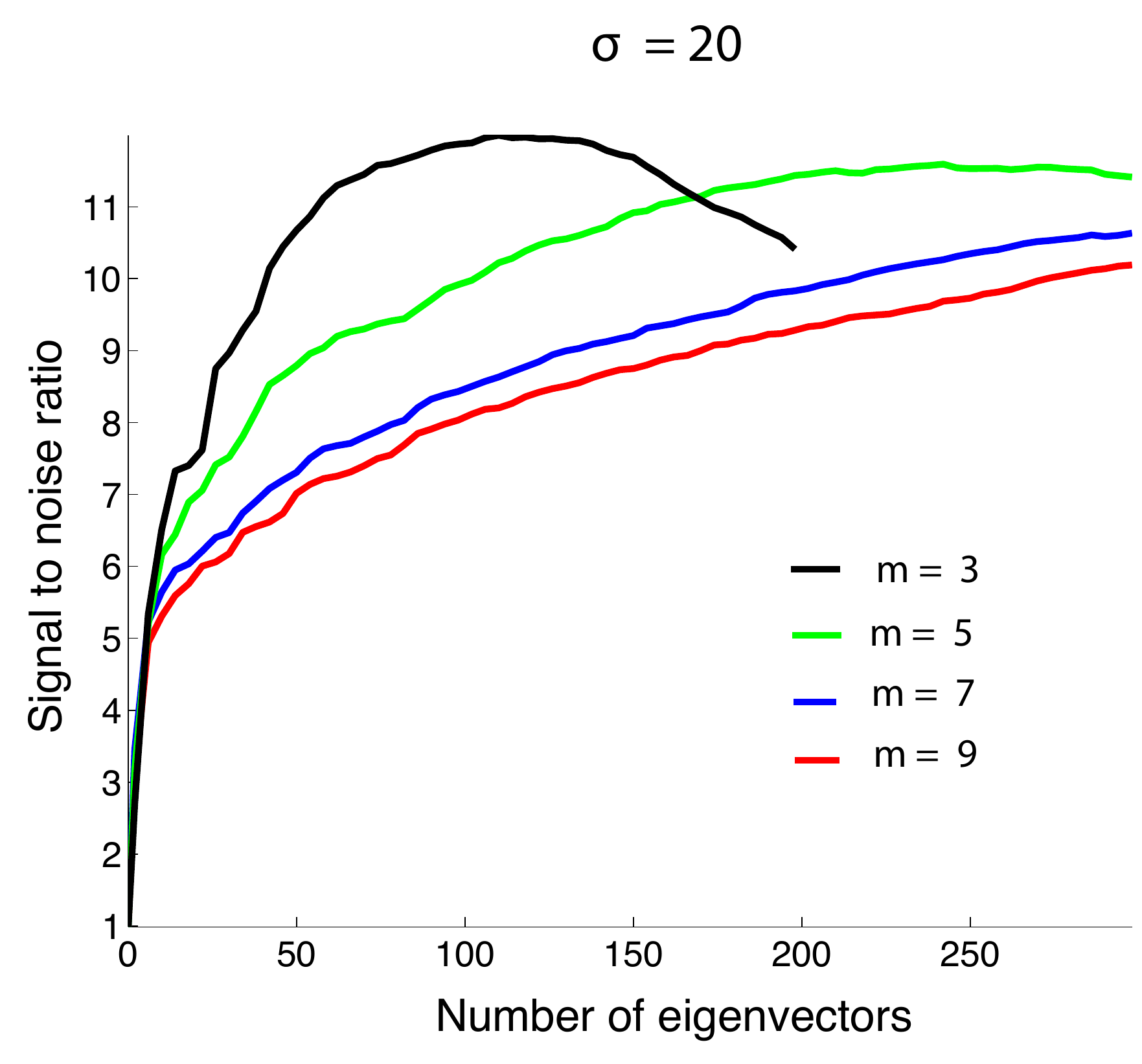}
    \includegraphics[width= 0.40\textwidth]{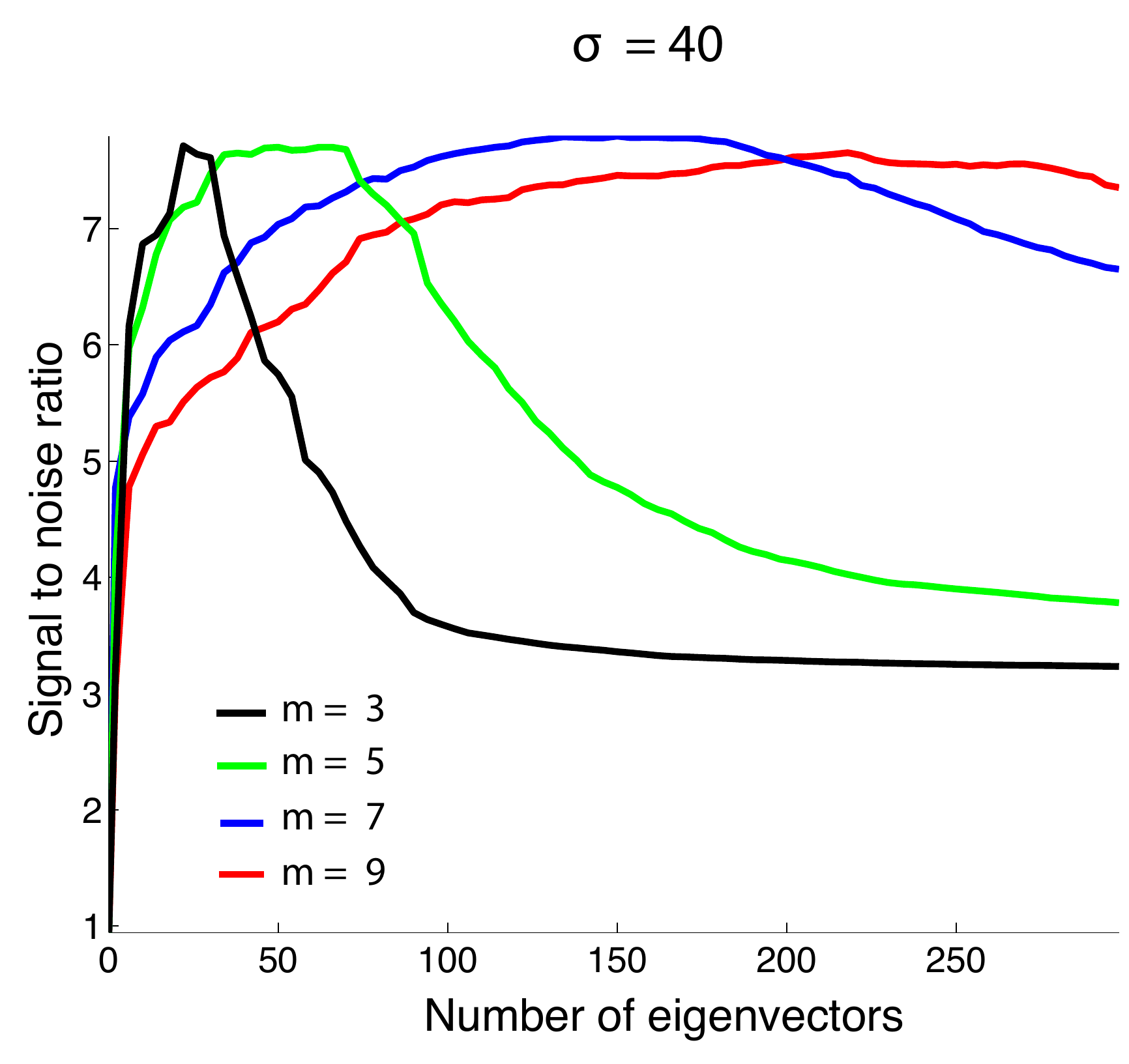}
  }
  \centerline{
    \includegraphics[width= 0.40\textwidth]{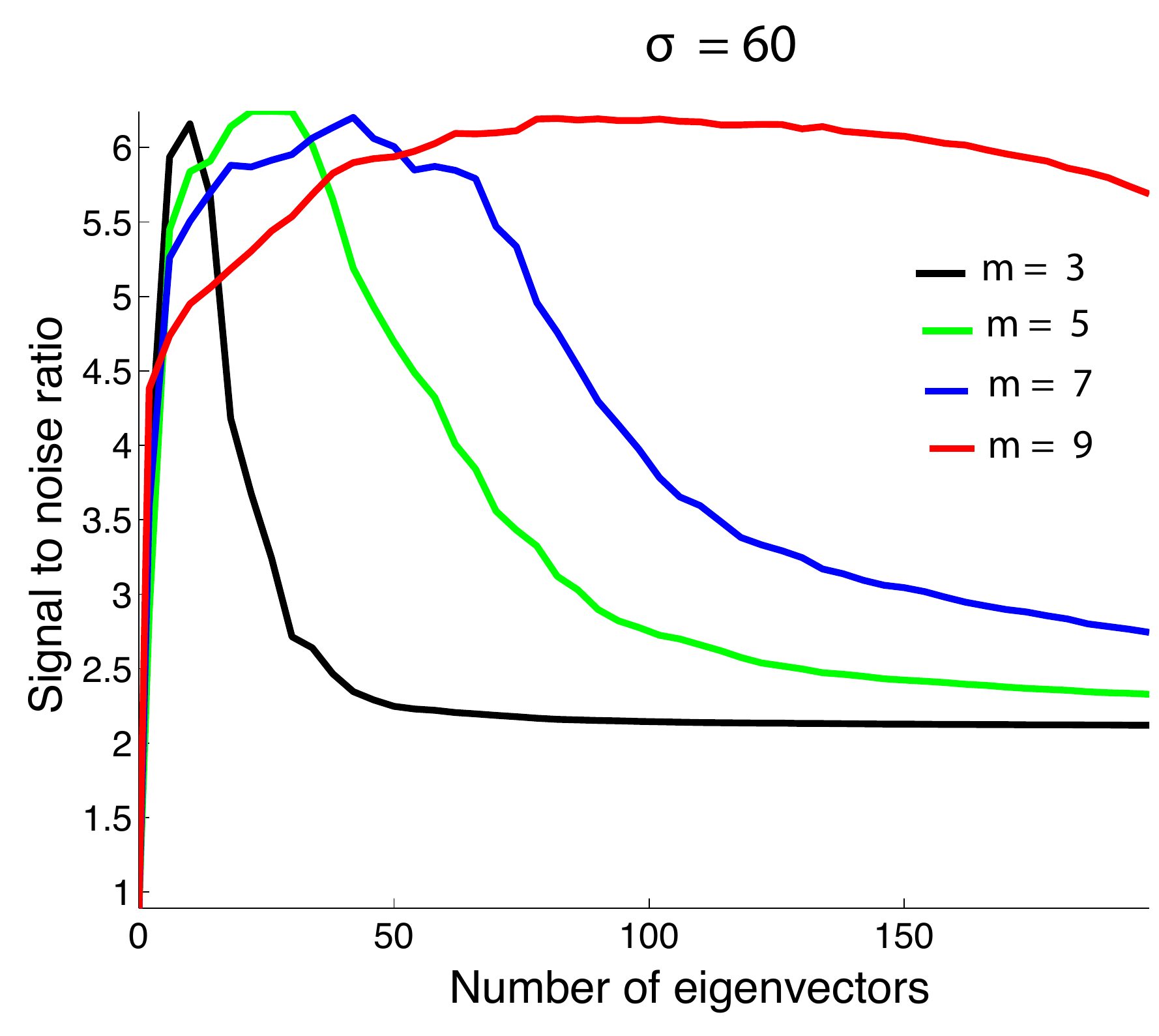}
    \includegraphics[width= 0.40\textwidth]{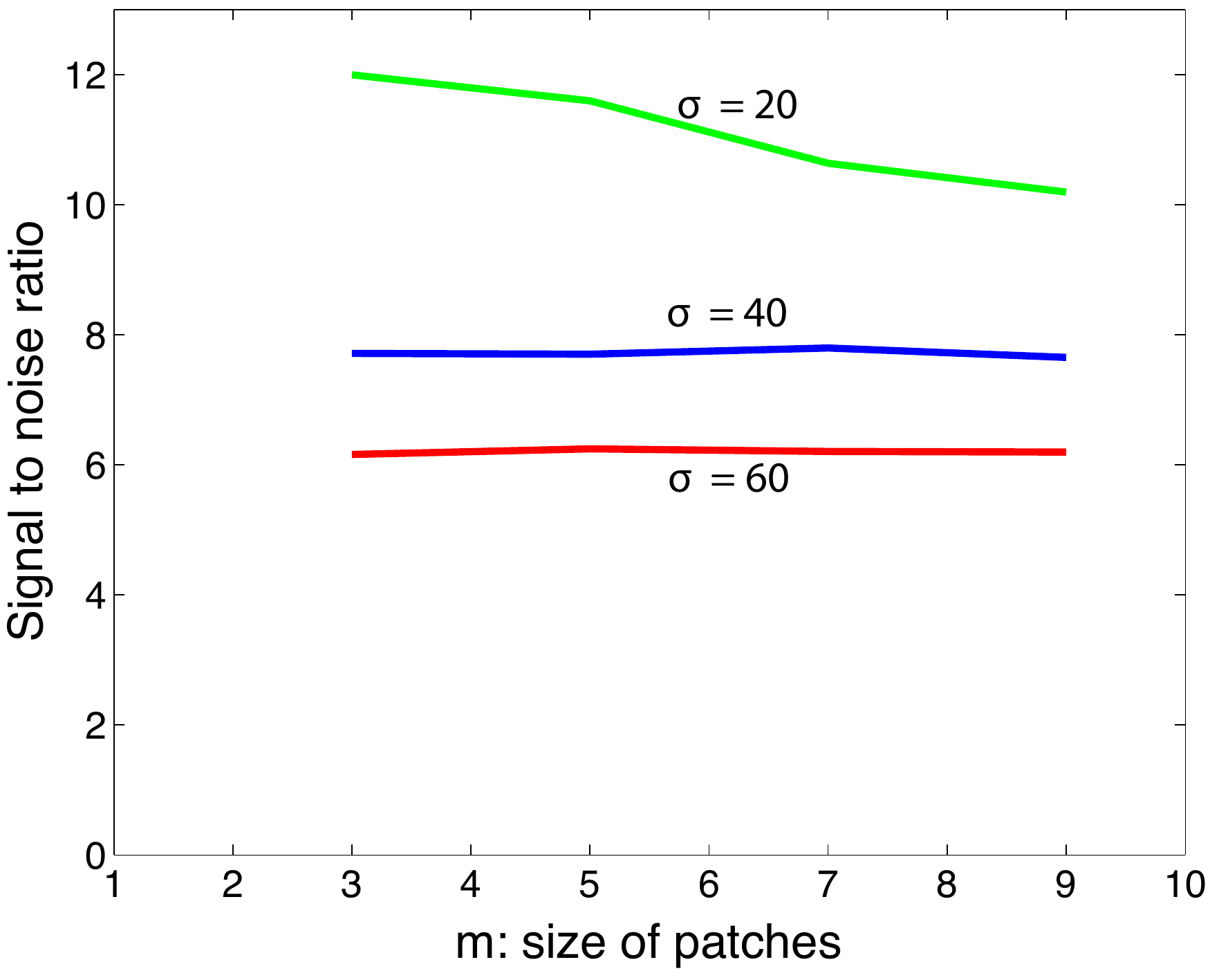}
  }
  \caption{ Top to bottom, left to right:   SNR as a function of the
    number of eigenvectors $K$ after a one pass reconstruction, for
    increasing noise levels. Each curve corresponds to a different
    patch size. Bottom right: optimal reconstruction error as a
    function of the patch size $m$ at different noise levels.
    \label{clown_vpatch}}
\end{figure}

We have compared the effect of the patch size on the quality of the
denoised image $\widehat{u}^{(2)}$ after the first pass (stage 1) of
the denoising algorithm (see Fig.~\ref{algo}). We define the signal to
noise ratio%
\begin{figure}[H]
  \makebox{\SF Algorithm: Two-pass denoising}\\
  \barre
  \begin{itemize}
  \item []{\SF Input:} noisy image $\widetilde{u}$; 
    \begin{itemize}
    \item [$\bullet$] number of nearest neighbors $\nu$;  patch sizes $m_1, m_2$;
      scale parameters $\delta_1, \delta_2$;
    \item  [$\bullet$] number of eigenvectors $K_1,K_2$;  $\gamma \in [0,1]$ 
    \end{itemize}
  \item []{\SF Stage 1}
    \begin{enumerate}
    \item build the graph from the image $\widetilde{u}$ and compute
      $\widetilde{\bW}$ and $\widetilde{\bD}$; 
    \item compute the first $K_1$ eigenvectors, $\tfi_k, \; k=1,\ldots, K_1$, of
      $\widetilde{\bD}^{-\frac{1}{2}}\widetilde{\bW}{\widetilde{\bD}}^{-\frac{1}{2}}$
    \item construct the nonlinear estimator $\hu^{(1)}(\bx)$ of the patch-set
      using (\ref{denoise}) 
    \item reconstruct the image $\widehat{u}^{(1)}$ by averaging the
      patches $\hu^{(1)}(\bx)$  using (\ref{average}) 
  \item  compute the denoised image $\widehat{u}^{(2)} = (1-\gamma)
      \widehat{u}^{(1)} +  \gamma \widetilde{u}$   
    \end{enumerate}
  \item []{\SF Stage 2}
    \begin{enumerate}
    \item build a second graph from the image $\widehat{u}^{(2)}$;
      compute $\widehat{\bW}^{(2)}$ and $\widehat{\bD}^{(2)}$;
    \item compute the first $K_2$ eigenvectors, $\hfi_k, k = 1,\ldots,
      K_2$ of  $\left[\widehat{\bD}^{(2)}\right]^{-\frac{1}{2}}\widehat{\bW}^{(2)}
      \left[\widehat{\bD}^{(2)}\right]^{-\frac{1}{2}}$
    \item construct the nonlinear estimator $\hu^{(3)}(\bx)$ of the patch-set
      using (\ref{denoise}) 
    \item reconstruct the image $\widehat{u}^{(3)}$ by averaging the
        patches $\hu^{(3)}(\bx)$  using (\ref{average})
    \end{enumerate}
  \item []{\SF Output:} the denoised image $\widehat{u}^{(3)}$.
  \end{itemize}
  \barre
  \caption{The two-stage denoising algorithm.
    \label{algo}}
\end{figure}
\noindent  (SNR) by
\begin{equation}
  \text{SNR} = \frac{ \|u \|}{\|u - \widehat{u}\|}.
  \label{snr_denoised}
\end{equation} 
Figure \ref{clown_vpatch} displays the signal-to-noise ratio as a
function of the number of eigenvectors $K$ used to reconstruct
$\widehat{u}^{(2)}$, for different patch sizes $m$, and at different
noise levels ($\sigma=20, 40,$ and $60$). For low noise levels
($\sigma =20$) the SNR is larger for smaller patches. For moderate
noise levels, the maximum SNR achieved with different patch size is
approximately the same (see Fig.~\ref{clown_vpatch}
bottom-right). Nevertheless, the visual quality of the denoised images
-- at the same SNR -- are quite%
\begin{figure}[H]
  \centerline{  
    \includegraphics[width= 0.25\textwidth]{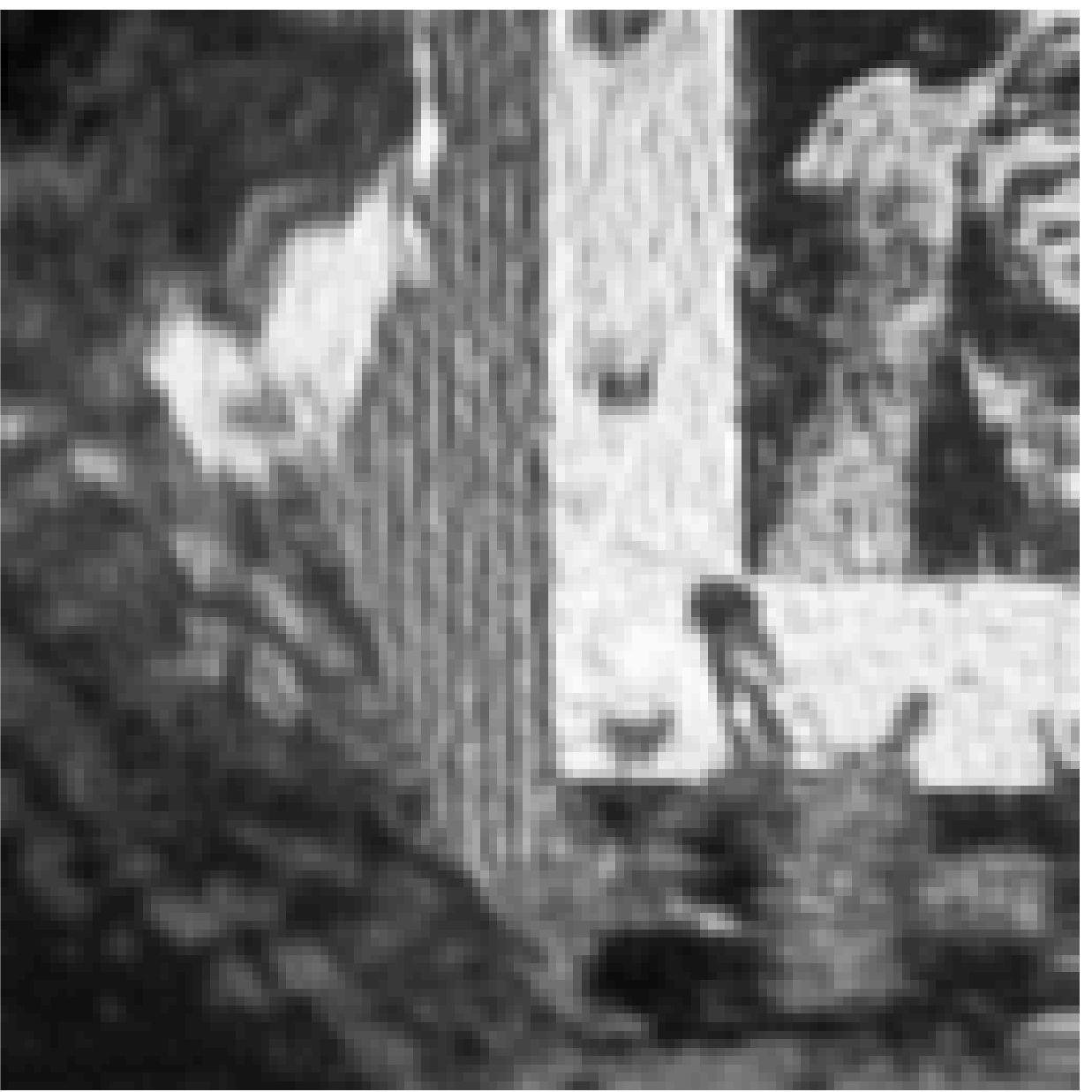}\hspace*{-0.25pc}  
    \includegraphics[width= 0.25\textwidth]{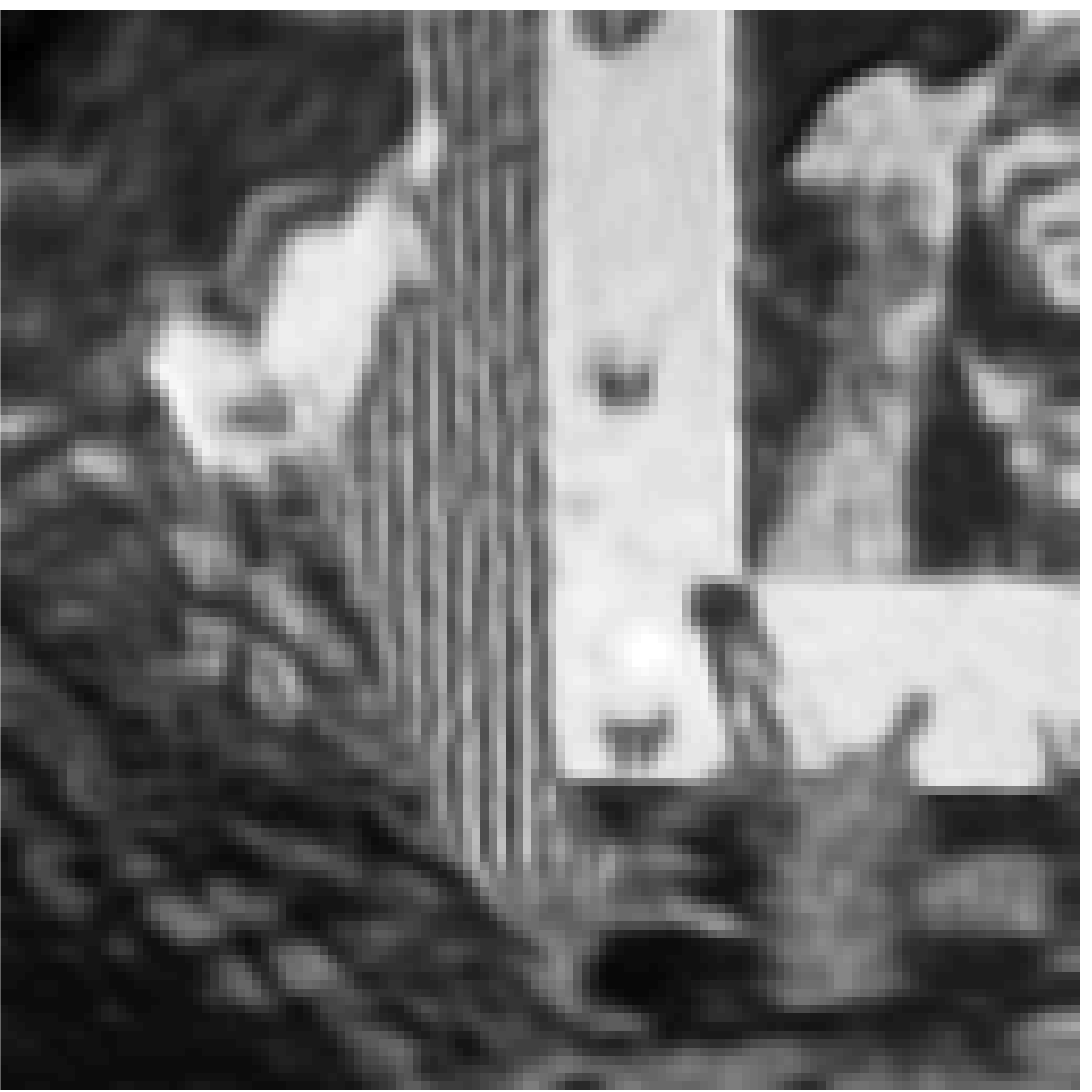}\hspace*{-0.25pc}  
    \includegraphics[width= 0.25\textwidth]{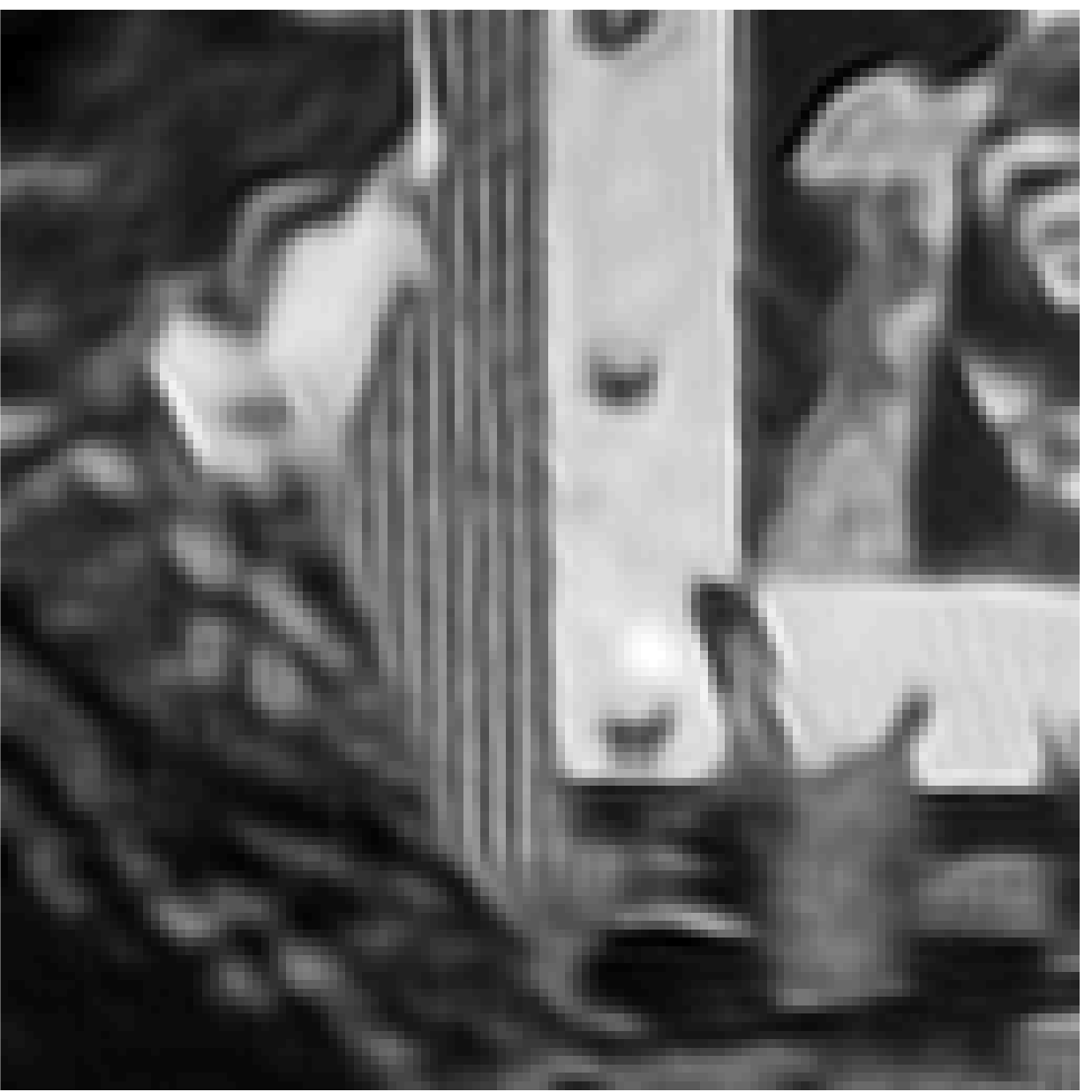}\hspace*{-0.25pc}  
    \includegraphics[width= 0.25\textwidth]{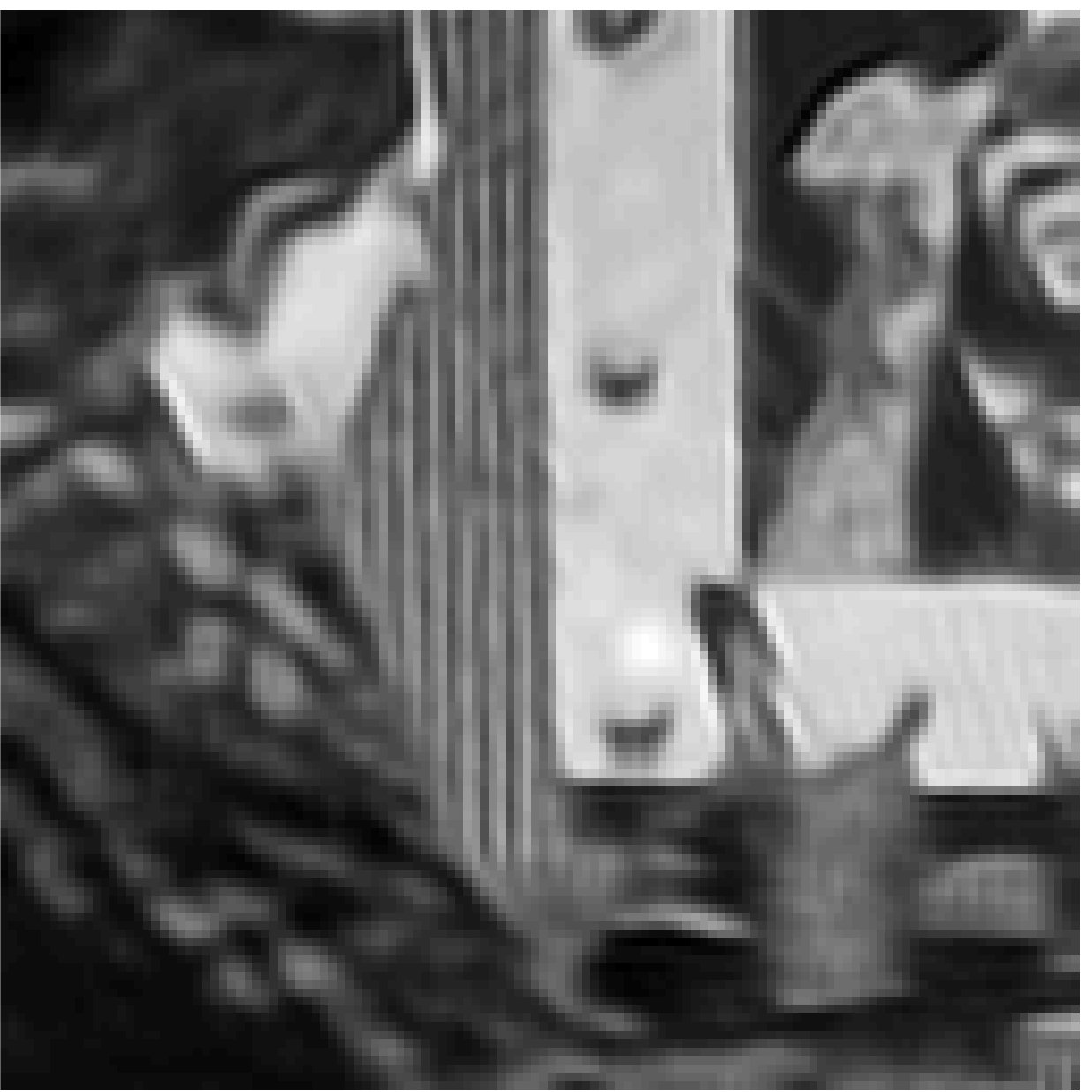}
  }
  \caption{Images with the same SNR denoised with  patch size
    $m=3,5,7,9$ (from left to right).
    \label{clown_vpatch_denoised}
  }
\end{figure}
\noindent different. Figure \ref{clown_vpatch_denoised} shows four
denoised images ($\sigma =40$) with very similar SNR, for $m=3,5,7$
and 9. With $m=3$, the artifact caused by the noise can still be seen
in the denoised image. However, with larger $m$, the denoised image is
much smoother, although some of the details in the original image are
lost. The choice of the patch size $m$ is in general based on the
noise level $\sigma$. In practice, we choose $m=3$ for $\sigma=20$,
$m=5$ for $\sigma = 40$, and $m=7$ for $\sigma = 60$.
\subsubsection{The number of eigenvectors $K$}
At both stages of the algorithm, the first $K$ eigenvectors are used
to compute a denoised image. In general, the eigenvectors degrade
faster as the variance of the noise level increases. We explain in the
following why, using a very small number of eigenvectors, we can still
reconstruct the structures of the patch-set that
contain patches extracted from smooth regions of the image.

{\noindent \bfseries The local dimensionality of the patch-set.}  In
this work we consider images that contain a mixture of smooth regions
and texture. Therefore, we only need a very small number of
eigenvectors to capture the patches extracted from smooth regions of
the image. Indeed, as explained by \citet{Taylor11b}, patches
extracted from smooth regions of the image align themselves along
smooth low-dimensional structures. Indeed, for smooth patches, spatial
proximity (in the image) implies proximity in $\R^{m^2}$. Conversely,
patches that contain large gradients and texture are usually scattered
across $\R^{m^2}$ \citep{Taylor11b}. Other researchers have made
similar experimental observations. For instance,
\citet{Chandler07,zontak11} observed that patches can be classified
into two classes according to their mutual distances. One class is
composed of patches extracted from smooth regions, and the other class
encompasses patches extracted from regions that contain texture, fast
oscillation, large gradient, etc. These authors observe that ``smooth
patches'' are at a small distance (in the patch-set) of one another,
and are also at a close spatial distance (in pixels). They also note
that proximity between ``texture patches'' is not dictated by spatial
proximity (in pixels). Texture patches are scattered in $\R^{m^2}$
because they are at a large distance of one another: \citet{zontak11}
estimate that the distance to the nearest neighbor of a patch grows
exponentially with the gradient within that patch.  Finally,
\citet{Taylor11b} have shown that the local dimensionality of the
patch-set is much lower around smooth patches, and much higher around
texture patches.

Consequently, the smooth low-dimensional structures formed by the
smooth patches can be accurately represented with a very small number
of eigenvectors. Because low-index eigenvectors are very stable, we
are able to reconstruct the smooth patches using the first few
eigenvectors $\tfi_k$ of $\widetilde{\bL}$. This partial
reconstruction can then be used to estimate a new patch graph. The
reconstruction%
\begin{figure}[H]
  \centerline{  
    \includegraphics[width= 0.40\textwidth]{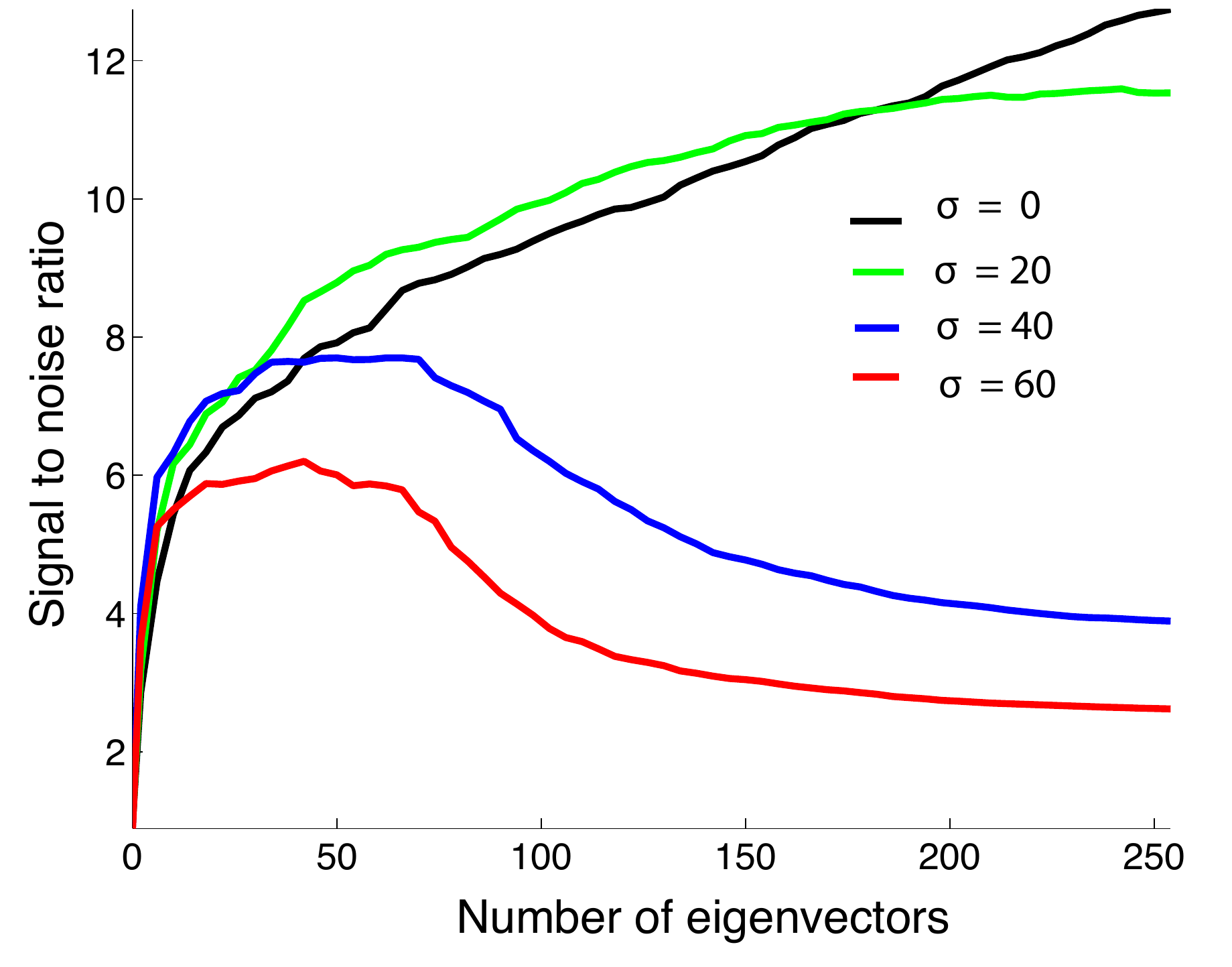}
  }
  \caption{Reconstruction error of the clown image, after a single pass of 
    denoising, as a function of the number of
    eigenvectors $K$, at various noise levels. The patch size $m$ was
    5 except for $\sigma =60$ where it was 7.
    \label{clown_vnoise}}
\end{figure}
 \noindent  of the texture patches requires many more eigenvectors
because that part of the patch-set has a much higher local
dimensionality. Using the second graph, one can then compute more
stable eigenvectors, and reconstruct the texture patches.

Figure \ref{clown_vnoise} displays the SNR of $\widehat{u}^{(2)}$ after a
one-stage reconstruction of the clown image as a function of the
number $K$ of eigenvectors, for several noise levels. The patch size
was $m=5$, except for $\sigma =60$ where it was $m=7$. As we increase
the number of eigenvectors $K$, the SNR initially increases. Except at
very low noise level ($\sigma \leq 20$), the SNR decreases after it
reaches its maximum. Indeed, at moderate and high noise levels, the
high index eigenmodes adapt to the incoherent structure of the noise
and become useless for the reconstruction. This is why we need a
second pass in the reconstruction. Figure \ref{clown_vnoise} can help
us determine the numbers $K_1$ and $K_2$ of eigenvectors needed for the first
and second denoising passes.
\section{Experiments
  \label{experiments}}
We implemented the two-stage denoising algorithm (see Fig.~\ref{algo})
and evaluated its performance against several denoising gold-standards.
For all experiments, we kept $K_1=35$ and $K_2=275$ eigenvectors during
the first and second passes of the algorithm, respectively. The patch
sizes for the first pass were $m_1=7$ for $\sigma=40$, and $m_1=9$ for
$\sigma=60$, respectively. We used $m_2 =5$ for all noise levels in
the second pass of the algorithm.

\begin{figure}[H]
  \centerline{  
    \includegraphics[width= 0.20\textwidth]{figure7a.pdf}\hspace*{-0.25pc}  
    \includegraphics[width= 0.20\textwidth]{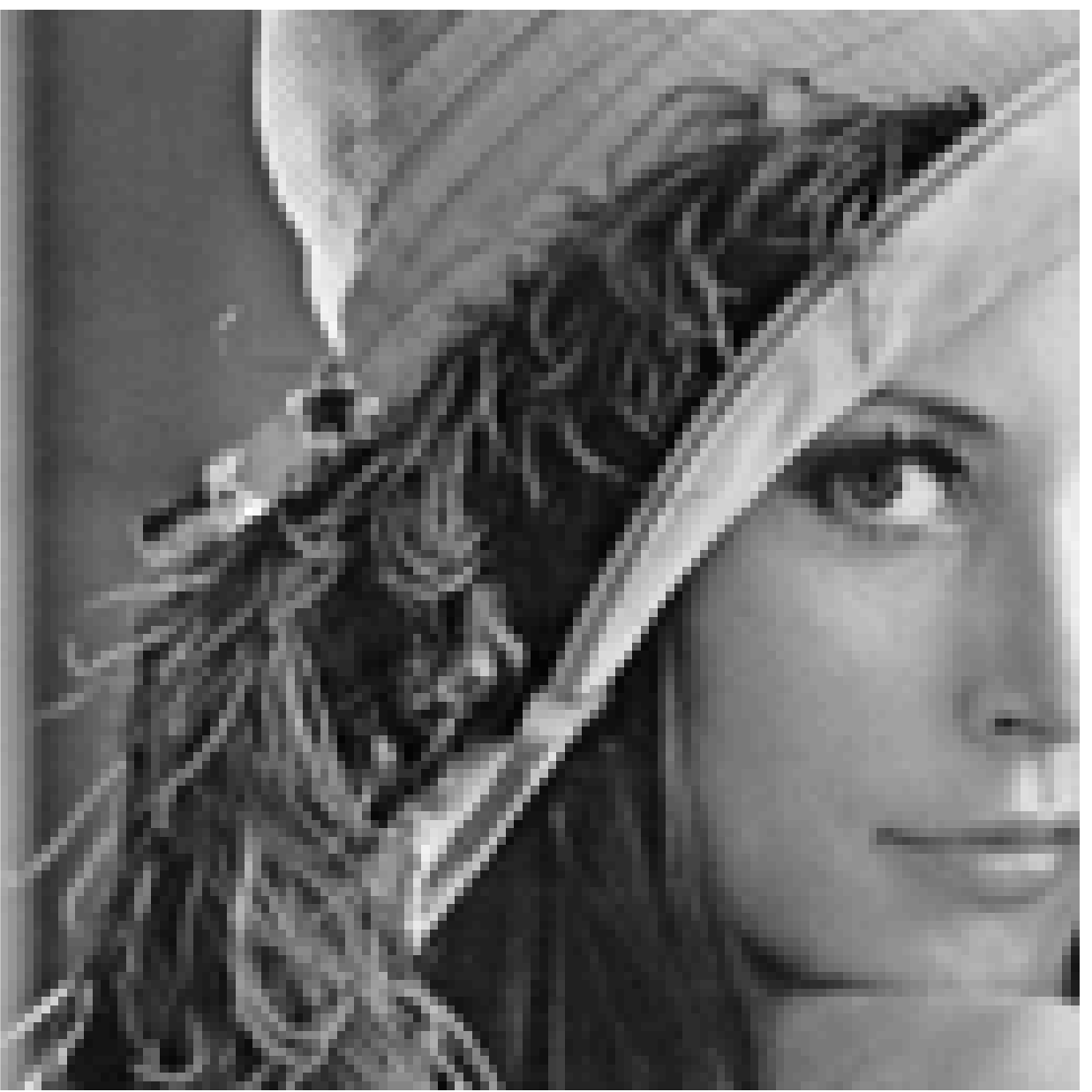}\hspace*{-0.25pc}
    \includegraphics[width= 0.20\textwidth]{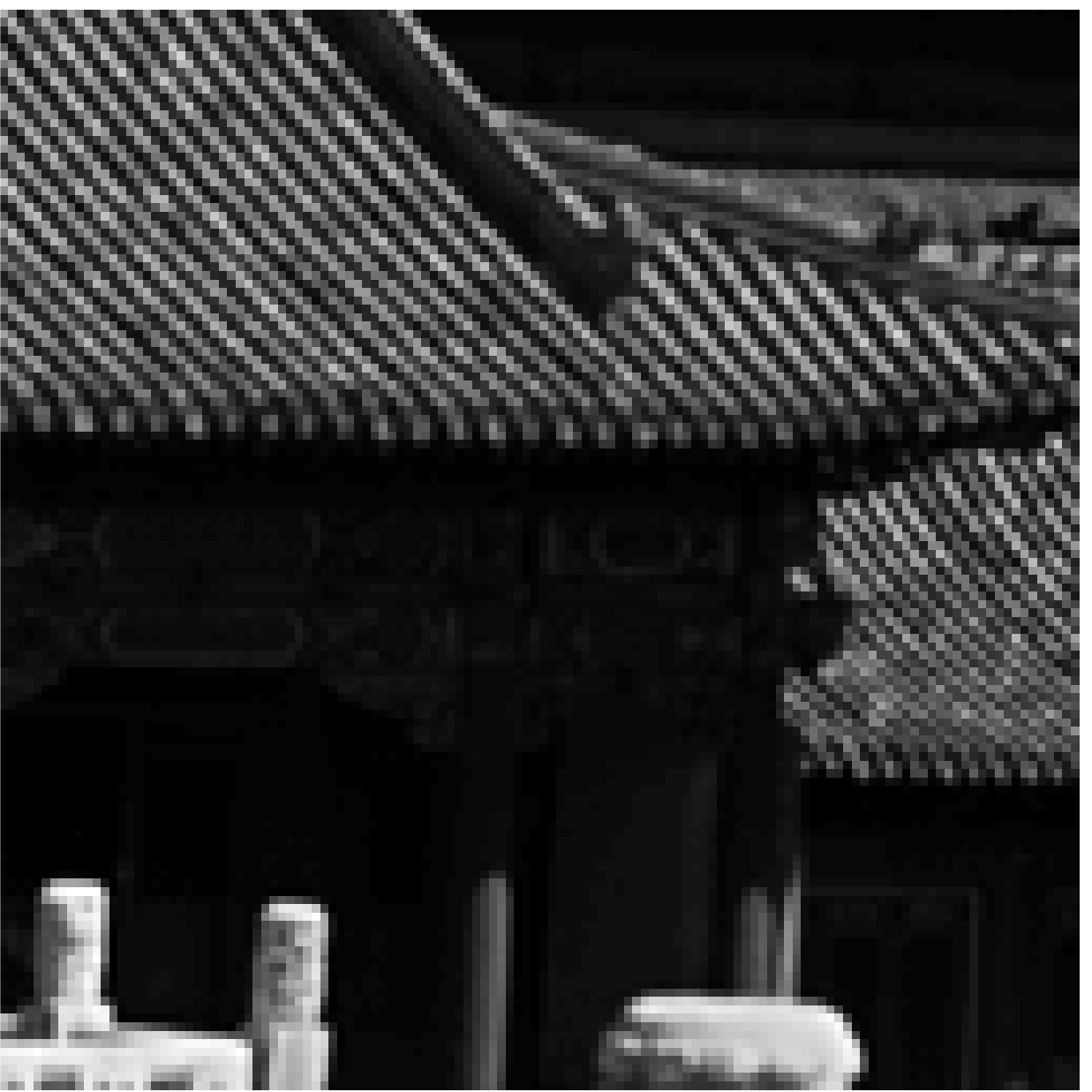}\hspace*{-0.25pc}  
    \includegraphics[width= 0.20\textwidth]{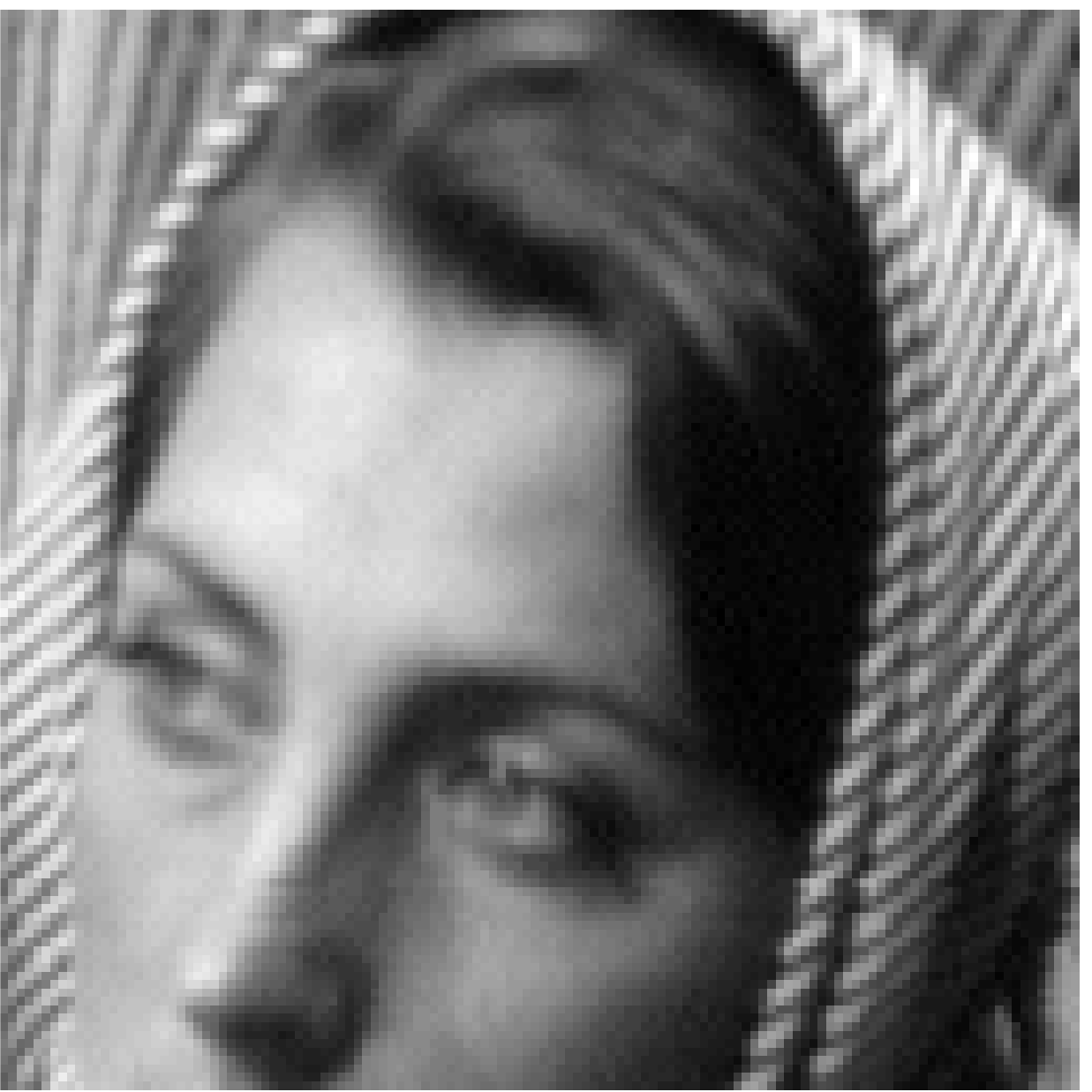}\hspace*{-0.25pc}  
    \includegraphics[width= 0.20\textwidth]{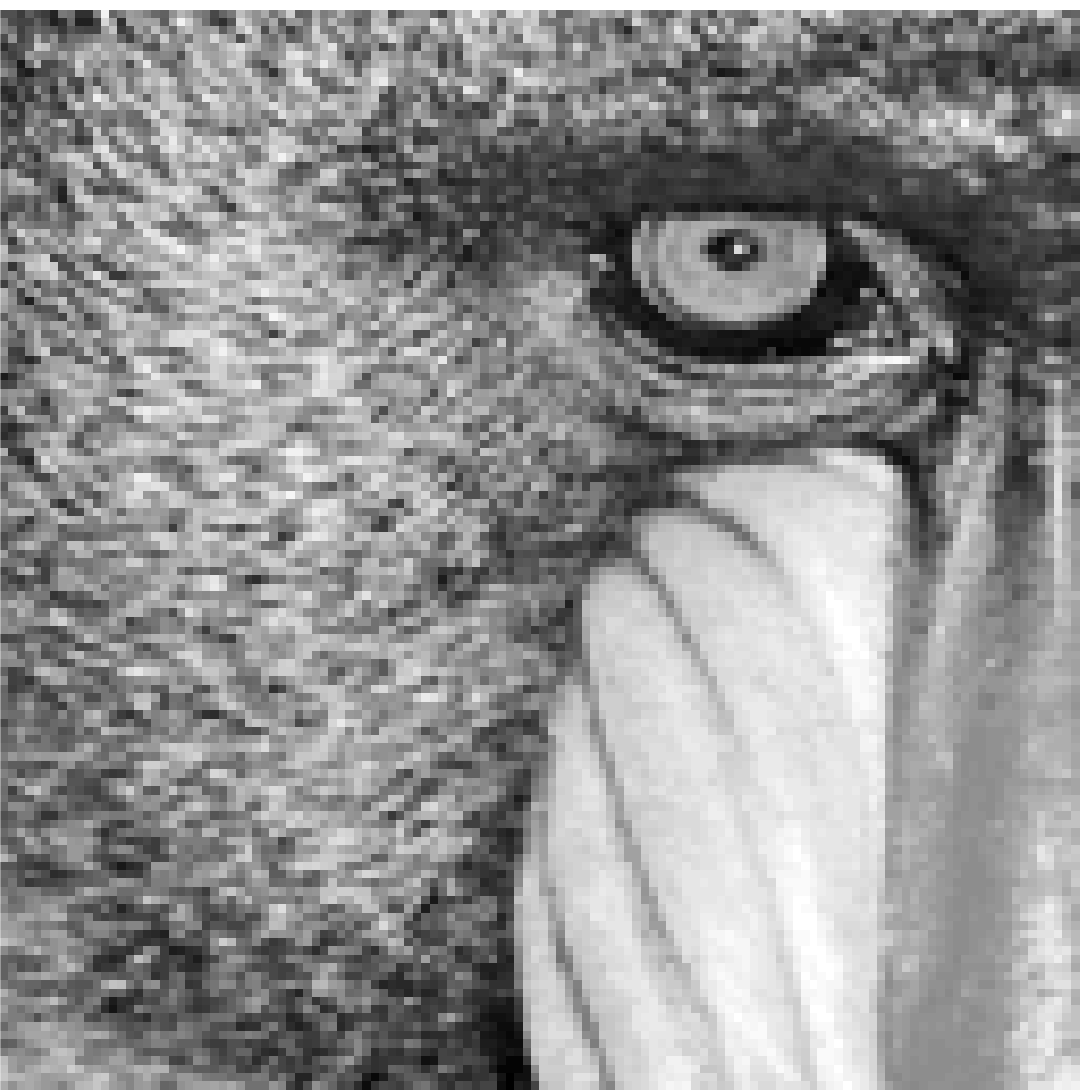}
  }
  \centerline{  
    \includegraphics[width= 0.20\textwidth]{figure7b.png}\hspace*{-0.25pc}  
    \includegraphics[width= 0.20\textwidth]{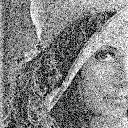}\hspace*{-0.25pc}  
    \includegraphics[width= 0.20\textwidth]{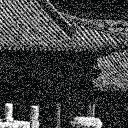}\hspace*{-0.25pc}  
    \includegraphics[width= 0.20\textwidth]{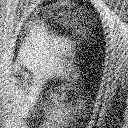}\hspace*{-0.25pc}  
    \includegraphics[width= 0.20\textwidth]{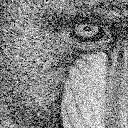}
  }
  \centerline{  
    \includegraphics[width= 0.20\textwidth]{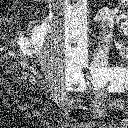}\hspace*{-0.25pc}  
    \includegraphics[width= 0.20\textwidth]{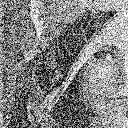}\hspace*{-0.25pc}  
    \includegraphics[width= 0.20\textwidth]{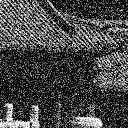}\hspace*{-0.25pc}  
    \includegraphics[width= 0.20\textwidth]{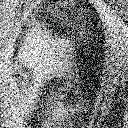}\hspace*{-0.25pc}  
    \includegraphics[width= 0.20\textwidth]{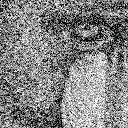}
  }
  \caption{Top: clean images ``clown'', ``lena'', ``roof'',
    ``barbara'', and ``mandrill''. White Gaussian noise with $\sigma
    =40$ (middle) and $\sigma=60$ (bottom), is added to the images.
    \label{test_img}}
\end{figure}

The evaluation was performed using five images that contain a mixture
of smooth regions, periodic and aperiodic texture  (see Fig.~\ref{test_img}). White Gaussian
noise was added to the images. We%
\noindent used two noise levels: moderate
$\sigma = 40$ (Fig.~\ref{test_img} middle row), and high $\sigma = 60$ (Fig.~\ref{test_img} bottom row).
We compared our approach to the following three denoising
algorithms:
\begin{enumerate}
\item translation invariant soft wavelet thresholding (TISWT) method
  \cite{Coifman95}. We used the implementation provided by Wavelab
(\url{http://www-stat.stanford.edu/~wavelab/}).
\item nonlocal means \cite{Buades05}. We used the implementation
  provided by  Jose Vicente Manj\'on Herrera and Antoni Buades
(\url{http://personales.upv.es/jmanjon/}).
\item k-SVD algorithm \cite{Aharon06b}. We used the implementation
  provided by Ron Rubinstein (\url{http://www.cs.technion.ac.il/~ronrubin}). 
\end{enumerate}
Table \ref{mse} displays the mean squared error between the
reconstructed image and the original images,
\begin{equation*}
\frac{1}{N^2} \sum_{i,j=1}^N |\widehat{u}^{(3)} (i,j) -u (i,j)|^2.
\end{equation*}
At all noise levels, and for all images our algorithm outperformed the
other algorithms in terms of mean squared error. In terms of visual
quality, the wavelet denoising yielded consistently the worst%
\begin{table}[H]
  \centerline{  
    \begin{tabular}{@{}lccccc@{}} \toprule%
      & noise level & \red Laplacian & k-SVD & NL-means & TISWT\\
      \midrule
      \multirow{4}*{clown}
      & 40 & \red 216 & 261 & 252 & 359 \\ 
      \cmidrule(l){2-6}
      & 60 & \red 354 & 509 & 447 & 585 \\ 
      \midrule
      \multirow{4}*{lena}
      & 40 & \red 184 & 236 & 248 & 329 \\ 
      \cmidrule(l){2-6}
      & 60 & \red 298 & 413 & 382 & 529 \\ 
      \midrule
      \multirow{4}*{roof}
      & 40 & \red 196 & 318 & 221  & 486 \\
      \cmidrule(l){2-6}
      & 60 & \red 310 & 542 & 395 & 767 \\
      \midrule
      \multirow{4}*{barbara}
      & 40 & \red 116 & 160 & 168 & 241 \\
      \cmidrule(l){2-6}
      & 60 & \red 191 & 300 & 290 & 436 \\ 
      \midrule
      \multirow{4}*{mandrill}
      & 40 & \red 314 & 366 & 357 & 418 \\ 
      \cmidrule(l){2-6}
      & 60 & \red 442 & 542 & 480 & 621 \\ 
      \bottomrule
    \end{tabular}
  }
  \caption{Mean squared error (smaller is better) for the four denoising algorithms.
    \label{mse}
  }
\end{table}
\noindent
reconstruction. Missing wavelet coefficients are very noticeable, even at
moderate noise level. K-SVD and nonlocal means yielded similar
results. The nonlocal means estimate was always more noisy than the
k-SVD estimate, which was often too smooth. In comparison, our
approach could restore the smooth regions and the texture (see
e.g. the mandrill image at $\sigma=40$ and $\sigma=60$), even at high
noise level.
\section{Discussion}
We proposed a two-stage algorithm to estimate a denoised set of
patches from a noisy image. The algorithm relies on the following two
observations: (1) the low-index eigenvectors of the diffusion, or
Laplacian, operators are very robust to random perturbations of the
weights and random changes in the connections of the graph; and (2)
patches extracted from smooth regions of the image are organized along
smooth low-dimensional structures of the patch-set, and therefore can be
reconstructed with few eigenvectors. Experiments demonstrate that our
denoising algorithm outperforms the  denoising gold-standards. This
work raises several questions that we address in the following.
\subsection{Fast computation of the eigenvectors}
A key tenet of this work is our ability to compute the eigenvectors of
the sparse matrix $\bL$, which has size $N^2 \times N^2$ but with only
$\nu$ non zero entries on each row. For the experiments we used the
restarted Arnoldi method for sparse matrices implemented by the Matlab
function {\tt eigs} to solve the eigenvalue problem.

There are several options for further speeding up the computation of
the eigenvectors. \citet{Saito08} proposes to modify the eigenvalue
problem, and compute the eigenvectors via the integral operator that
commutes with the Laplacian. This approach leads to fast algorithms.
The recent work of Kushnir et al. \cite{Kushnir10} indicates that
multigrid methods yield an immediate improvement over Krylov subspace
projection methods (e.g. Arnoldi's method). Another appealing approach
involves fast randomized methods \cite{Halko11}. We note that the
application of these methods still necessitates the computation of the
entire matrix $\bW$. We believe that more efficient methods that
randomly sample the patch-set should be developed. While the sampling
density should clearly guarantee that short wavelength (high-index)
eigenvectors are adequately sampled, we can probably estimate the
low-index eigenvectors with a very coarse sampling.  Similar ideas
have been recently proposed for the reconstruction of manifolds
\cite{Baraniuk09}.

\subsection{Extension}
This work opens the door to a new generation of image processing
algorithms that use the eigenvectors of the graph Laplacian to filter
the patch-set. We note that some of the most successful inpainting and
super-resolution algorithms already operate locally on the patch-set
\citep[e.g.,][and references therein]{Criminisi04}. These algorithms
currently do not take advantage of the existence of global basis
functions to represent the patch-set.  Additional extensions include
the construction of patch-sets from large collections of images. Lee
and Mumford \cite{Lee03} demonstrated that high-contrast patches
extracted from optical images were organized around 2-dimensional
smooth ($C^1$) sub-manifold. Simple models of synthetic images were
constructed in \cite{Peyre08} and were shown to lie close to
low-dimensional manifolds. We note that the set of all $m\times m$
image patches includes the sub-manifold of $m \times m$ patches
constructed from a single image, which is the sub-manifold studied in
this paper.
\subsection{Open questions}
This work provides experimental evidence of the stability of the
low-indices eigenvectors of the Laplacian  $\bL$ defined on a
graph. Similar results exist in the literature in the context of the
stability%
\begin{figure}[H]
  \vspace*{-3pc}
  \centerline{\hfill Laplacian \hfill k-SVD \hfill  NL-means \hfill  TISWT \hfill}
  \centerline{  
    \includegraphics[width = 0.25\textwidth]{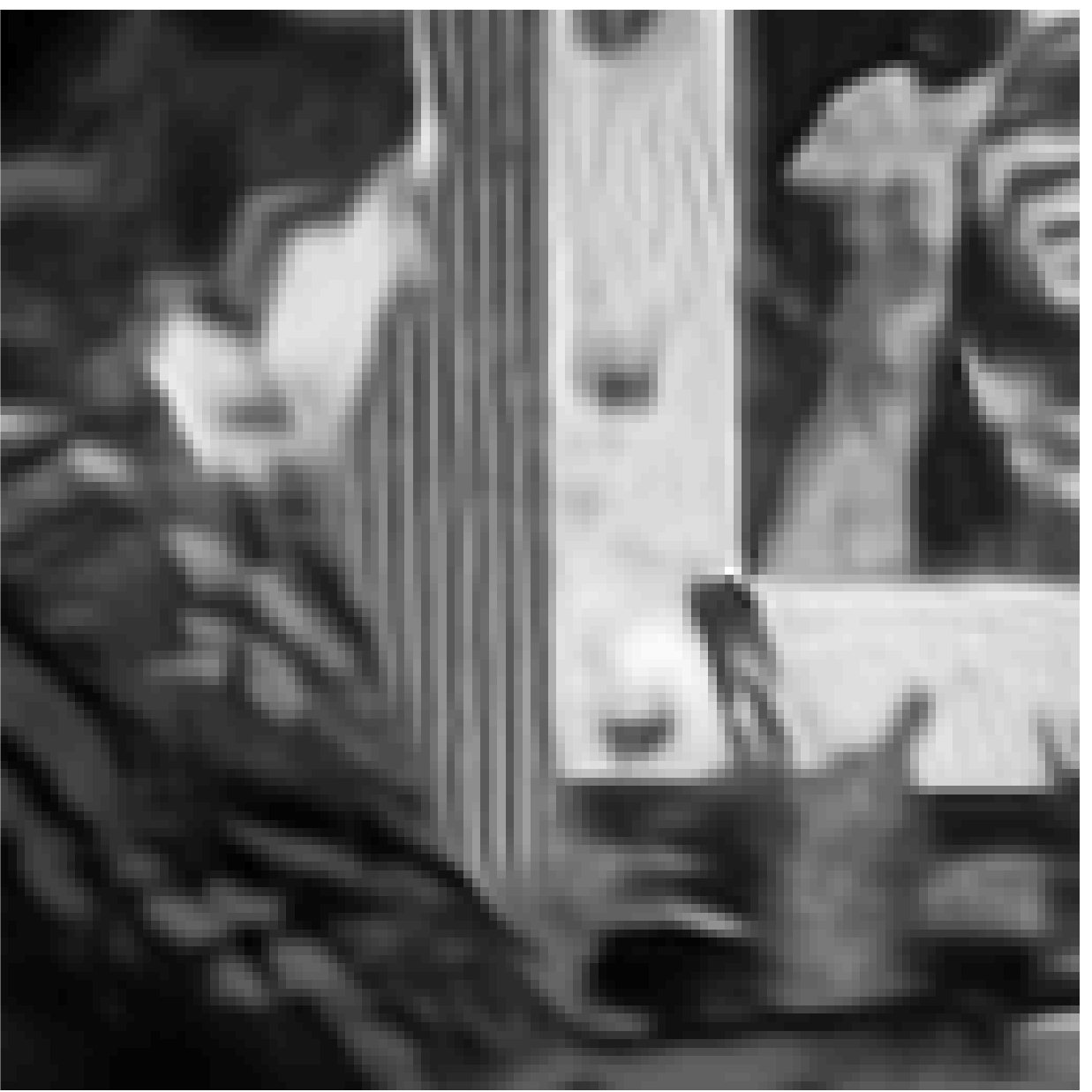}\hspace*{-0.25pc}  
    \includegraphics[width = 0.25\textwidth]{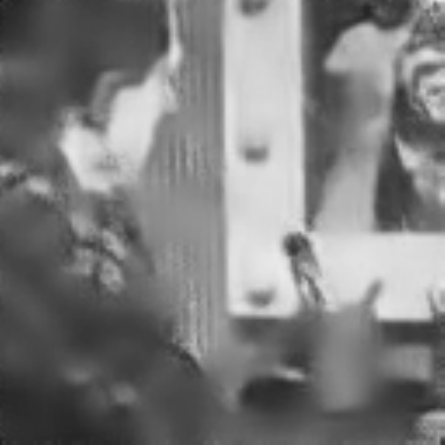}\hspace*{-0.25pc}  
    \includegraphics[width = 0.25\textwidth]{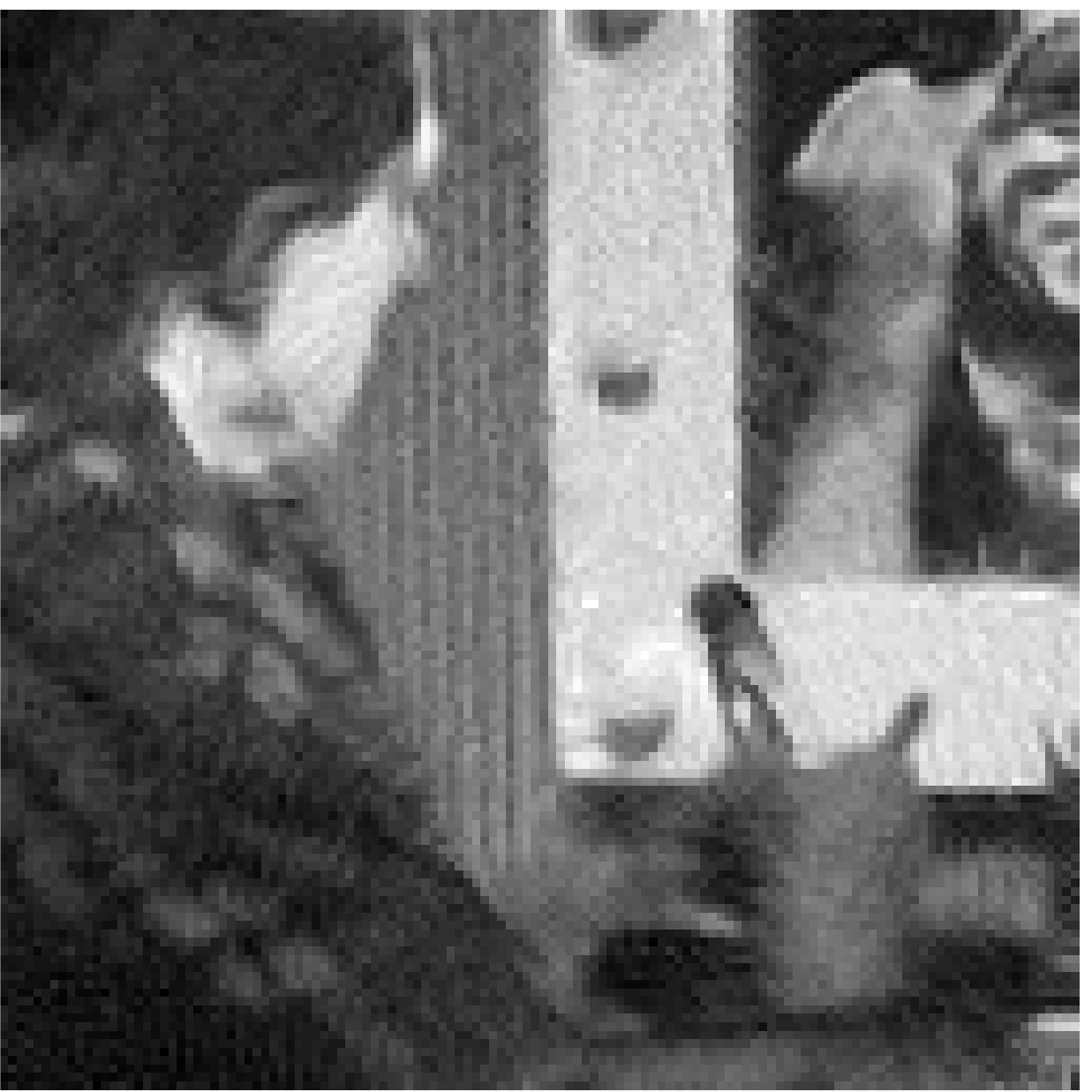}\hspace*{-0.25pc}  
    \includegraphics[width = 0.25\textwidth]{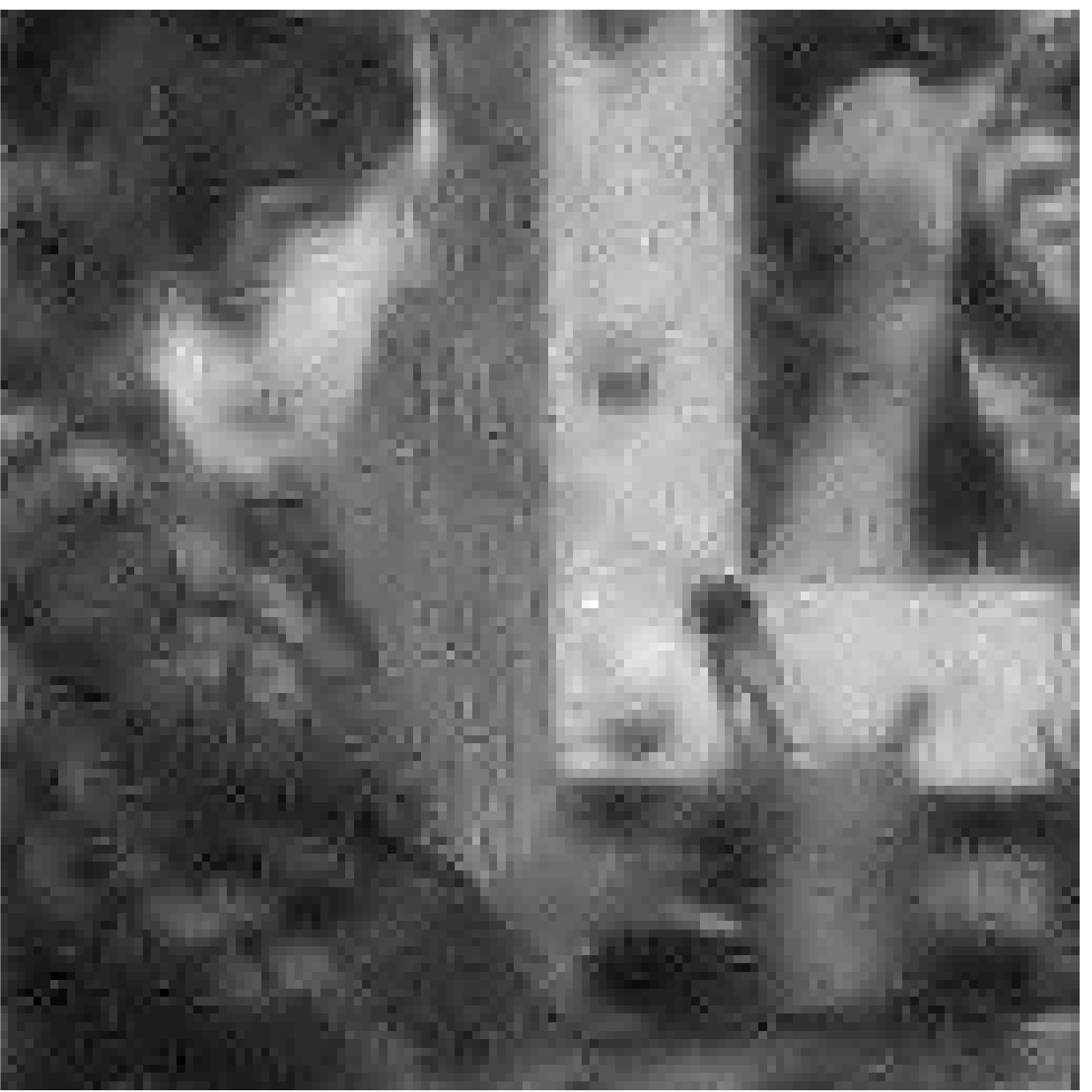}  
  }
  \centerline{  
    \includegraphics[width = 0.25\textwidth]{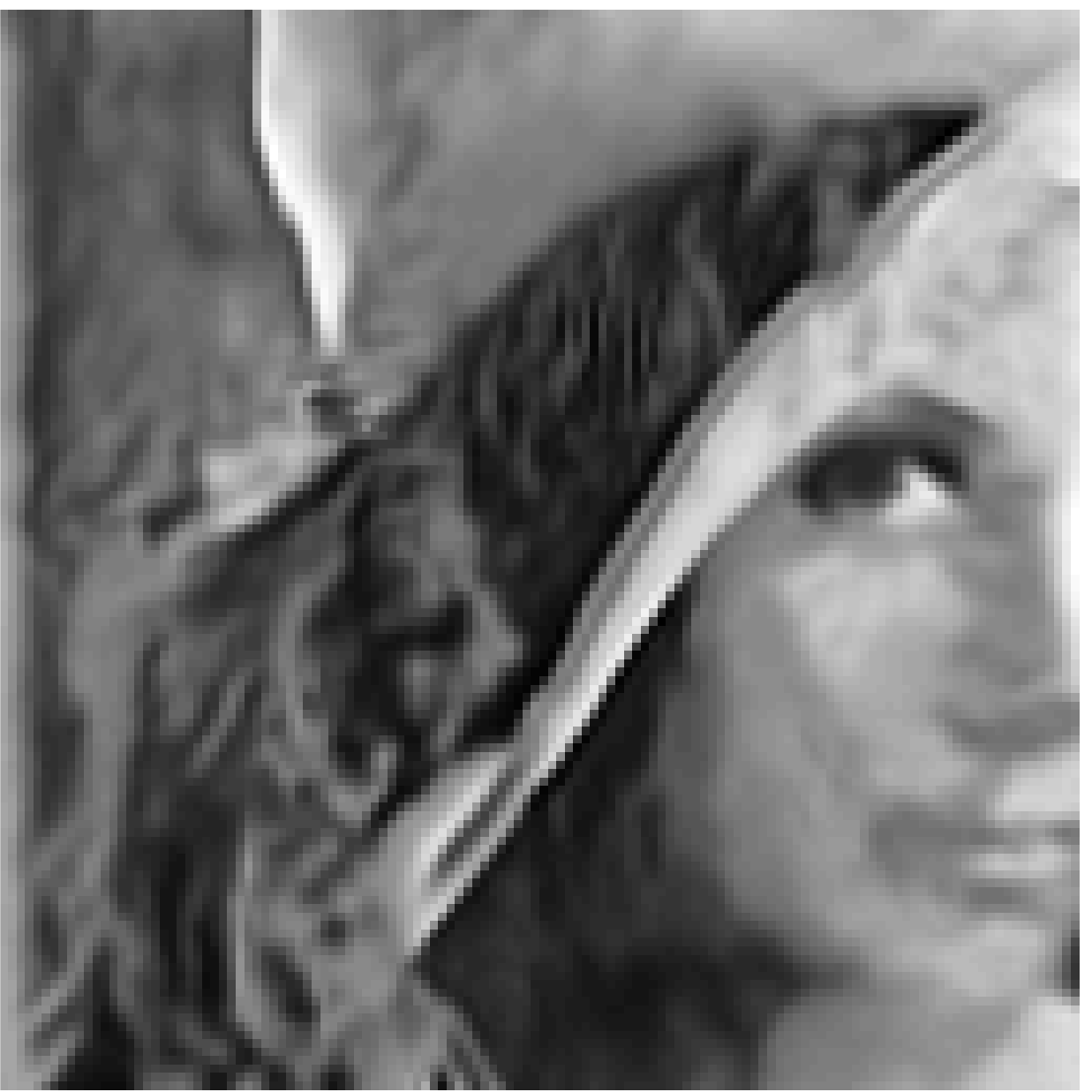}\hspace*{-0.25pc}  
    \includegraphics[width = 0.25\textwidth]{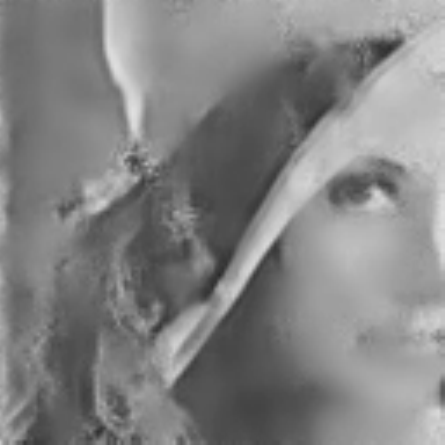}\hspace*{-0.25pc}  
    \includegraphics[width = 0.25\textwidth]{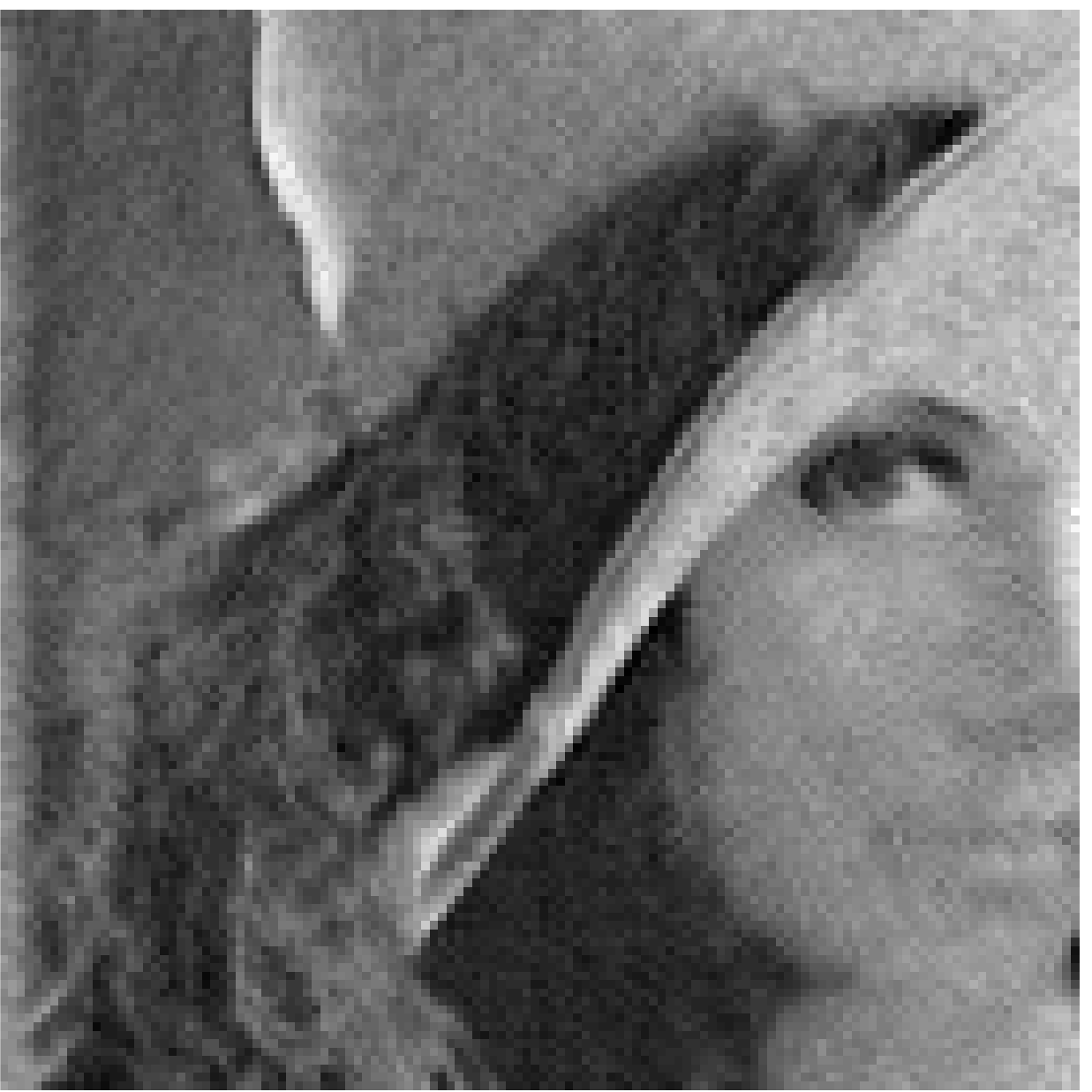}\hspace*{-0.25pc}  
    \includegraphics[width = 0.25\textwidth]{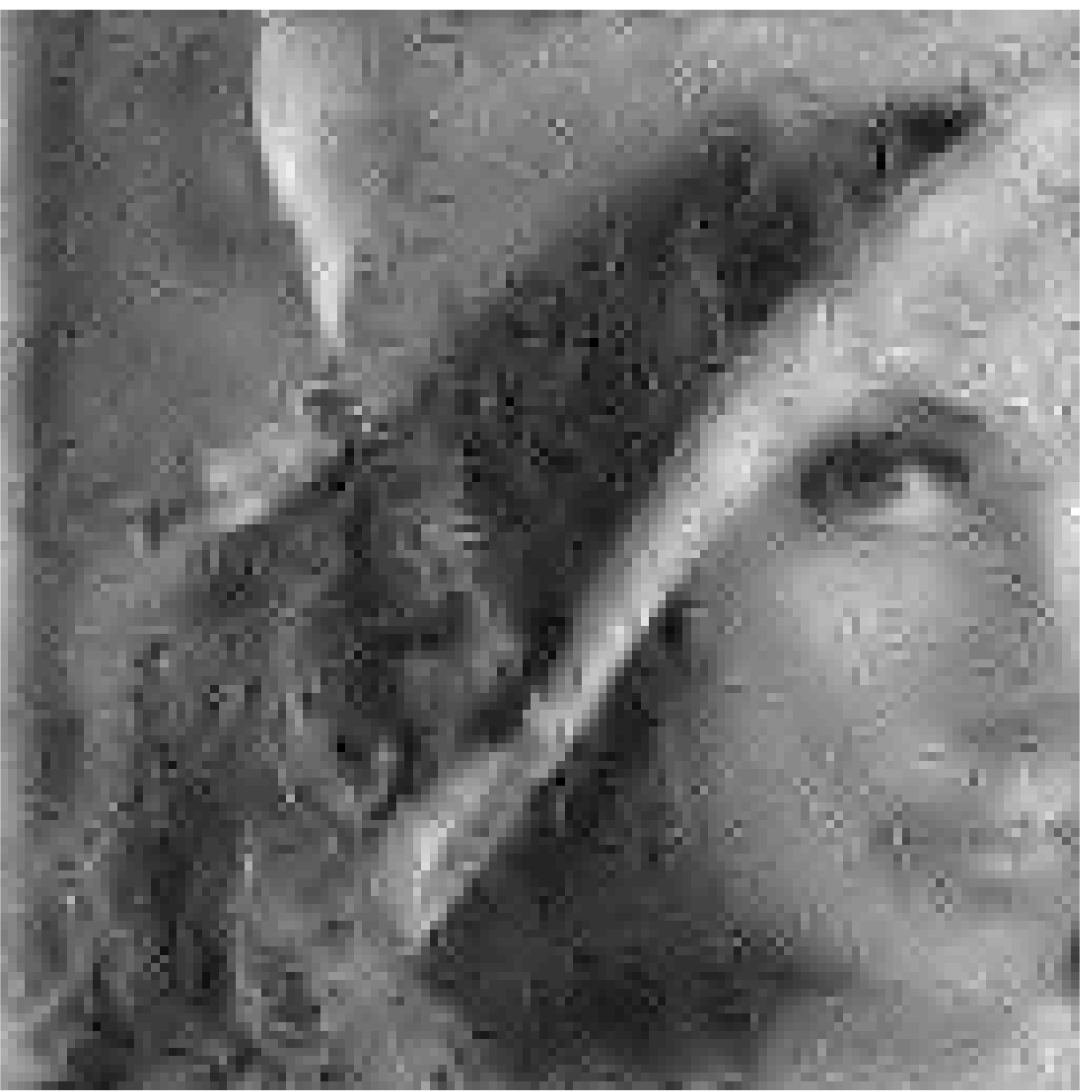}  
  }
  \centerline{  
    \includegraphics[width = 0.25\textwidth]{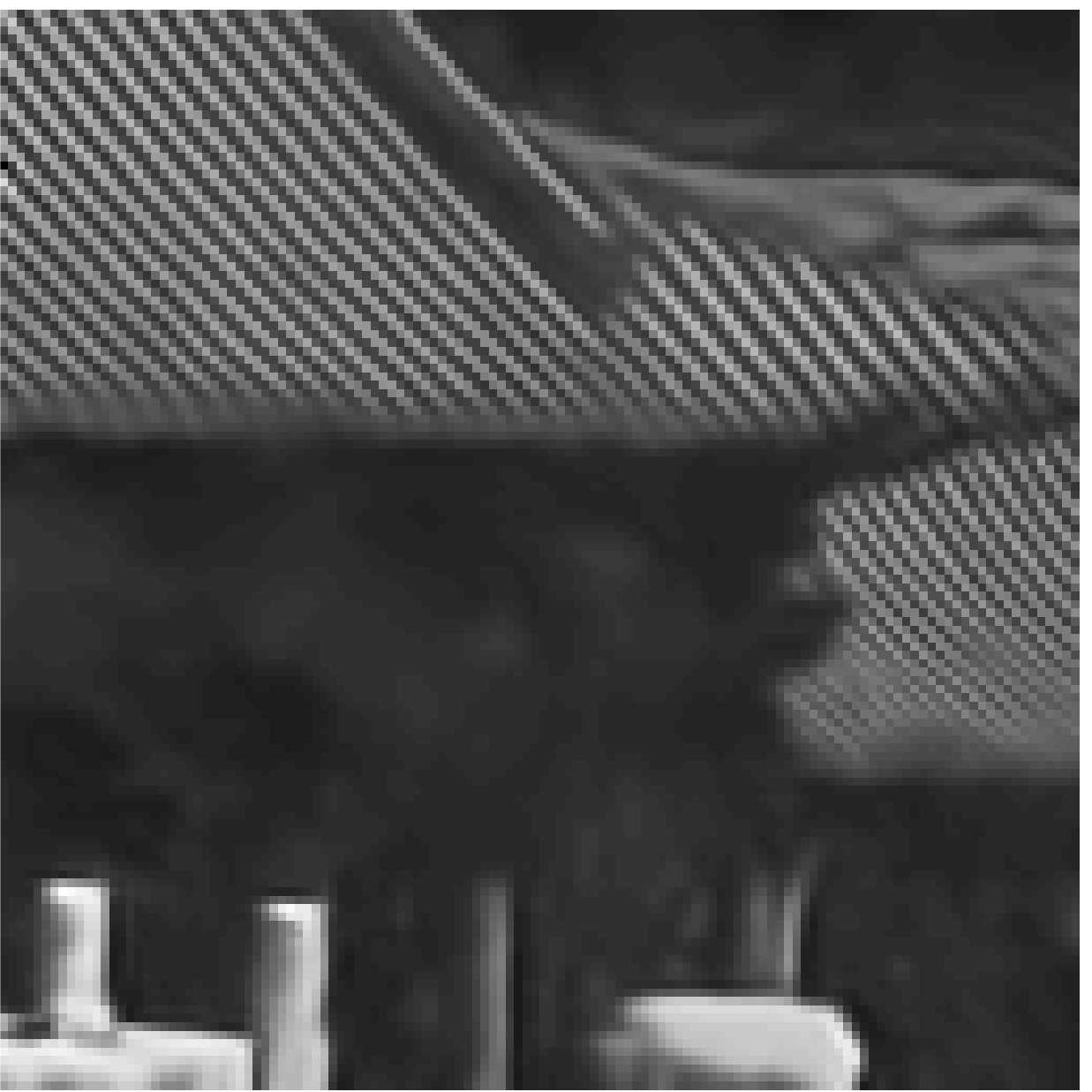}\hspace*{-0.25pc}  
    \includegraphics[width = 0.25\textwidth]{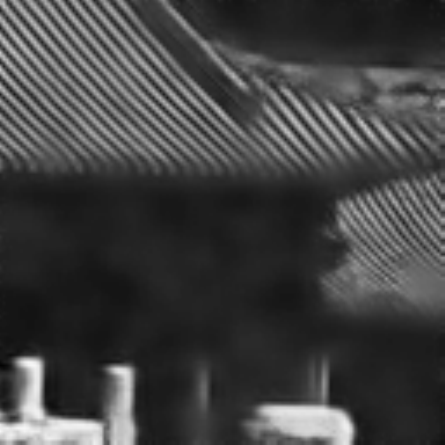}\hspace*{-0.25pc}  
    \includegraphics[width = 0.25\textwidth]{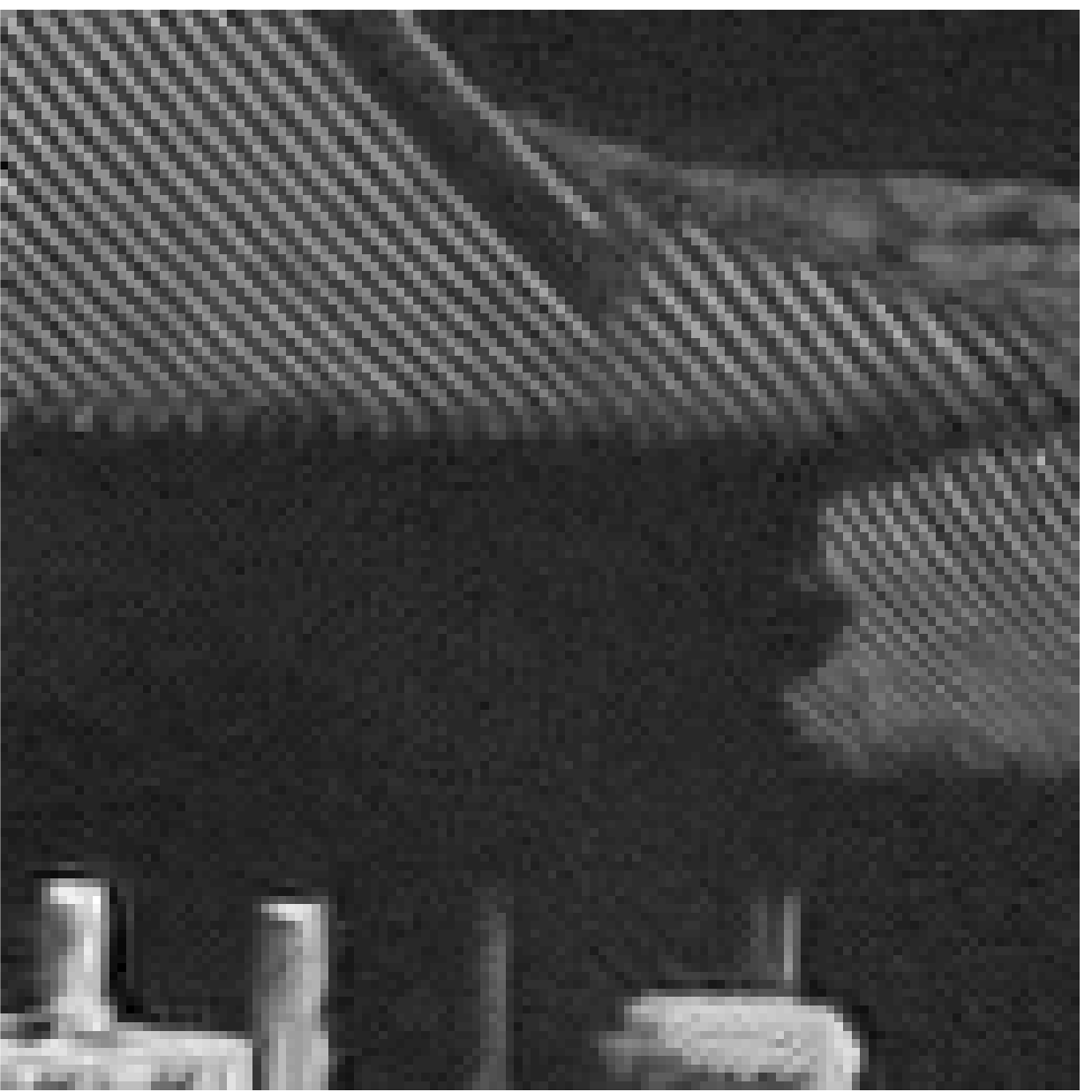}\hspace*{-0.25pc}  
    \includegraphics[width = 0.25\textwidth]{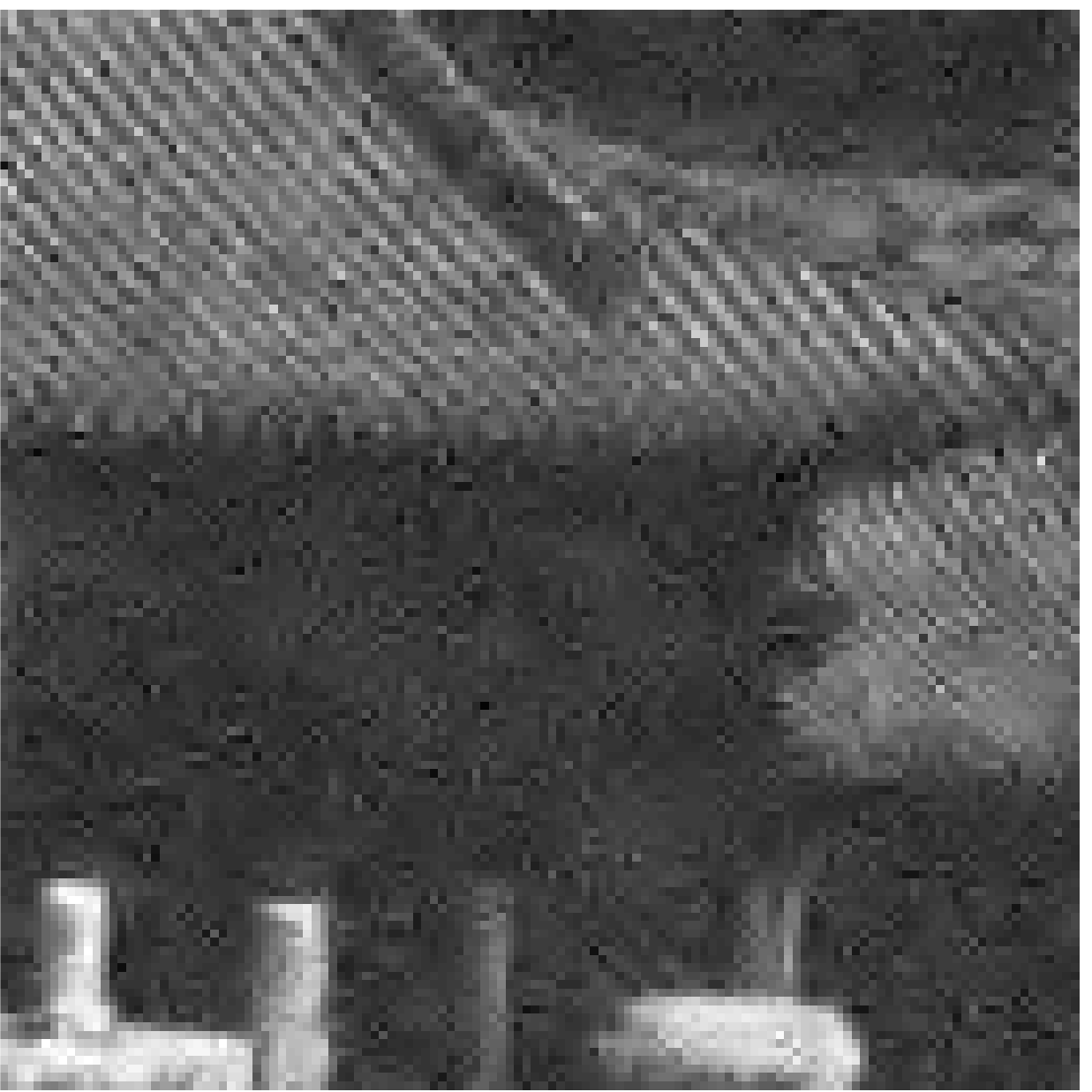}  
  }
  \centerline{  
    \includegraphics[width = 0.25\textwidth]{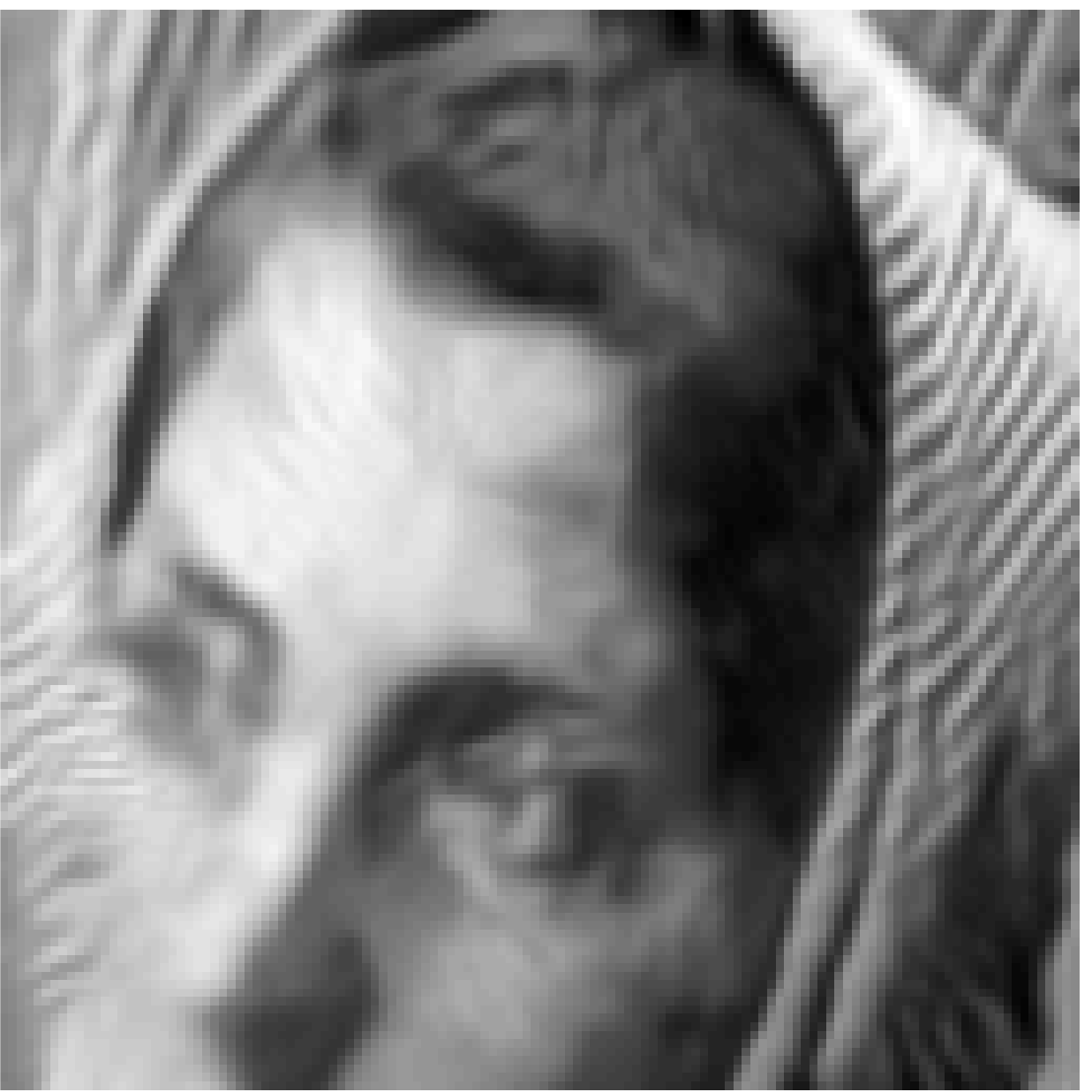}\hspace*{-0.25pc}  
    \includegraphics[width = 0.25\textwidth]{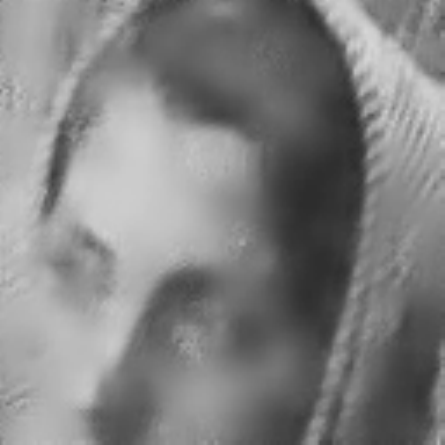}\hspace*{-0.25pc}  
    \includegraphics[width = 0.25\textwidth]{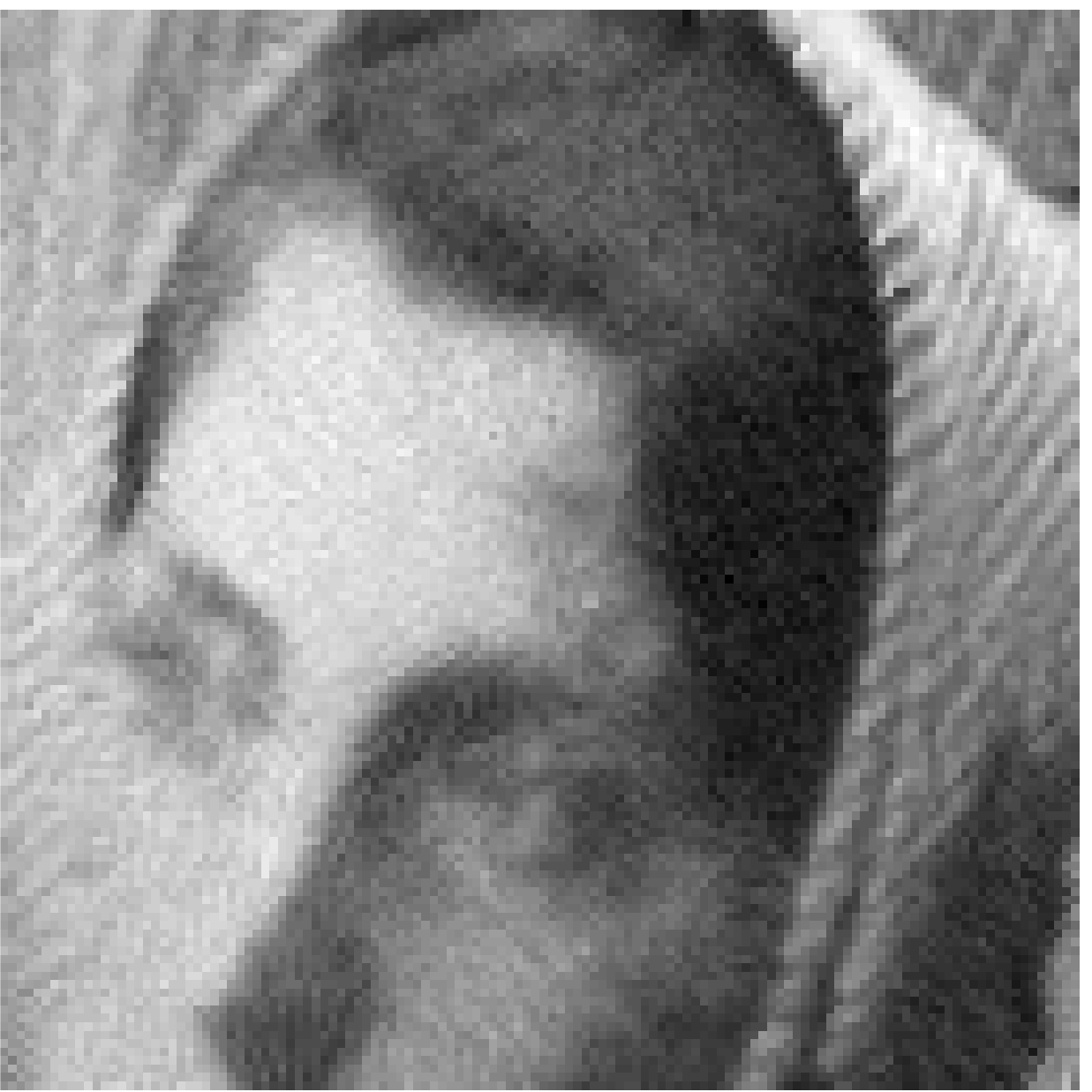}\hspace*{-0.25pc}  
    \includegraphics[width = 0.25\textwidth]{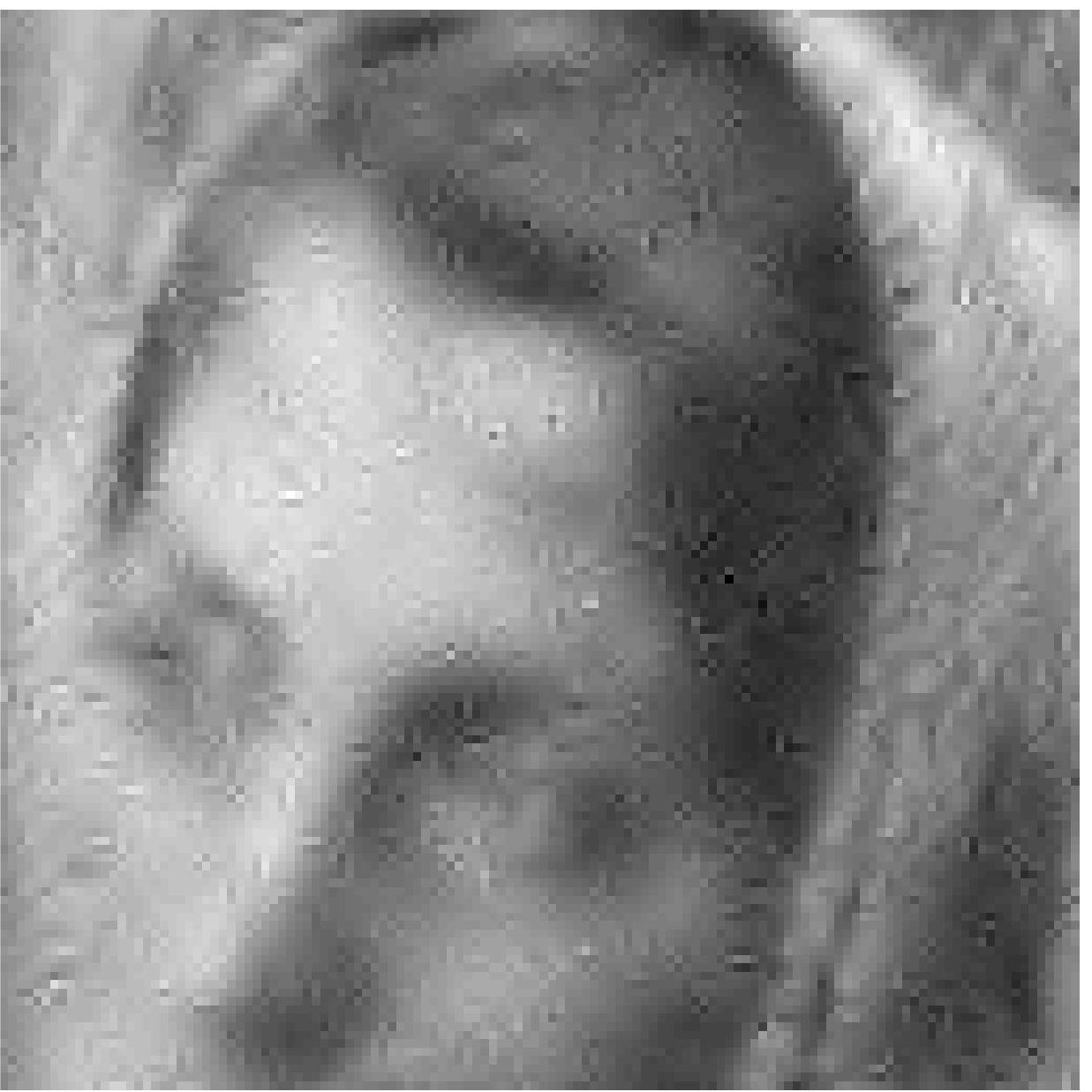}  
  }
  \centerline{  
    \includegraphics[width = 0.25\textwidth]{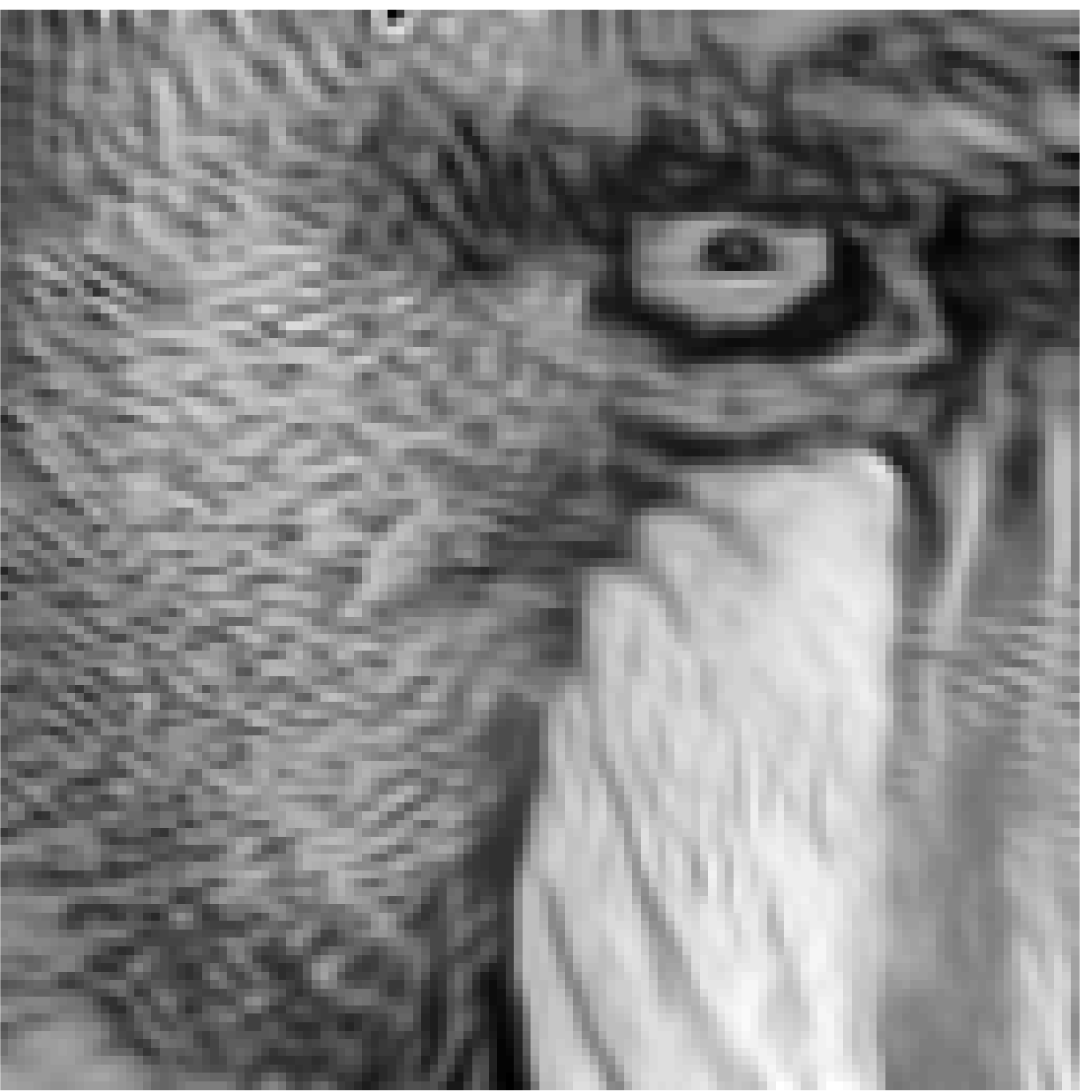}\hspace*{-0.25pc}  
    \includegraphics[width = 0.25\textwidth]{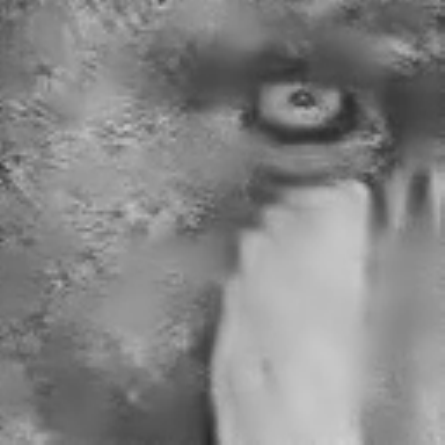}\hspace*{-0.25pc}  
    \includegraphics[width = 0.25\textwidth]{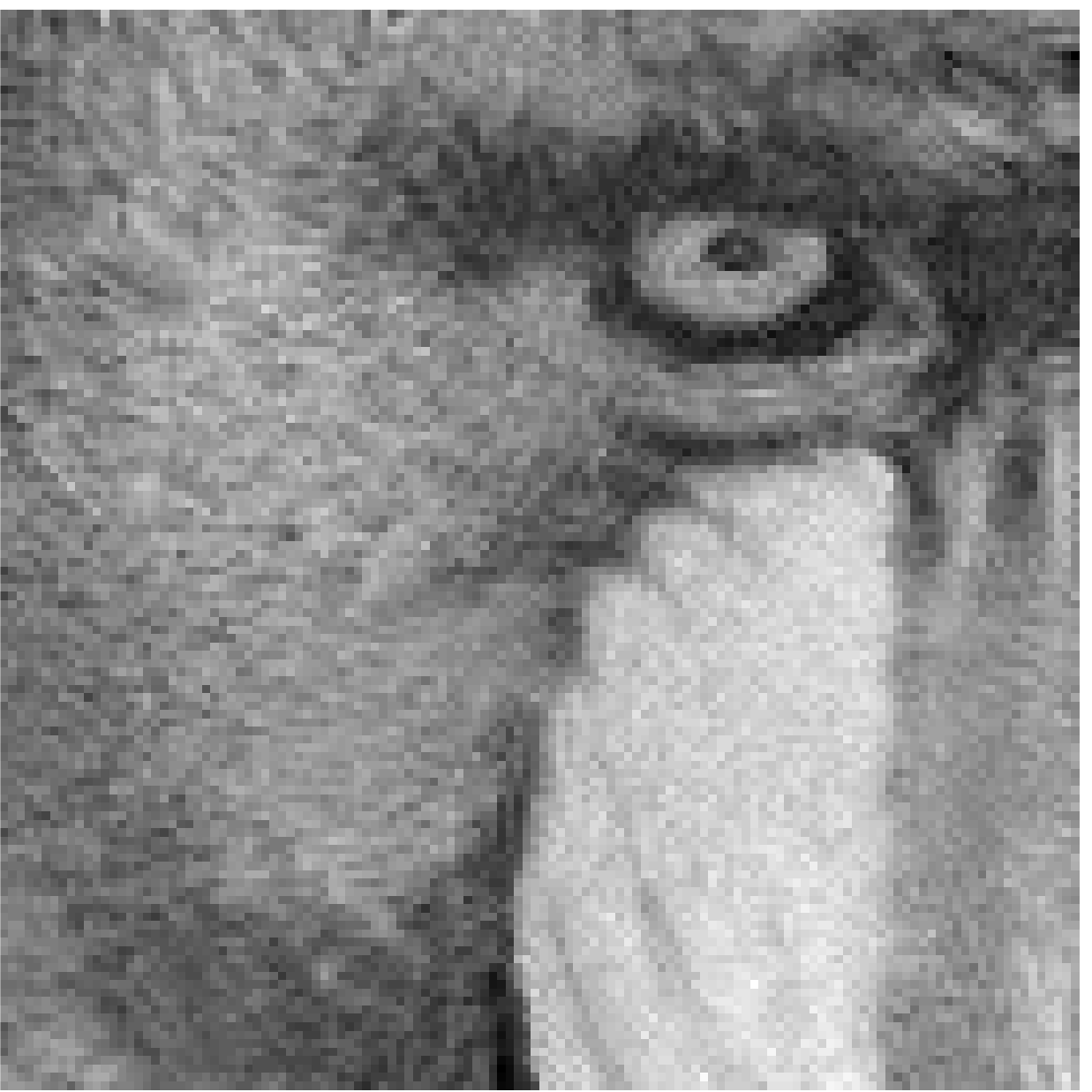}\hspace*{-0.25pc}  
    \includegraphics[width = 0.25\textwidth]{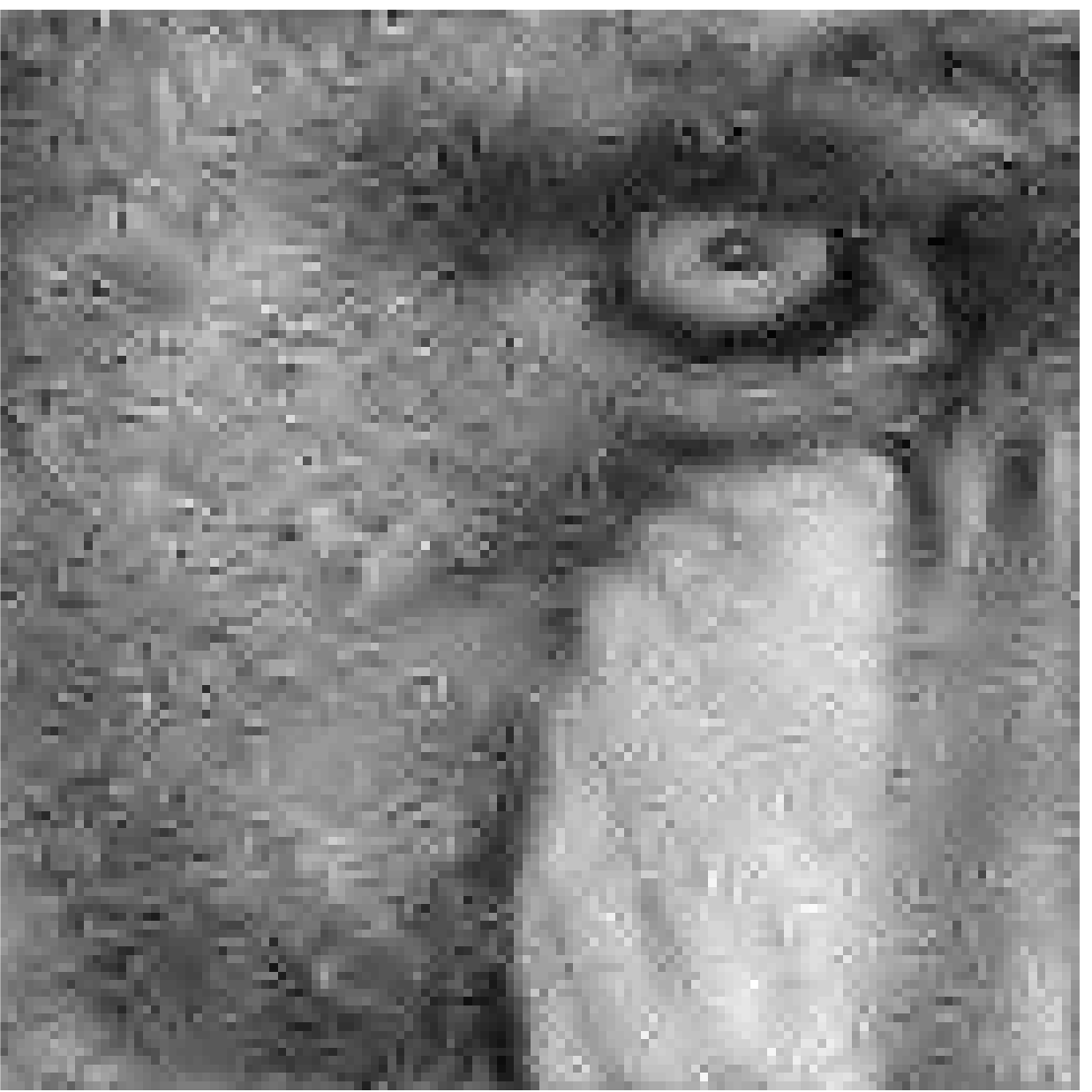}  
  }
  \centerline{\hfill Laplacian \hfill k-SVD \hfill  NL-means \hfill  TISWT \hfill}
  \caption{Noise level $\sigma = 40$. From left to right: our
    two-stage approach, k-SVD, nonlocal means, translation-invariant
    wavelet denoising.
    \label{denoised_img}}
\end{figure}
\begin{figure}[H]
  \vspace*{-3pc}
  \centerline{\hfill Laplacian \hfill k-SVD \hfill  NL-means \hfill  TISWT \hfill}
  \centerline{  
    \includegraphics[width = 0.25\textwidth]{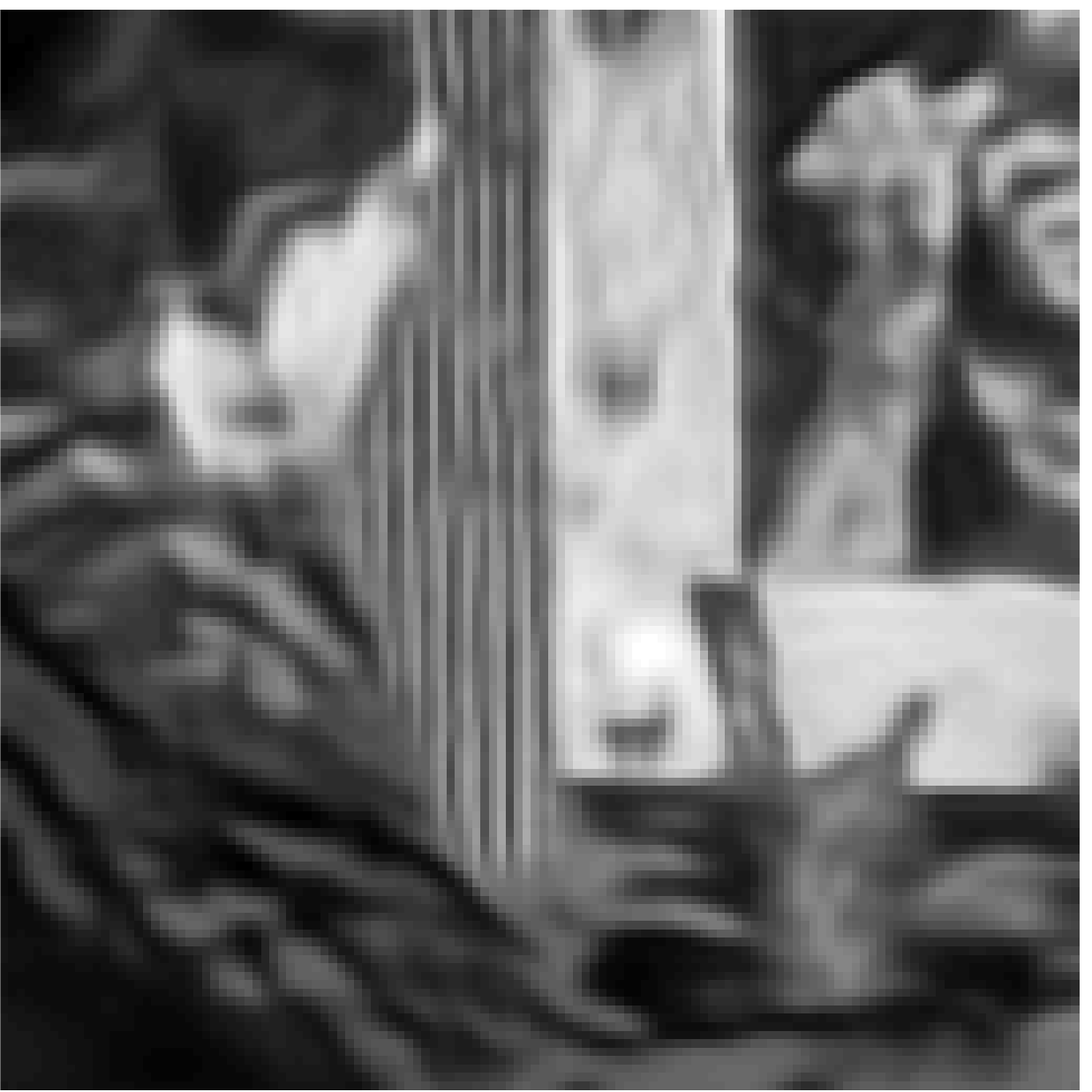}\hspace*{-0.25pc}    
    \includegraphics[width = 0.25\textwidth]{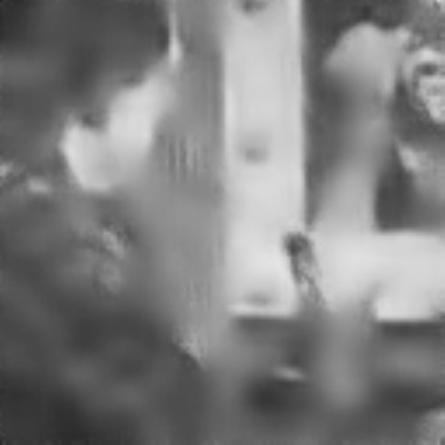}\hspace*{-0.25pc}  
    \includegraphics[width = 0.25\textwidth]{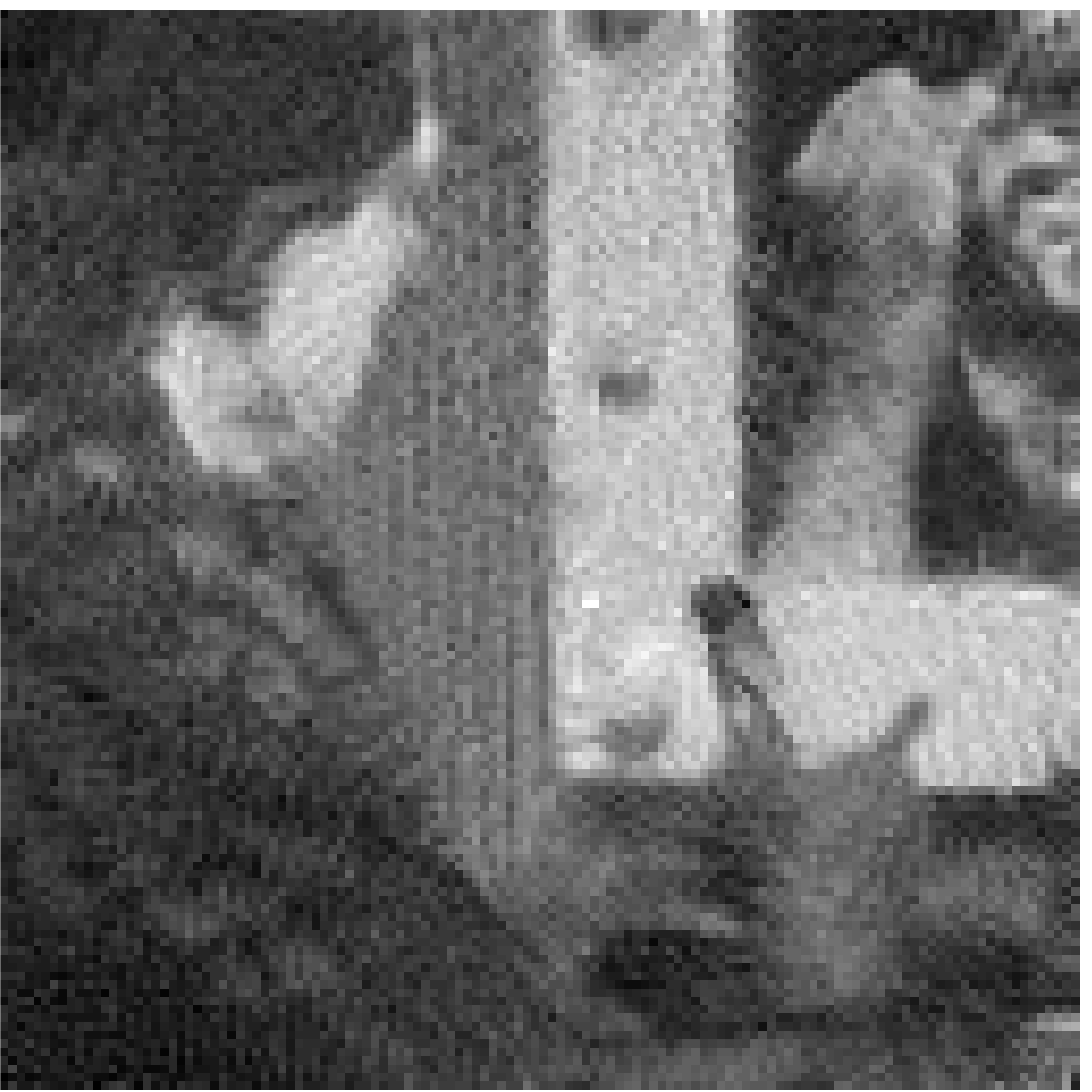}\hspace*{-0.25pc}  
    \includegraphics[width = 0.25\textwidth]{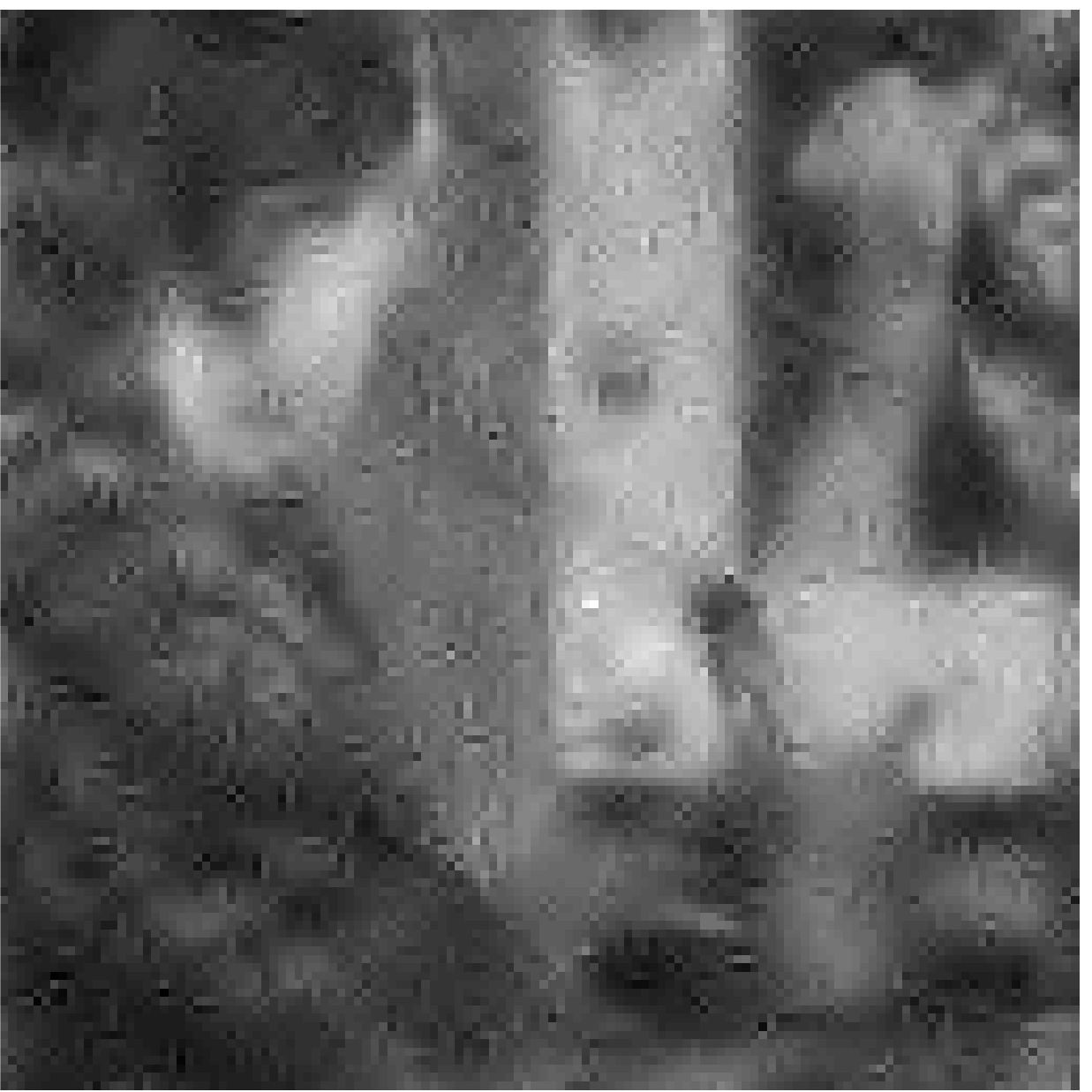}  
  }
  \centerline{  
    \includegraphics[width = 0.25\textwidth]{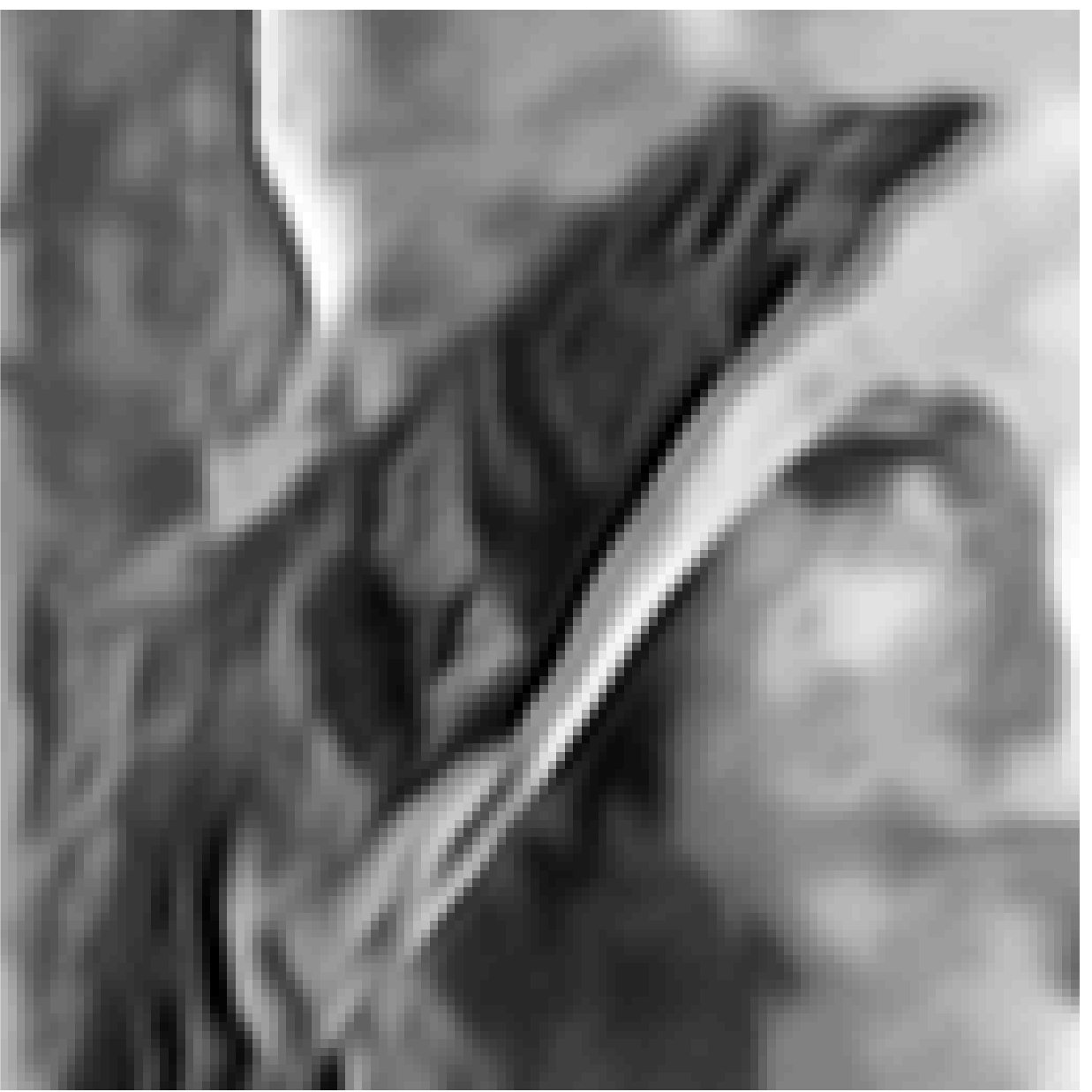}\hspace*{-0.25pc}  
    \includegraphics[width = 0.25\textwidth]{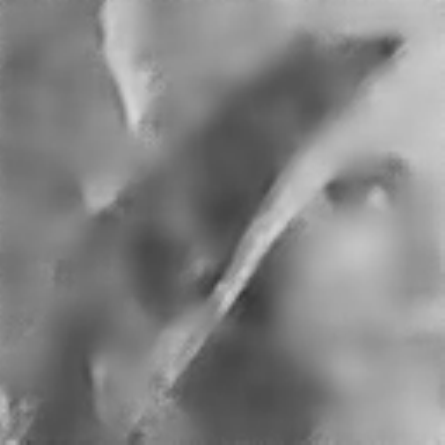}\hspace*{-0.25pc}  
    \includegraphics[width = 0.25\textwidth]{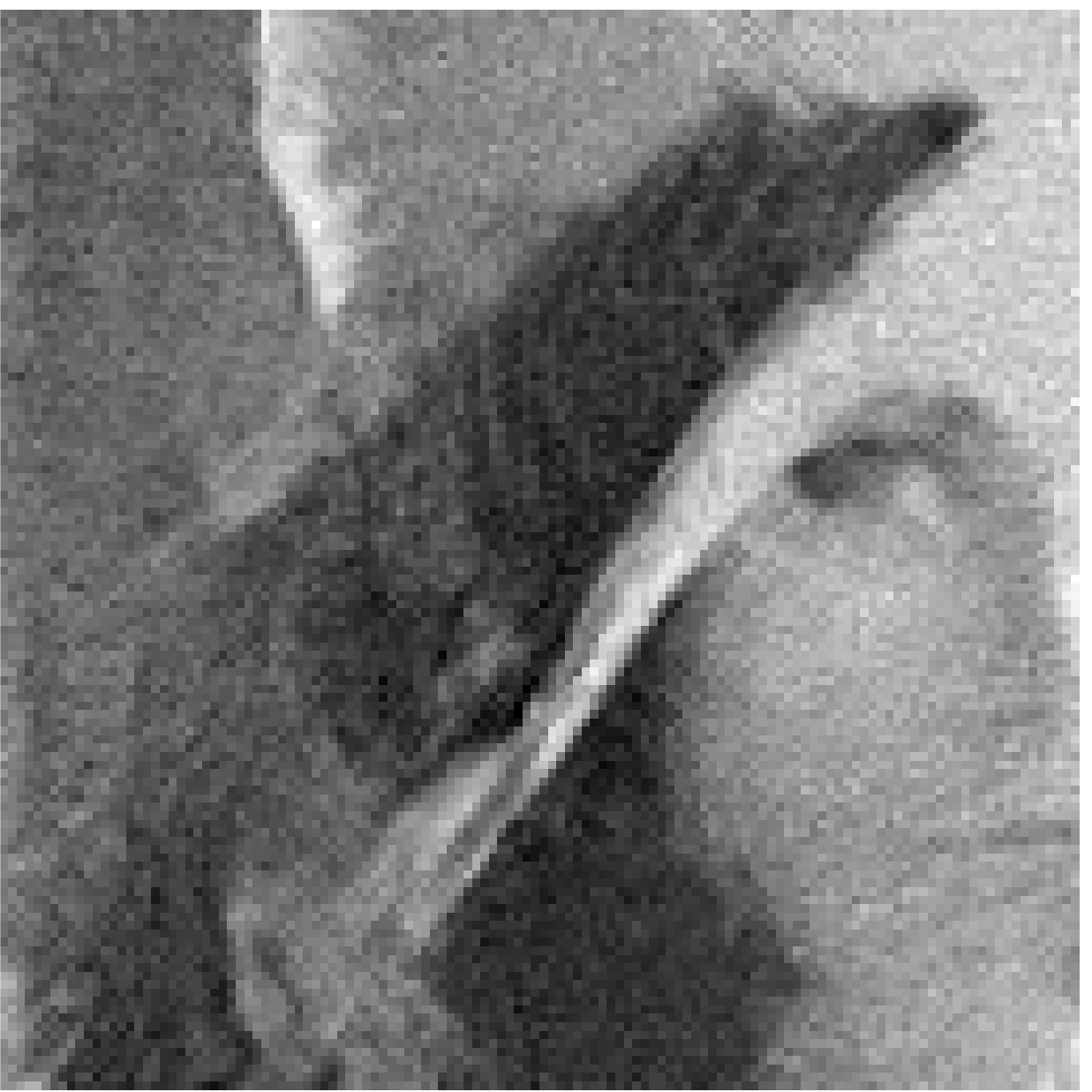}\hspace*{-0.25pc}  
    \includegraphics[width = 0.25\textwidth]{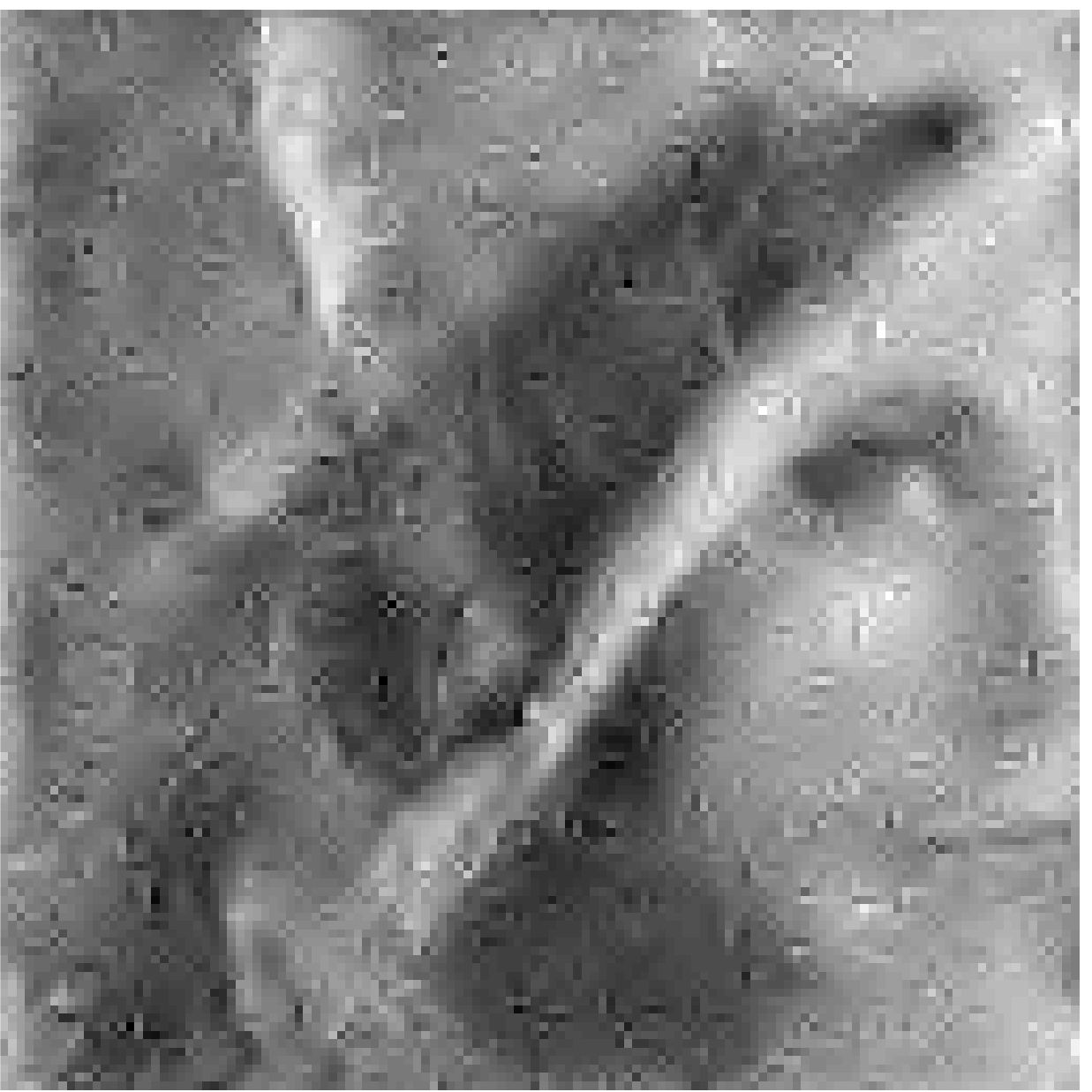}  
  }
  \centerline{  
    \includegraphics[width = 0.25\textwidth]{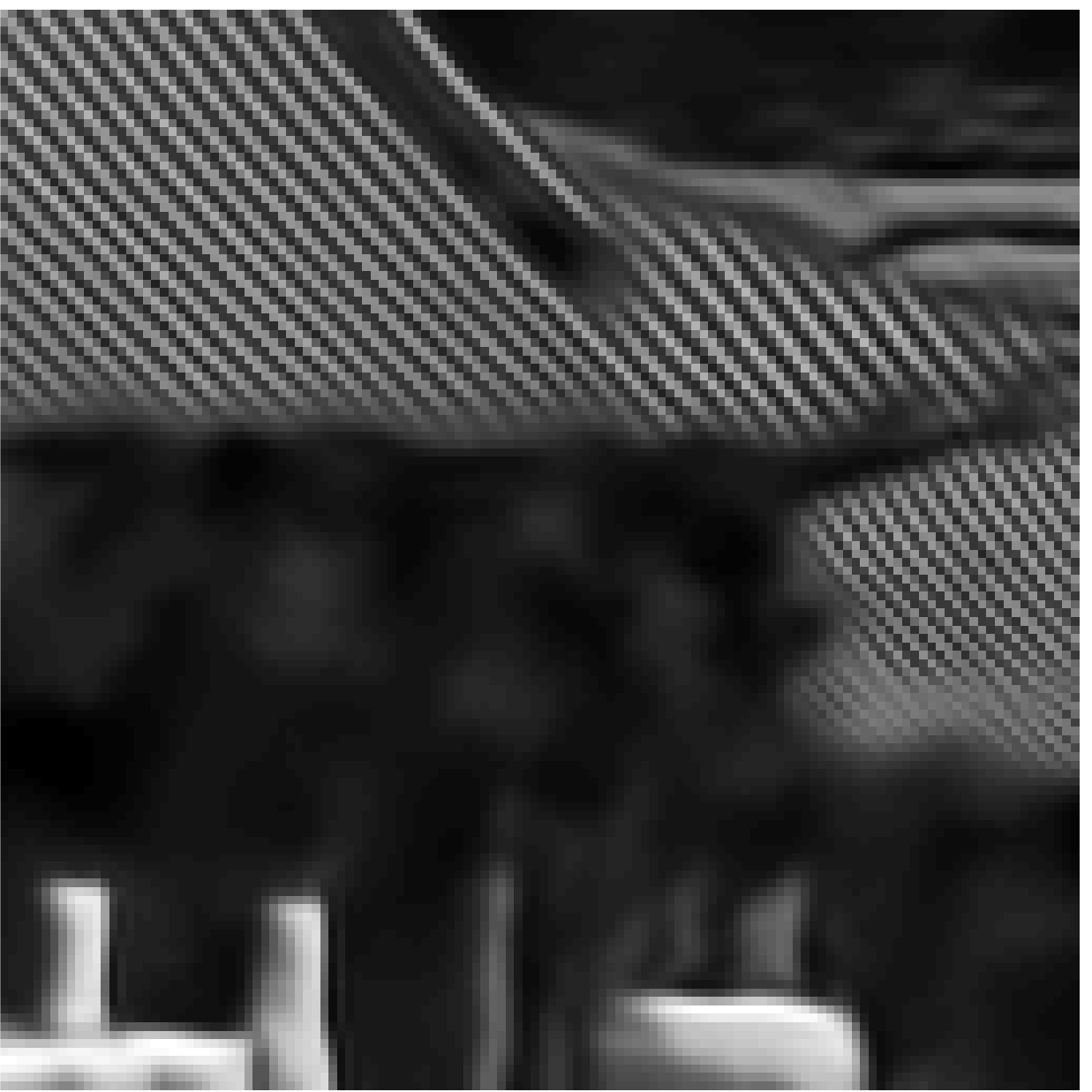}\hspace*{-0.25pc}  
    \includegraphics[width = 0.25\textwidth]{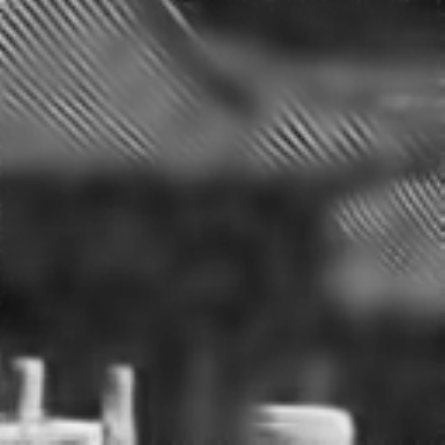}\hspace*{-0.25pc}  
    \includegraphics[width = 0.25\textwidth]{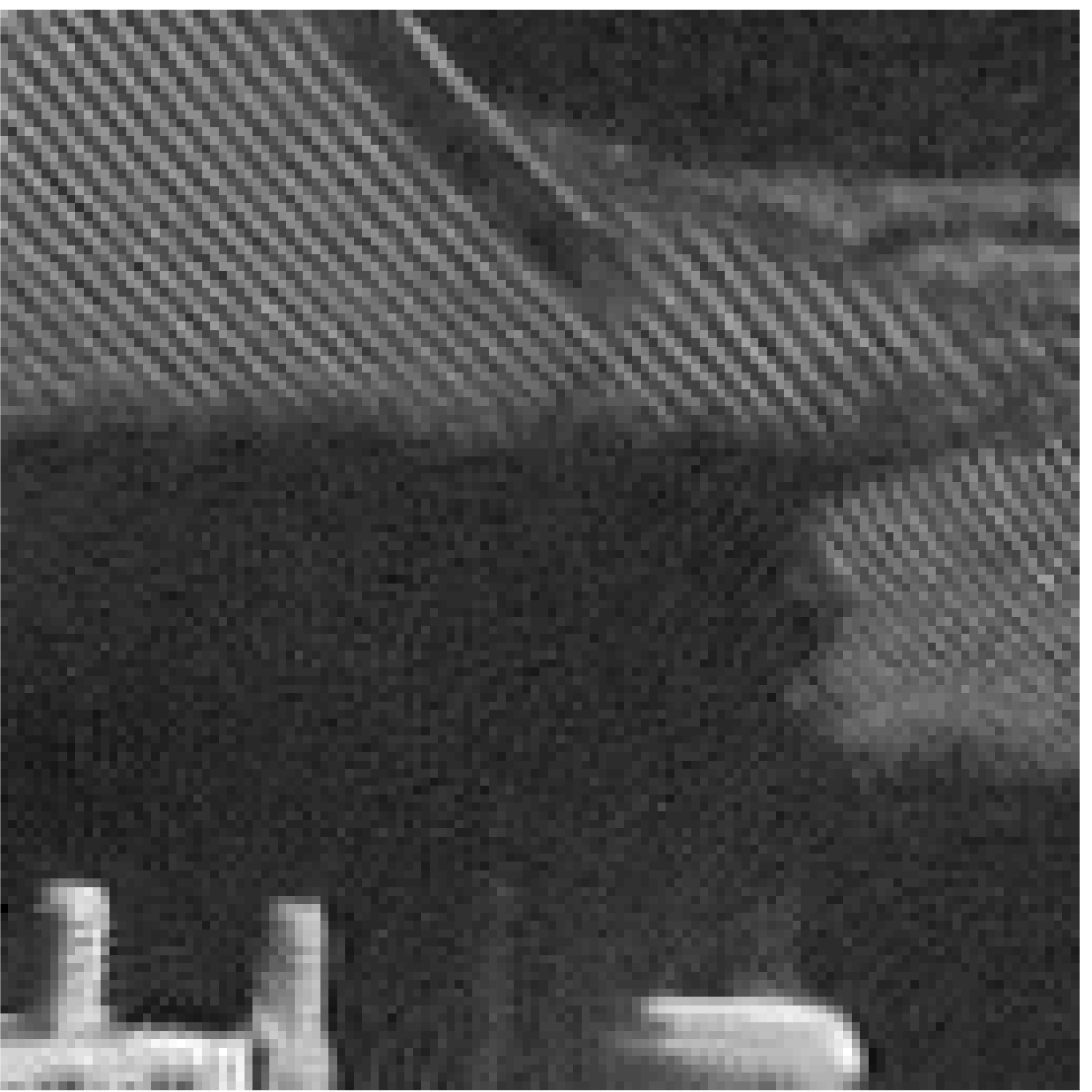}\hspace*{-0.25pc}  
    \includegraphics[width = 0.25\textwidth]{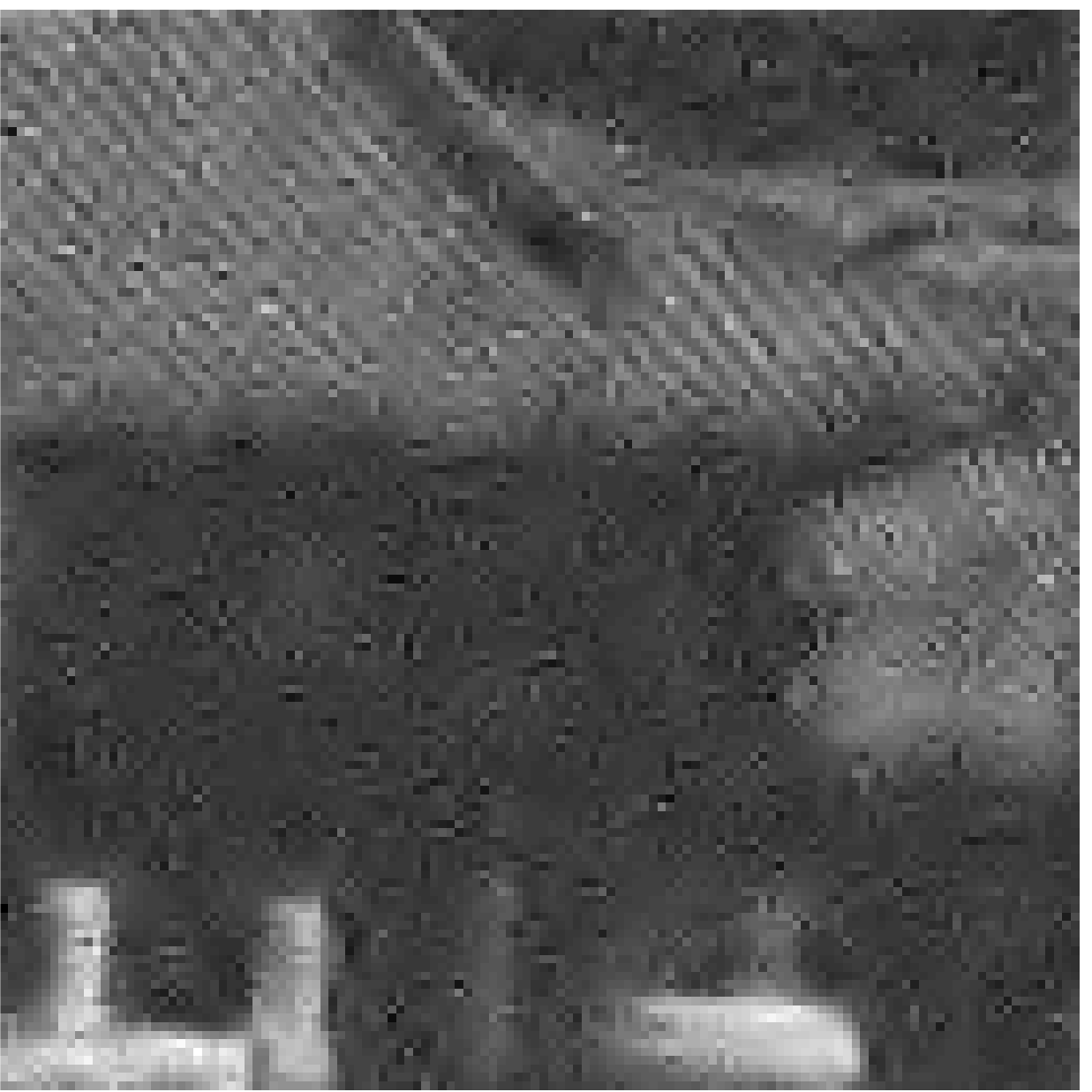}  
  }
  \centerline{  
    \includegraphics[width = 0.25\textwidth]{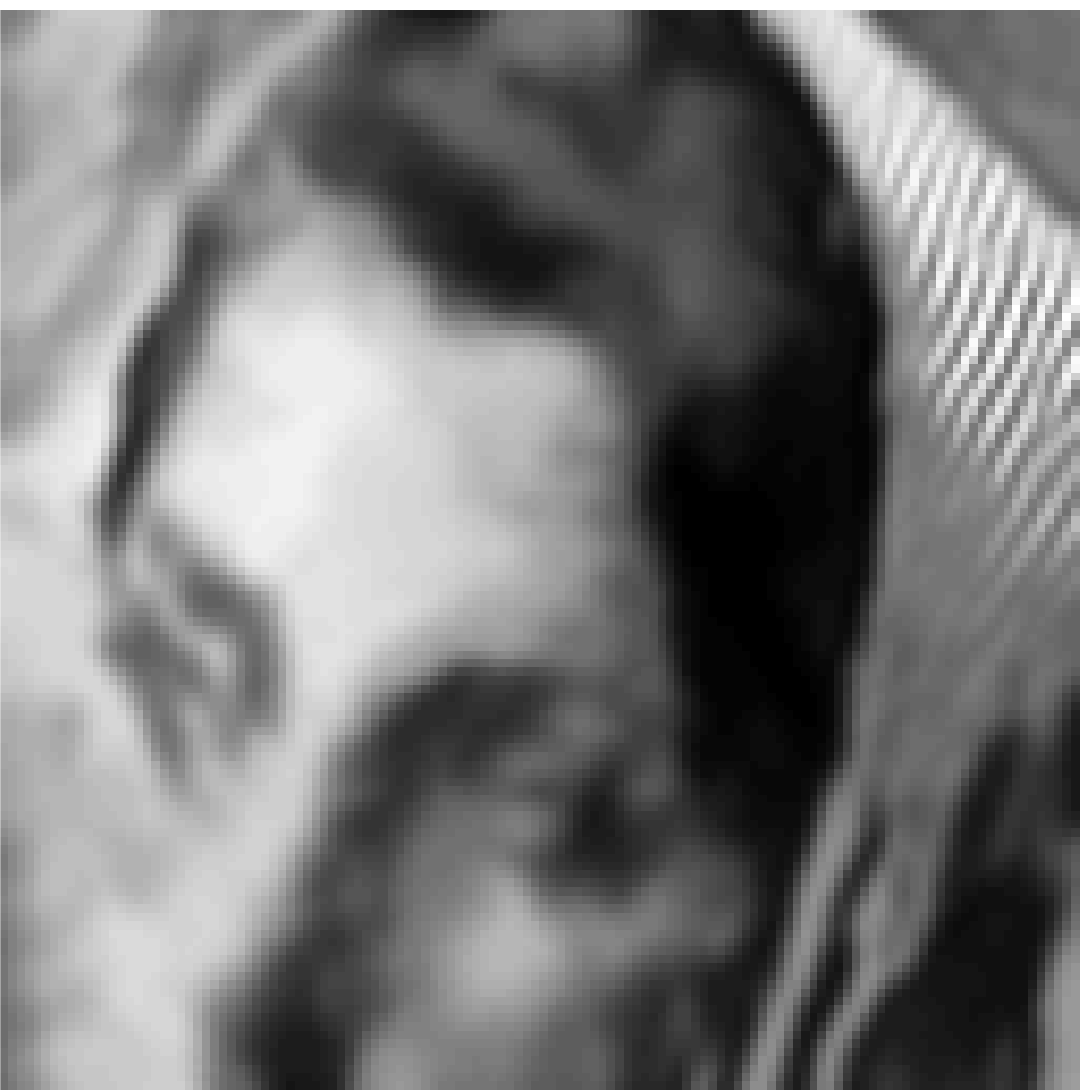}\hspace*{-0.25pc}  
    \includegraphics[width = 0.25\textwidth]{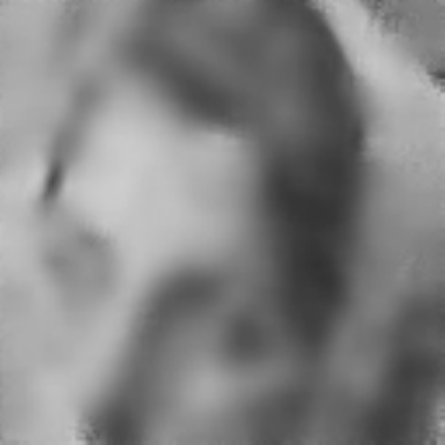}\hspace*{-0.25pc}  
    \includegraphics[width = 0.25\textwidth]{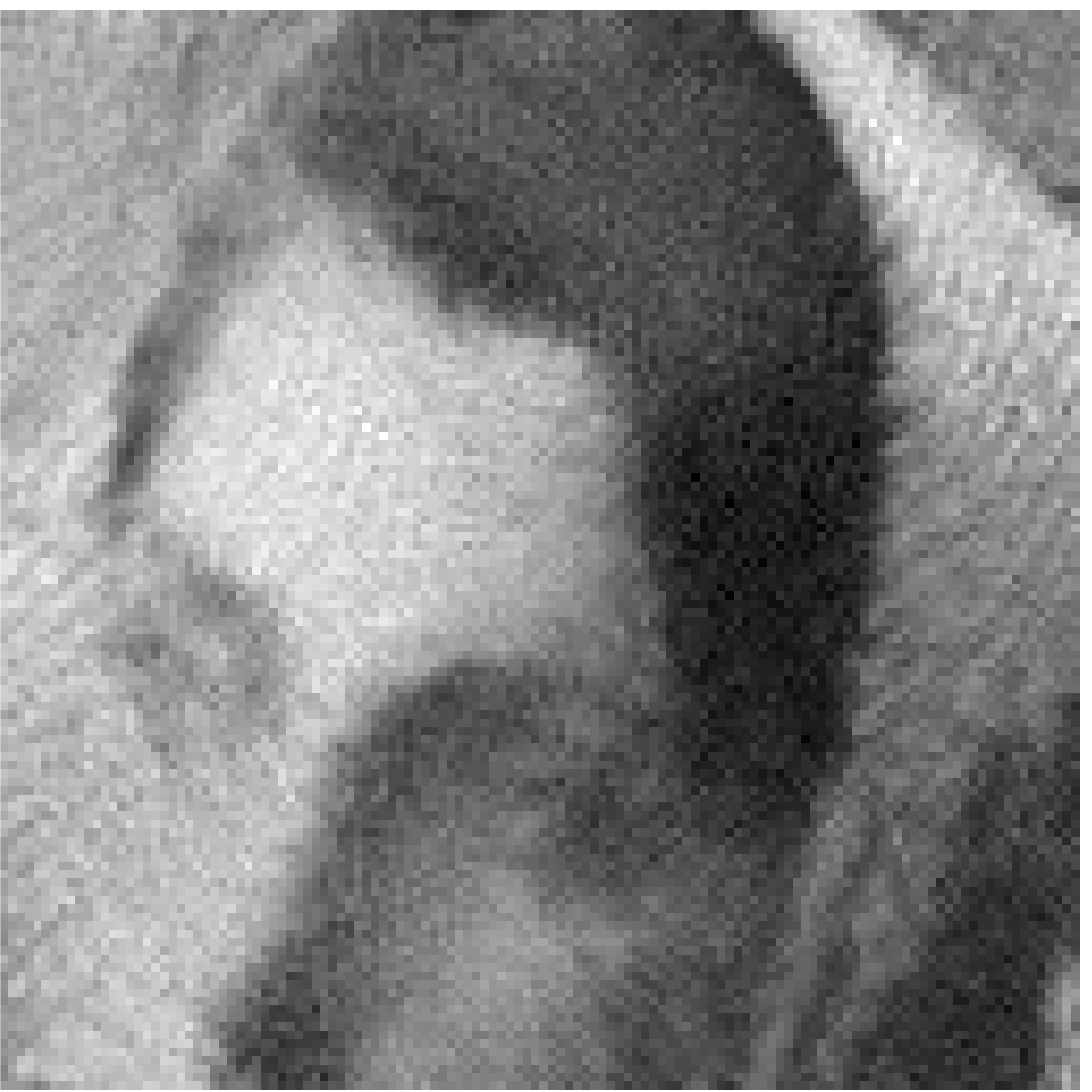}\hspace*{-0.25pc}  
    \includegraphics[width = 0.25\textwidth]{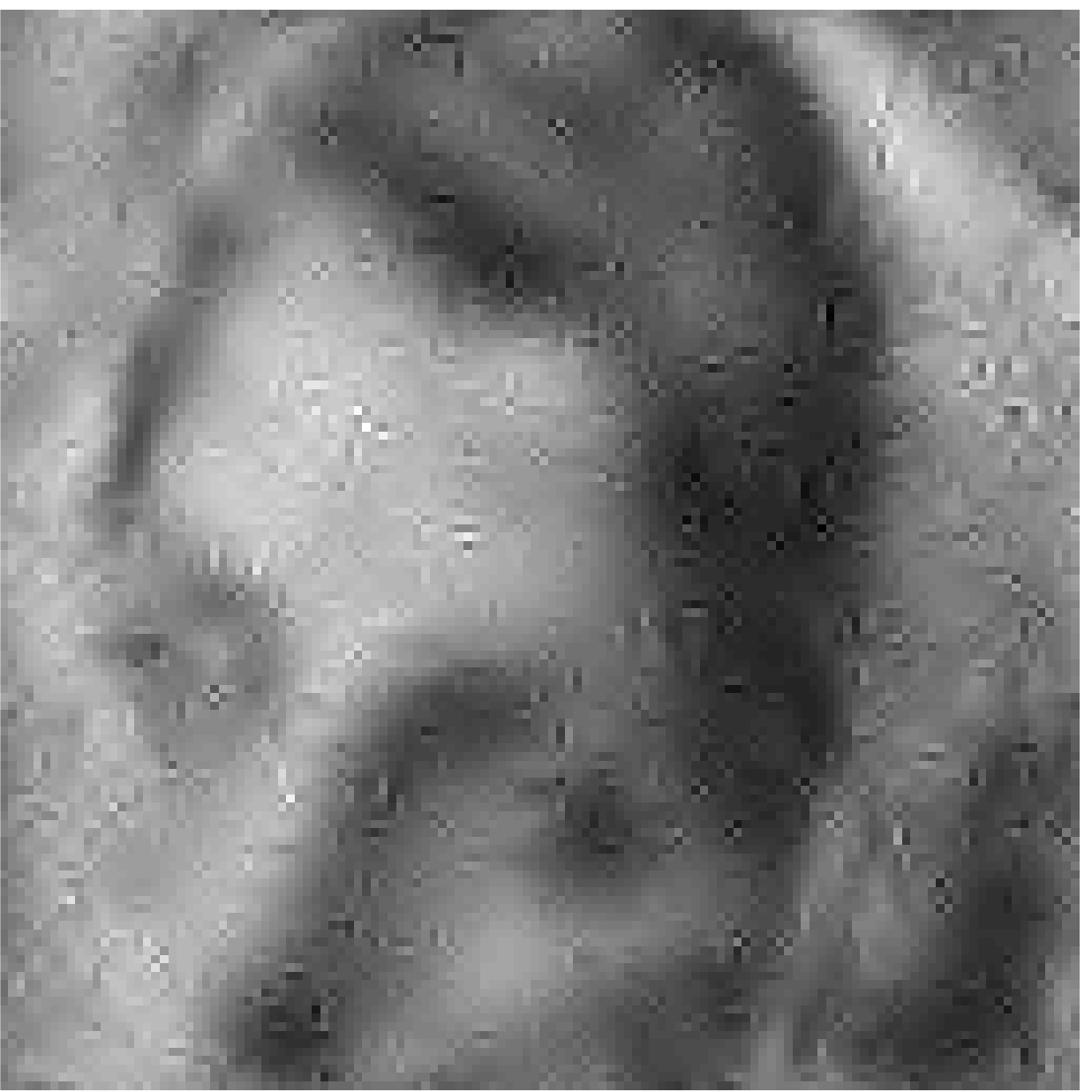}  
  }
  \centerline{  
    \includegraphics[width = 0.25\textwidth]{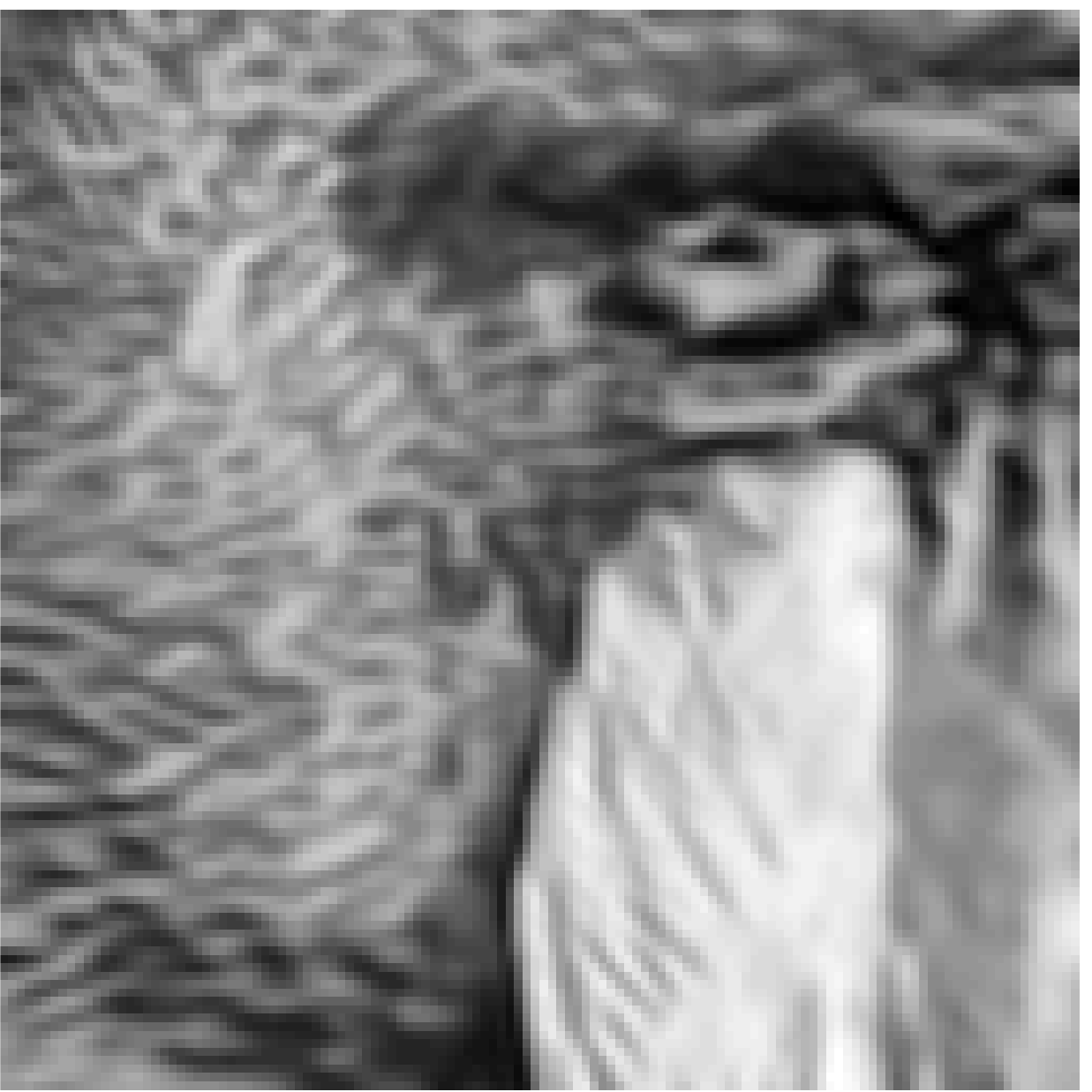}\hspace*{-0.25pc}  
    \includegraphics[width = 0.25\textwidth]{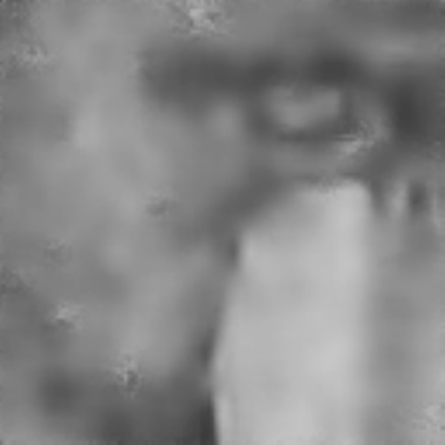}\hspace*{-0.25pc}  
    \includegraphics[width = 0.25\textwidth]{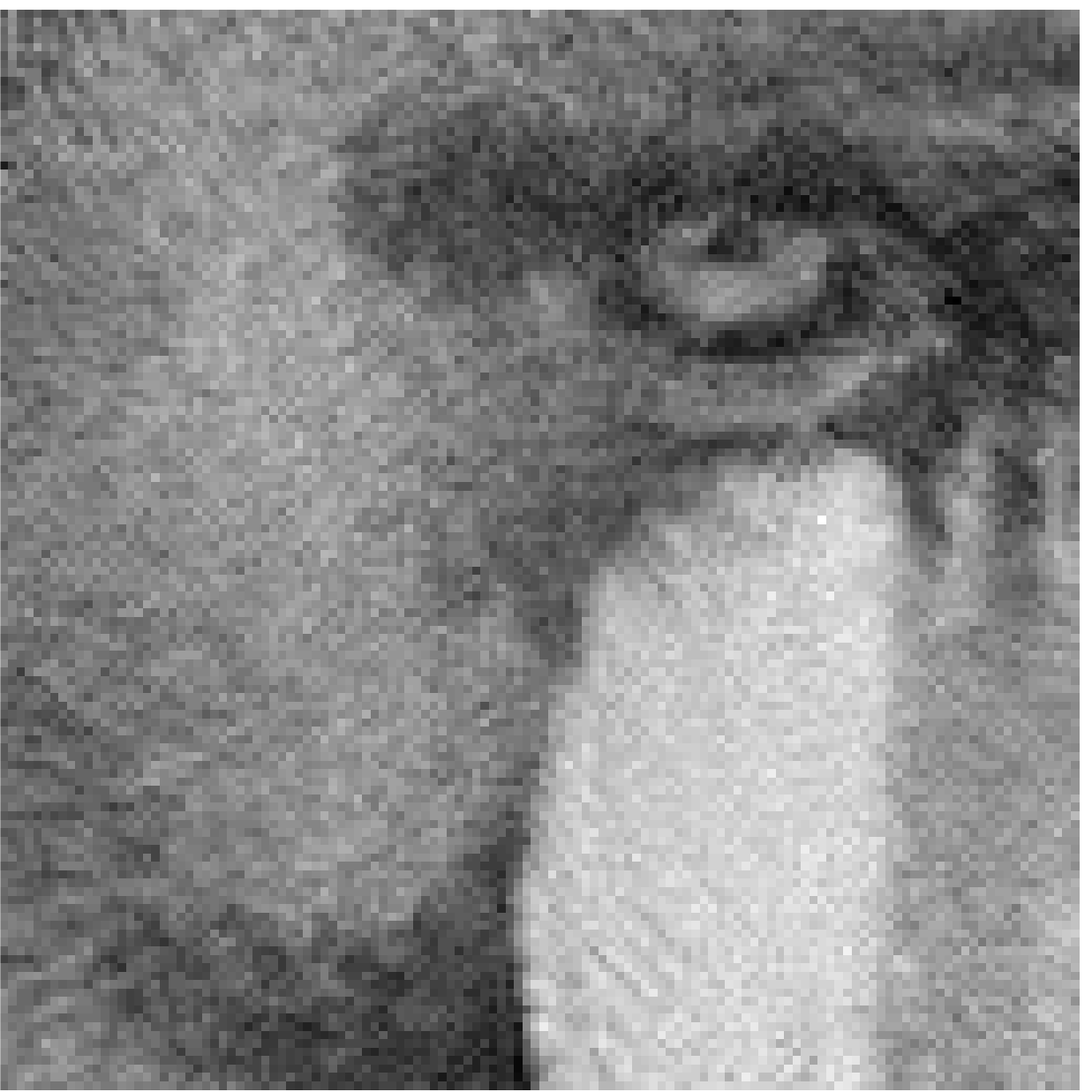}\hspace*{-0.25pc}  
    \includegraphics[width = 0.25\textwidth]{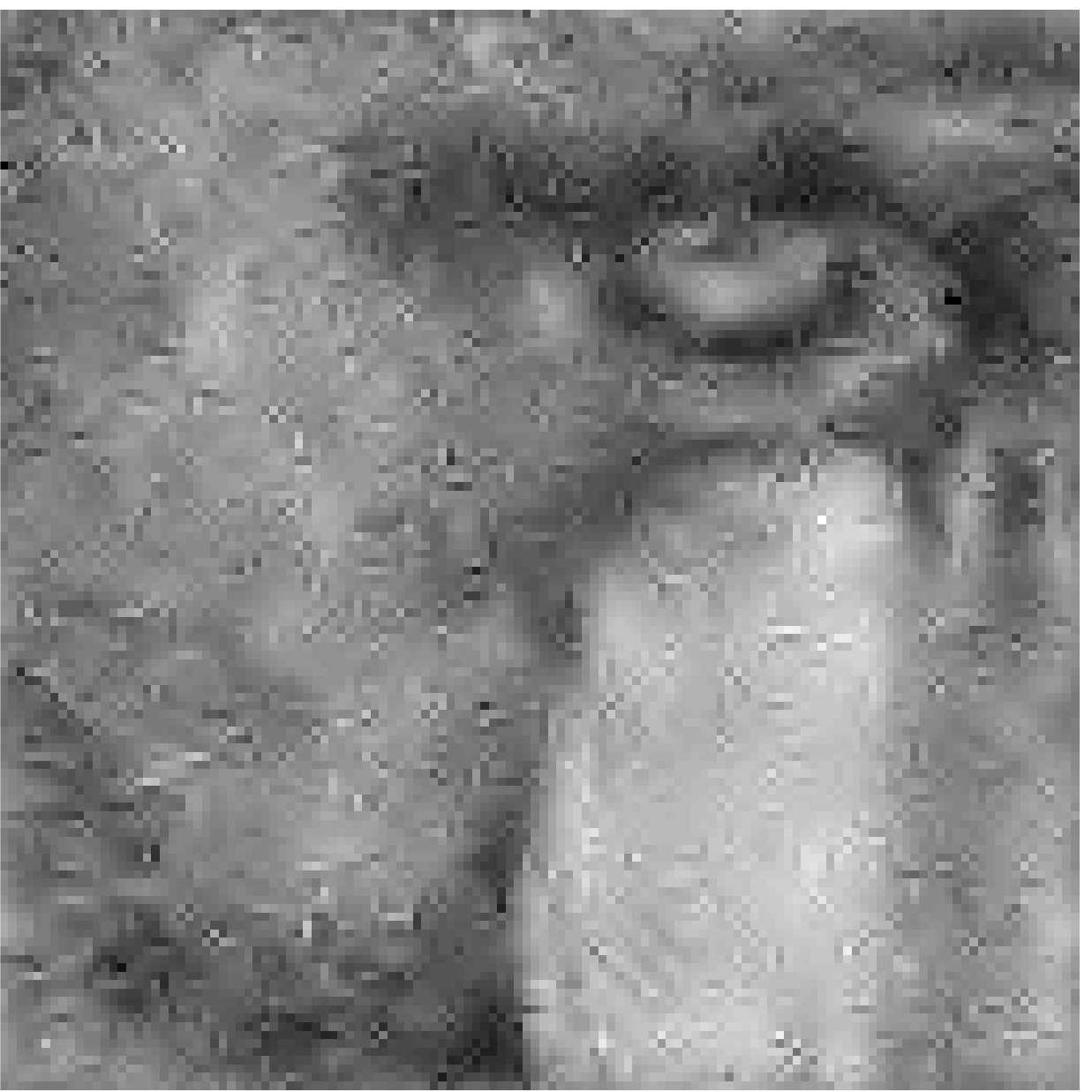}  
  }
  \centerline{\hfill Laplacian \hfill k-SVD \hfill  NL-means \hfill  TISWT \hfill}
  \caption{Noise level $\sigma = 60$. From left to right: our
    two-stage approach, k-SVD, nonlocal means, translation-invariant
    wavelet denoising.
    \label{denoised_img60}}
\end{figure}
\noindent of the eigenvectors of the Laplace-Beltrami operator when
the domain is perturbed \cite{Barbatis10}, or when the metric is
perturbed \cite{Barbatis96,Davies90}. To the best of our knowledge,
there appears to be little work on more precise analysis of the
perturbation of the eigenspaces of the graph Laplacian. \citet{Yan09}
acknowledge that standard bounds (e.g. \cite{Stewart90}) overestimate
the perturbations of the first eigenvector of the matrix $\bL$.
\section*{Acknowledgments}
FGM was partially supported by National Science Foundation Grants DMS 0941476,
ECS 0501578, and DOE award DE-SCOO04096. This work benefited from
fruitful  discussions with Raphy Coifman,  Peter Jones, and Arthur Szlam, 
during the 2009 IPAM Workshop on ``Laplacian Eigenvalues and
Eigenfunctions: Theory, Computation, Application''.

\end{document}